\documentclass[astrosymb, tighten, twocolumn]{aastex631}
\usepackage[utf8]{inputenc}
\usepackage{graphicx}
\usepackage{wrapfig}
\usepackage{amsmath}
\usepackage[capitalize]{cleveref}

\graphicspath{{./}{figures/}}

\usepackage{mathptmx,txfonts,tikz,bm}

\renewcommand{\edit}[2]{{\ifnum#1<5%
#2%
\else%
\textbf{#2}%
\fi}}

\begin{document}

\correspondingauthor{Christopher Lindsay}
\email{christopher.lindsay@yale.edu}

\title{Fossil Signatures of Main-sequence Convective Core Overshoot Estimated through Asteroseismic Analyses }

\author[0000-0001-8722-1436]{Christopher J. Lindsay}
\affiliation{Department of Astronomy, Yale University, PO Box 208101, New Haven, CT 06520-8101, USA}
\author[0000-0001-7664-648X]{J. M. Joel Ong}
\affiliation{Institute for Astronomy, University of Hawai‘i, 2680 Woodlawn Drive, Honolulu, HI 96822, USA}
\affiliation{Hubble Fellow}
\author[0000-0002-6163-3472]{Sarbani Basu}
\affiliation{Department of Astronomy, Yale University, PO Box 208101, New Haven, CT 06520-8101, USA}

\newcommand{\remark}[1]{{\color{red}{#1}}}

\shortauthors{Lindsay, Ong, \& Basu}
\shorttitle{Fossil Signatures of Convective Overshoot}

\begin{abstract}
Some physical processes that occur during a star’s main-sequence evolution also affect its post main-sequence evolution. It is well known that stars with masses above approximately 1.1 $M_{\odot}$ have well-mixed convective cores on the main sequence, however, the structure of the star in the neighborhood of the convective core regions is currently underconstrained. We use asteroseismology to study the properties of the stellar core, in particular, convective boundary mixing through convective overshoot, in such intermediate mass stars. These core regions are poorly constrained by the acoustic (p) mode oscillations observed for cool main sequence stars. Consequently, we seek fossil signatures of main sequence core properties during the subgiant and early first-ascent red giant phases of evolution. During these stages of stellar evolution, modes of mixed character that sample the deep interior, can be observed. These modes sample the regions of the stars that are affected by the main-sequence structure of these regions. We model the global and near-core properties of 62 \edit{1}{subgiant and early first-ascent red giant branch stars} observed by the \textit{Kepler}, K2, and TESS space missions. We find that the effective overshoot parameter, $\alpha_{\text{ov, eff}}$, increases from $M = 1.0M_{\odot}$ to $M = 1.2 M_{\odot}$ before flattening out, \edit{3}{although we note that the relationship between $\alpha_{\text{ov, eff}}$ and mass will depend on the incorporated modelling choices of internal physics and nuclear reaction network}. We also situate these results within existing studies of main-sequence convective core boundaries.

\end{abstract}




\section{Introduction} 
\label{sec:intro}

\begin{figure}
    \centering
    \includegraphics[width=0.45\textwidth]{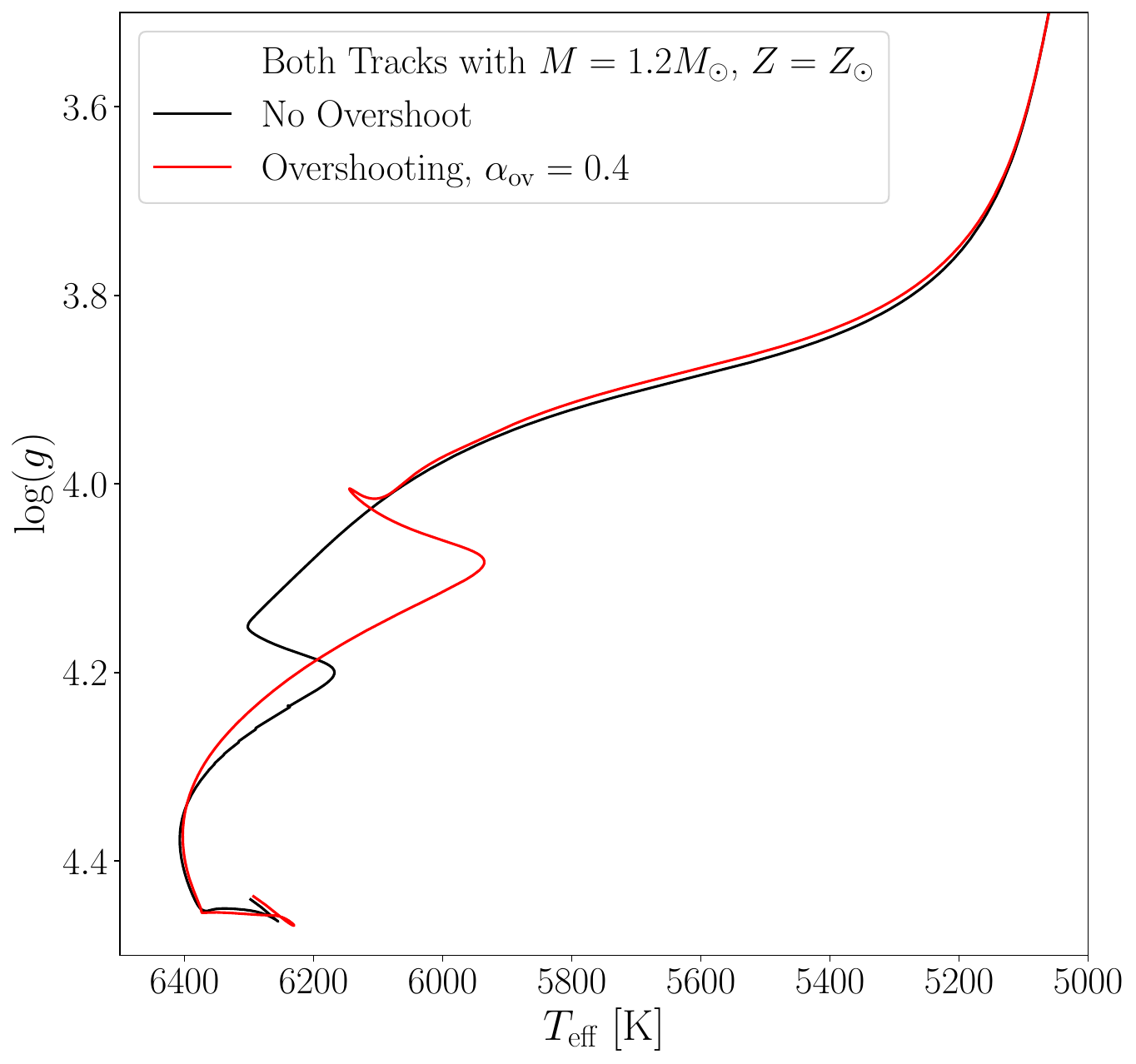}
    \caption{Two evolutionary tracks showing the effects of incorporating core overshooting into a stellar model. Both tracks show $M = 1.2 M_{\odot}$, $Z = Z_{\odot}$ models evolved using MESA version r12778 from the pre-main sequence to the red giant branch. The track with overshooting parameter $\alpha_{\text{ov}} = 0.4$, shown in red, shows that incorporating overshoot changes the evolution leading up the subgiant branch.}
    \label{fig:core_overshoot_effect}
\end{figure}

Physical processes occurring within a star during its main-sequence dictate its future evolutionary history. Stars with masses above about 1.1 $M_{\odot}$ host well-mixed convective cores, but the structure of the regions near the outside of the convective core is currently not well understood. Physical processes such as convective overshooting, the process by which parcels of convective fluid pass the classical convective boundary because of their momentum, are thought to extend the well-mixed convective cores past the classical boundary, usually defined by the Schwartzchild or Ledoux criterion. Determining how best to model overshooting in 1-D stellar evolutionary codes is important, since incorporating core overshoot into stellar models increases the amount of hydrogen available to a main-sequence star, thereby increasing its main-sequence lifetime and altering the star's evolution, see \autoref{fig:core_overshoot_effect}. A better understanding of this convective boundary mixing will better anchor our calibration of absolute ages in stellar modelling, in turn clarifying the ages of other astrophysical systems of interest, such as Milky Way progenitors \citep{Chaplin2020} and exoplanet hosts \citep{Huber2019}. 

\edit{1}{In addition to changing a star's internal structure and evolutionary history,} convective overshoot from the stellar core \edit{1}{also changes the observable global properties of stars, such as their Radius and Effective Temperature. Previous studies used observed global properties of stars and have shown that convective overshooting is} necessary in order for stellar models to reproduce observations of eclipsing binaries \citep[e.g.][]{Schroder1997, Pols1997, Ribas2000, Claret2007, ClaretandTorres2018, ClaretTorres2019, Claret2021, Constantino2018, Costa2019} and the color-magnitude diagrams of clusters \citep[e.g.][]{Maeder1981, Aparicio1990, Bertelli1992, Demarque1994, VandenBerg2006, Rosenfield2017}. Extra mixing beyond the convective core is also known to emerge from numerical hydrodynamics, although internal gravity waves also contribute to the convective boundary mixing \citep[e.g.][]{Higl2021}. See \citet{AndersPedersen2023} for a recent review of observational and theoretical constraints on mixing processes at the convective boundaries in main-sequence stars.  

Different investigations find that the amounts of core overshoot needed to match a star's observable quantities depend on global stellar properties. For example, \citet{ClaretTorres2016} studied a sample of eclipsing binary stars with observed masses, radii, temperatures, and elemental abundances and found that the size of the overshoot region has a positive dependence on the stellar mass. This result remains debated due to uncertainties in calibrating convective overshoot using eclipsing binaries \citep{Constantino2018, ClaretTorres2019} and the exact relationship between overshoot and stellar properties such as mass, metallicity, and evolutionary state remains uncertain. Even so, published grids of stellar evolution models also frequently make use of core overshooting prescriptions, typically scaling the amount of overshooting with stellar mass \citep[][]{Demarque2004, Pietrinferni2004, Bressan2012}. 

Asteroseismology, or the study of stellar oscillations, has provided the means to directly study the interiors of stars. The global oscillation properties of solar-like oscillators, the frequency of maximum oscillation power($\nu_{\text{max}}$), and the large frequency separation ($\Delta \nu$ ), have been used extensively to determine global stellar parameters \citep{Yu2018_16000}. However, they cannot be used to probe the deep stellar interiors we are interested in. Additionally, even modelling individual p-mode (pressure-mode) oscillation frequencies cannot give us detailed insights into the near-core structure of main-sequence stars with convective cores. In a previous paper \citep{Lindsay2023}, we have shown that pure p-modes are not able to sample the near-core regions directly, making inferences about the amount of core overshooting occurring in main-sequence stars difficult for some targets. Instead of analyzing the global asteroseismic properties ($\nu_{\text{max}}$ or $\Delta \nu$) or the p-mode oscillations of main-sequence stars, we study dipolar oscillation modes which exhibit mixed character, which only arise after a star leave the main sequence. 

After a star depletes its reserves of hydrogen in the core, its core begins to contract while its envelope begins to expand. At this stage of evolution, hydrogen is burned in a thin shell surrounding the now inert helium core. Since the convective motions in the core have also ceased by the time the star becomes a subgiant, the chemical composition gradient in these overshooting regions, having been frozen in at the main-sequence turnoff, serve as a fossil signature of the main sequence structure around the convective core. At the same time, the contraction of the stellar core is accompanied by an expansion of the envelope outside the hydrogen burning shell, leading to a large density contrast between the stellar core and envelope, allowing the study of these inner layers through mixed-mode asteroseismology \citep[see][for a review of evolved star asteroseismology]{2017A&ARv..25....1H}. Thus, studying the structure of subgiant \edit{1}{and early first-ascent red giant branch} stars can answer questions about main-sequence processes occurring above convective cores, since the main-sequence structural details of the star will be frozen into the \edit{1}{star's} inert core and the \edit{1}{star's} structure render their inspection observationally feasible through the asteroseismic analysis of mixed modes. \edit{1}{This idea of using mixed modes observed in evolved stars to study interior processes occurring during the main sequence was first proposed in \citet{Deheuvels2011}.} 

\begin{figure*}[!tbp]
\centering

\begin{minipage}{0.49\textwidth}
  \centering
  \includegraphics[width=\linewidth]{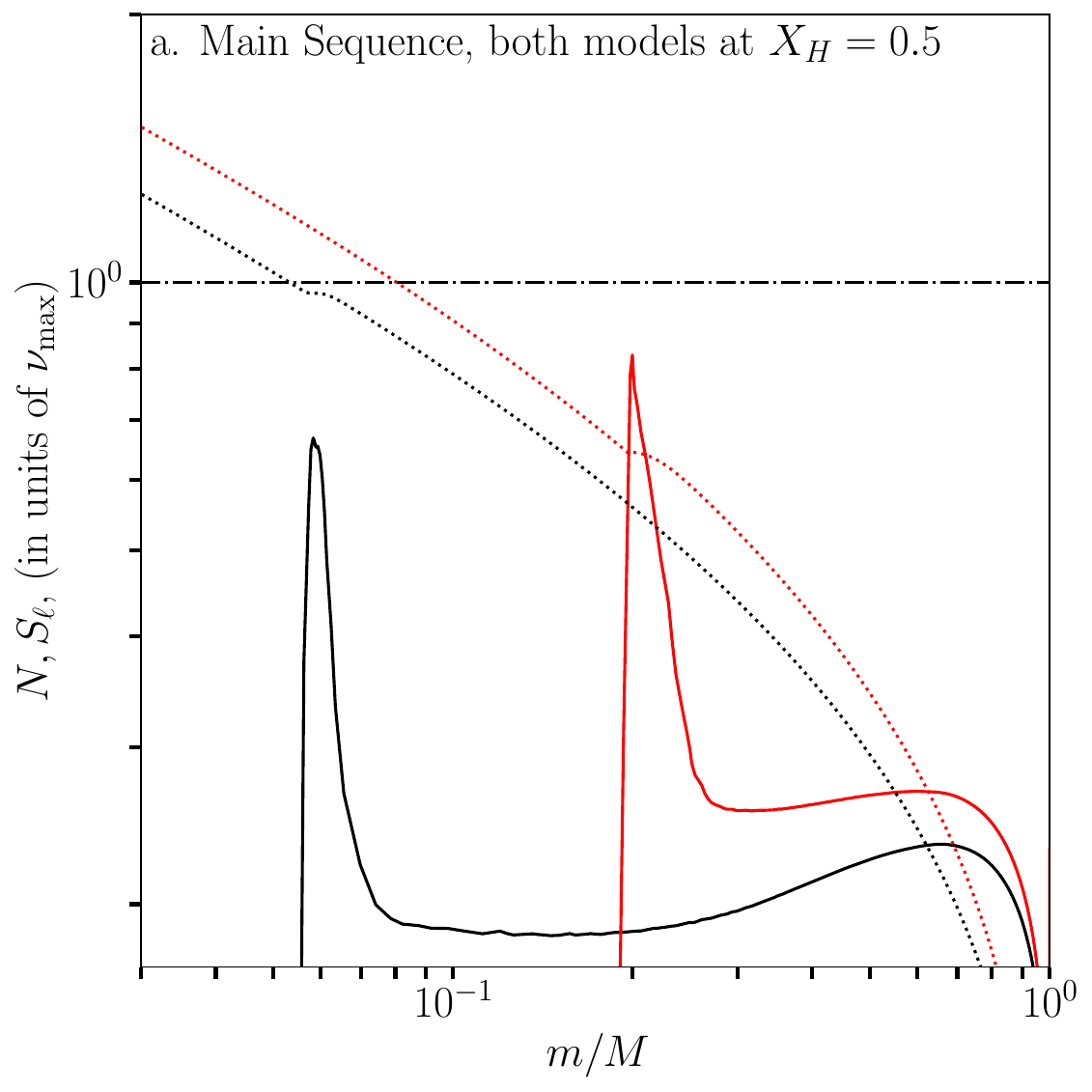}
\end{minipage}
\begin{minipage}{0.49\textwidth}
  \centering
  \includegraphics[width=\linewidth]{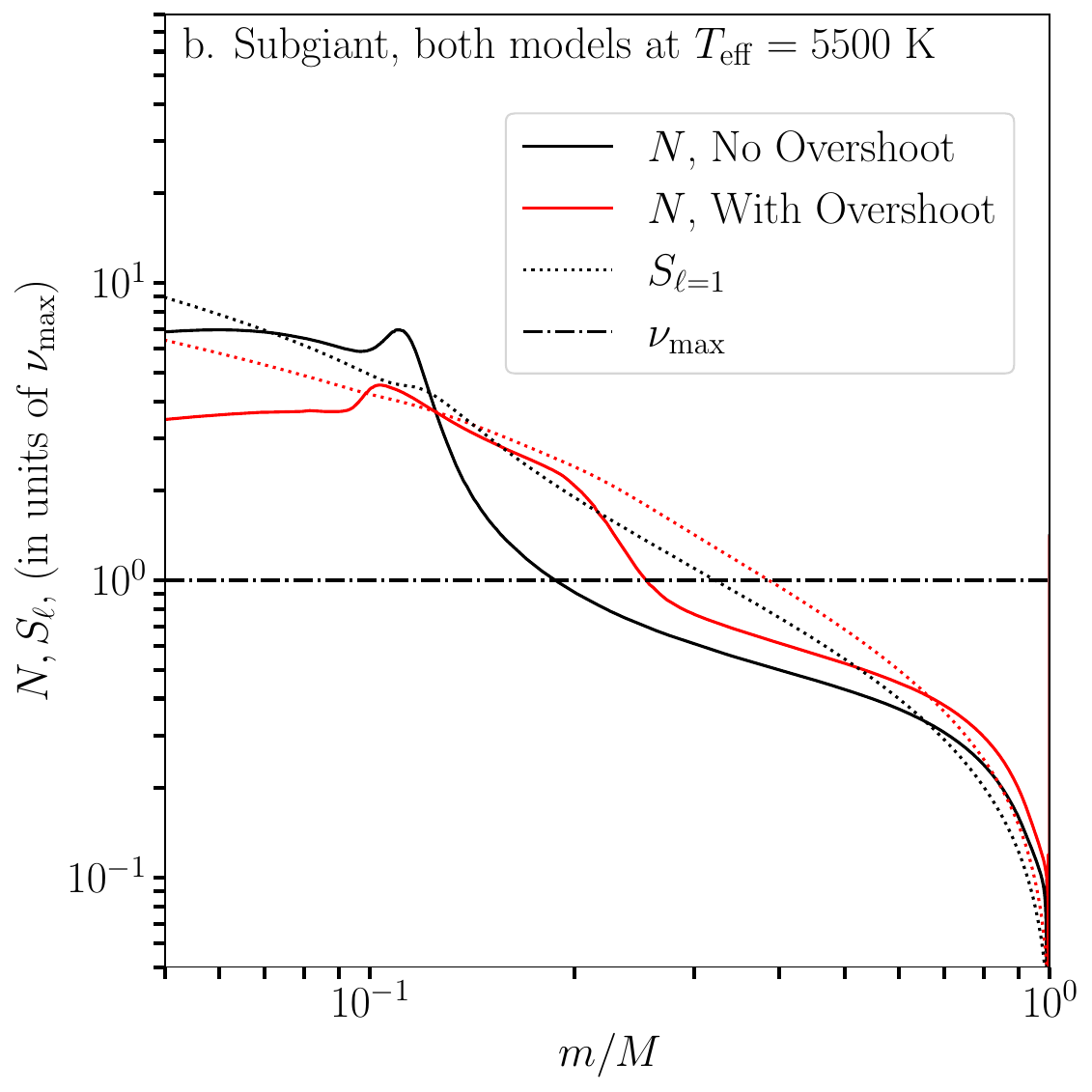}
\end{minipage}
\begin{minipage}{0.49\textwidth}
  \centering
  \includegraphics[width=\linewidth]{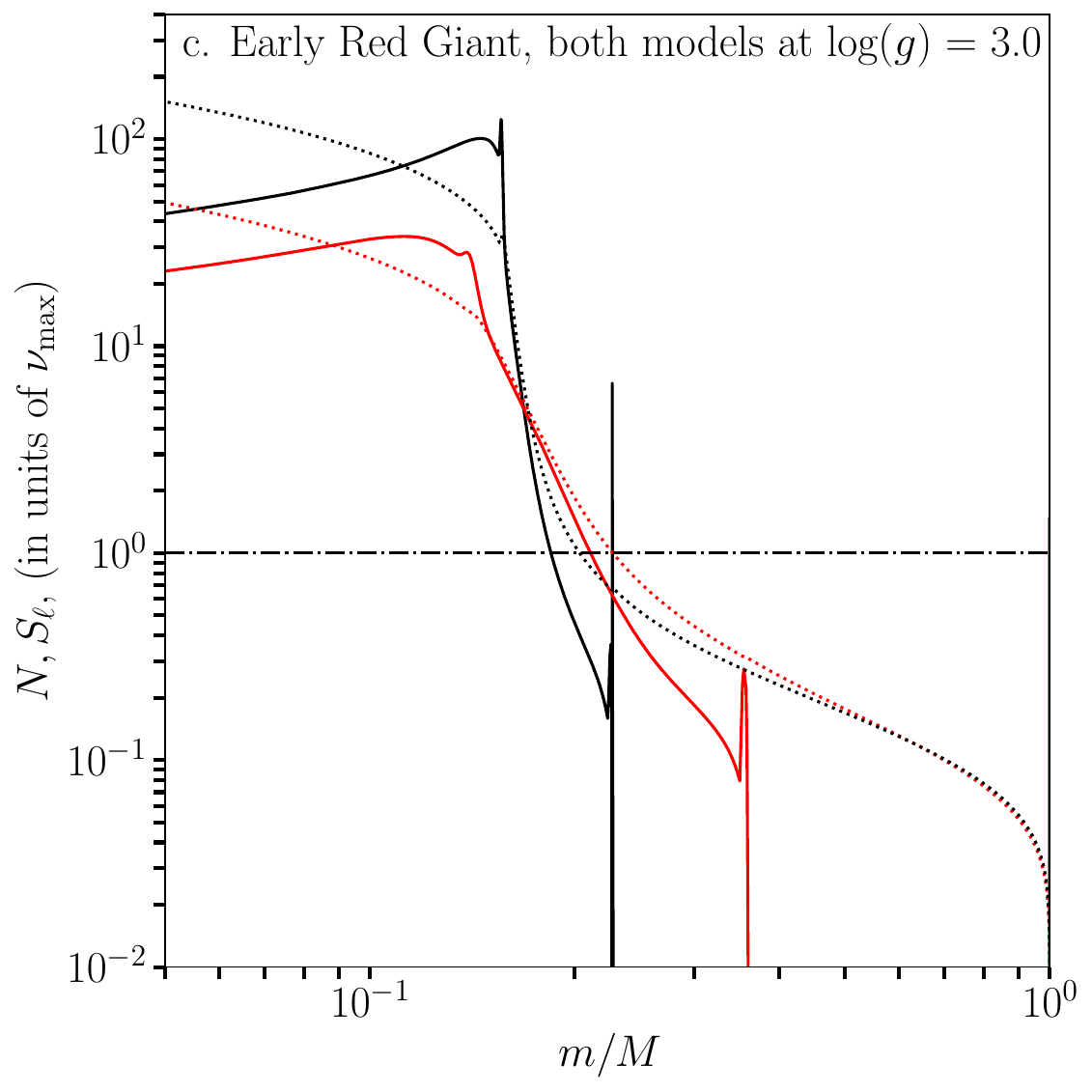}
\end{minipage}
\begin{minipage}{0.49\textwidth}
  \centering
  \includegraphics[width=\linewidth]{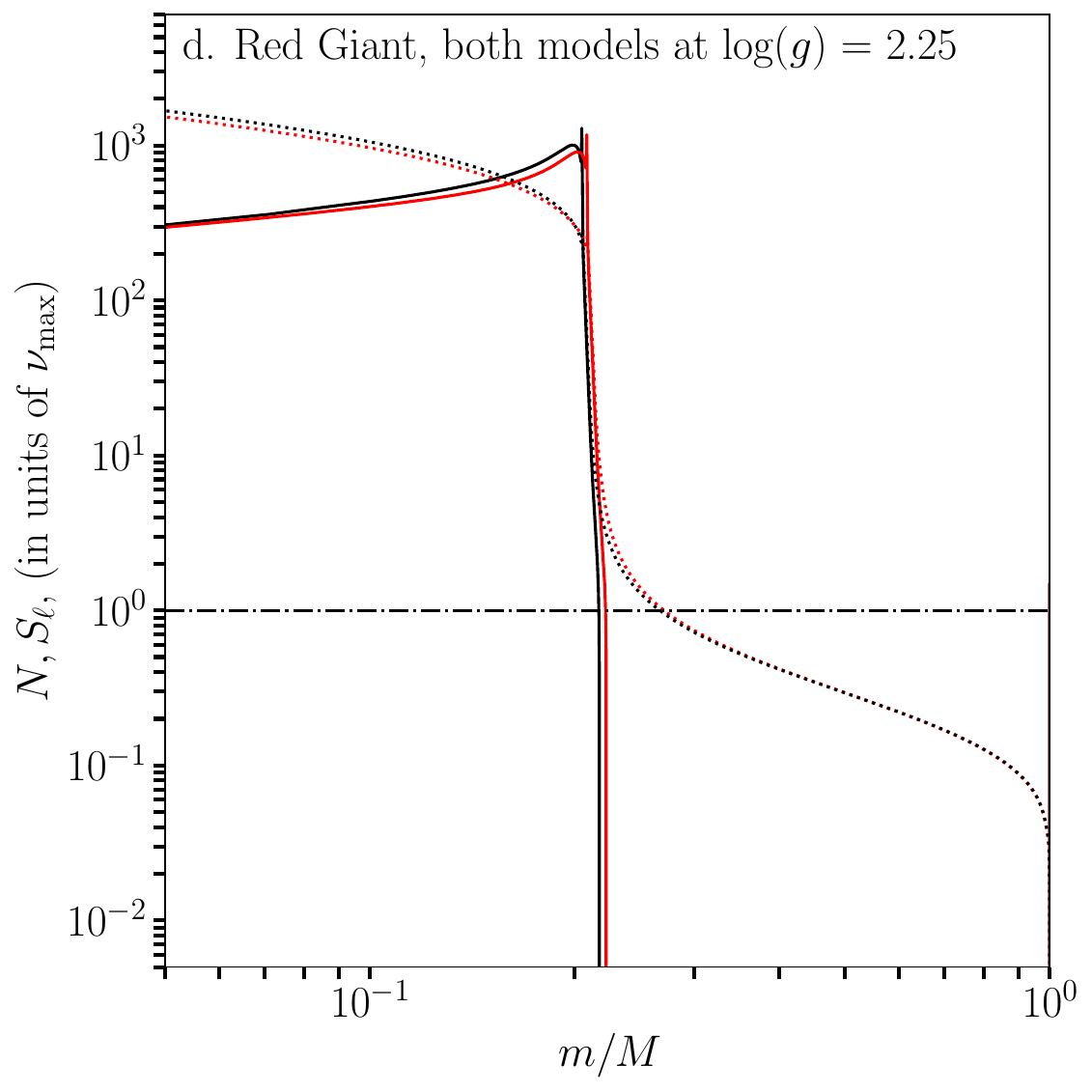}
\end{minipage}

\caption{Propagation diagrams showing \edit{1}{the interior structural evolution of} two 1.3$M_{\odot}$ stellar models with and without overshooting. \edit{1}{The main-sequence structure of the models (at a central hydrogen fraction of 0.5) are shown in panel a}. Both models' future subgiant structures \edit{1}{(at $T_{\text{eff}} = 5500$ K)} are show in \edit{1}{panel b}. \edit{1}{The models' structure on during their first-ascent red giant branch stage (log($g$) = 3.0) are shown in panel c, where there is still a significant difference between the stellar models with and without overshooting. The models' structure after the red giant branch bump (log($g$) = 2.25) is shown in panel d.} \edit{1}{In all panels,} the Brunt–Väisälä (buoyancy) frequency ($N$) is indicated with the solid lines, while the $\ell=1$ Lamb frequency ($S_{\ell = 1}$) is indicated with the dotted lines. The horizontal dot-dashed lines show the frequency of maximum oscillation power, $\nu_{\text{max}}$. The models incorporating convective core overshoot with overshoot parameter $\alpha_{\text{ov}} = 0.3$ are shown in red, while the models without overshoot are shown in black. Both models have the same initial metallicity ($[\mathrm{Fe/H}]_0 = -0.025$), initial helium abundance ($Y_0 = 0.284$), and mixing length parameter $\alpha_{\text{MLT}} = 1.75 $. }
\label{fig:propagation_evolution}
\end{figure*}

The effects of main-sequence core overshoot on the structural properties of subgiants can be seen in \autoref{fig:propagation_evolution}. \edit{1}{Panel a} of the \autoref{fig:propagation_evolution} shows the propagation diagrams of two main-sequence stellar models with the same mass and chemical compositions, but with different amounts of convective core overshoot. While significant difference can be seen in the deep, near-core layers, these are inaccessible to inspection using non-radial p-modes. After a main-sequence star evolves to a subgiant (\edit{1}{panel b} of \autoref{fig:propagation_evolution}), a large density contrast develops between the stellar core and envelope. \edit{1}{Some of the stars we study in this work are early red giant branch stars (see \autoref{fig:sample}) but, as shown in panel c of \autoref{fig:propagation_evolution}, structural differences between stellar models with and without core overshooting remain for stars with log($g$) values larger than 3.0, which is the case for all the stars in our sample. These differences don't remain forever, as after the stars evolve past the red giant branch luminosity bump, the structures of the stellar models with and without core overshooting are the same (panel d of \autoref{fig:propagation_evolution}).} 

\edit{1}{During the subgiant and red giant stages of evolution,} $\ell = 1$ oscillation modes with frequencies near $\nu_{\text{max}}$ may propagate in two different regions: the envelope, which supports pressure modes (p-modes where the restoring force is pressure, with frequencies $\nu \geq N \text{ and } S_{\ell = 1}$) and the core, which supports gravity modes (g-modes where the restoring force is buoyancy, with frequencies $\nu \leq N \text{ and } S_{\ell = 1}$). \edit{1}{Panels b, c, and d in \autoref{fig:propagation_evolution} show stellar model structures which may support $\ell = 1$ mixed modes with frequencies near $\nu_{\text{max}}$, with the p-mode region above the $N$ and $S_{\ell = 1}$ curves, and the g-mode region below the $N$ and $S_{\ell = 1}$ curves. } The observed mixed modes couple these two regions, with g-like character in the core and p-like character in the envelope \citep{Scuflaire1974, Aizenman1977}. \edit{1}{The observed stellar oscillations will depend strongly on the details of how the boundry between the core g-mode region and envelope p-mode region is configured, so we use} individual mixed-mode oscillations to investigate the properties and structures around stellar cores in this work.

Space based photometry missions such as CoRoT \citep{Baglin2006}, \textit{Kepler} \citep{Kepler_inst}, and \textit{TESS} \citep{TESS_inst} which observe many stars over long temporal baselines have made possible the observational detection of mixed modes in many stars possible. Since these mixed modes sample the interior layers of evolved stars near the core-envelope boundary, they can, and have been, used to constrain the amplitude of overshooting above convective cores. In particular, \citet{Deheuvels2011} found from the CoRoT data of the solar-like oscillator HD49385 that the oscillation spectrum of the star can only be properly explained by an avoided crossing (a characteristic feature of on-resonance mixed modes), whose shape constrains the amount of core overshooting above the stellar core \edit{1}{during the main sequence} to either very small or moderate values. \citet{Deheuvels2016} also made seismic estimates of the extent of convective cores in 8 low mass main sequence stars using data from \textit{Kepler}. \citet{Viani2020} examined 9 intermediate mass main-sequence stars from the \textit{Kepler} LEGACY sample \citep{Lund2017, SilvaAguirre2017} and found through asteroseismic modelling that the amplitude of convective overshoot from the main-sequence core increases with stellar mass. \citet{Noll2021} then studied the Kepler subgiant, KIC10273246, and found again that accounting for core overshooting improved their models' agreement with the observed oscillation mode frequencies. Building on these studies, we now make similar measurements from an analysis using a grid based modelling approach to determine the convective core boundary properties of a larger sample of subgiants \edit{1}{and early first-ascent red giant branch stars} observed by \textit{Kepler} and TESS. 

In this paper, we analyze a sample of 62 subgiants \edit{1}{and early first-ascent red giant branch stars} to determined how much convective core overshoot was present in these stars during their main sequence. Our objective in doing so is to determine relationships between global stellar properties and near core-mixing processes. \edit{1}{Incorporating convective overshooting also changes the global properties of stars, such as their effective temperature, so we combine the asteroseismic data for our sample of stars with spectroscopic observables avaliable from the literature.} The rest of this paper is organized as follows. We discuss our sample of subgiant stars in \autoref{sec:sample} and describe our grid of models in \autoref{sec:modelgrid}. In \autoref{sec:methods} we explain how we compare the observed spectroscopic and asteroseismic properties of the subgiant target stars to our model grid. In \autoref{sec:results} we show how the main-sequence core properties of the stars in our sample depend on the global stellar parameters and place our results into the context of other studies focused on determining the relationship between core overshoot properties and stellar parameters. 

\section{The Sample}
\label{sec:sample}
\begin{figure}
    \centering
    \includegraphics[width=0.45\textwidth]{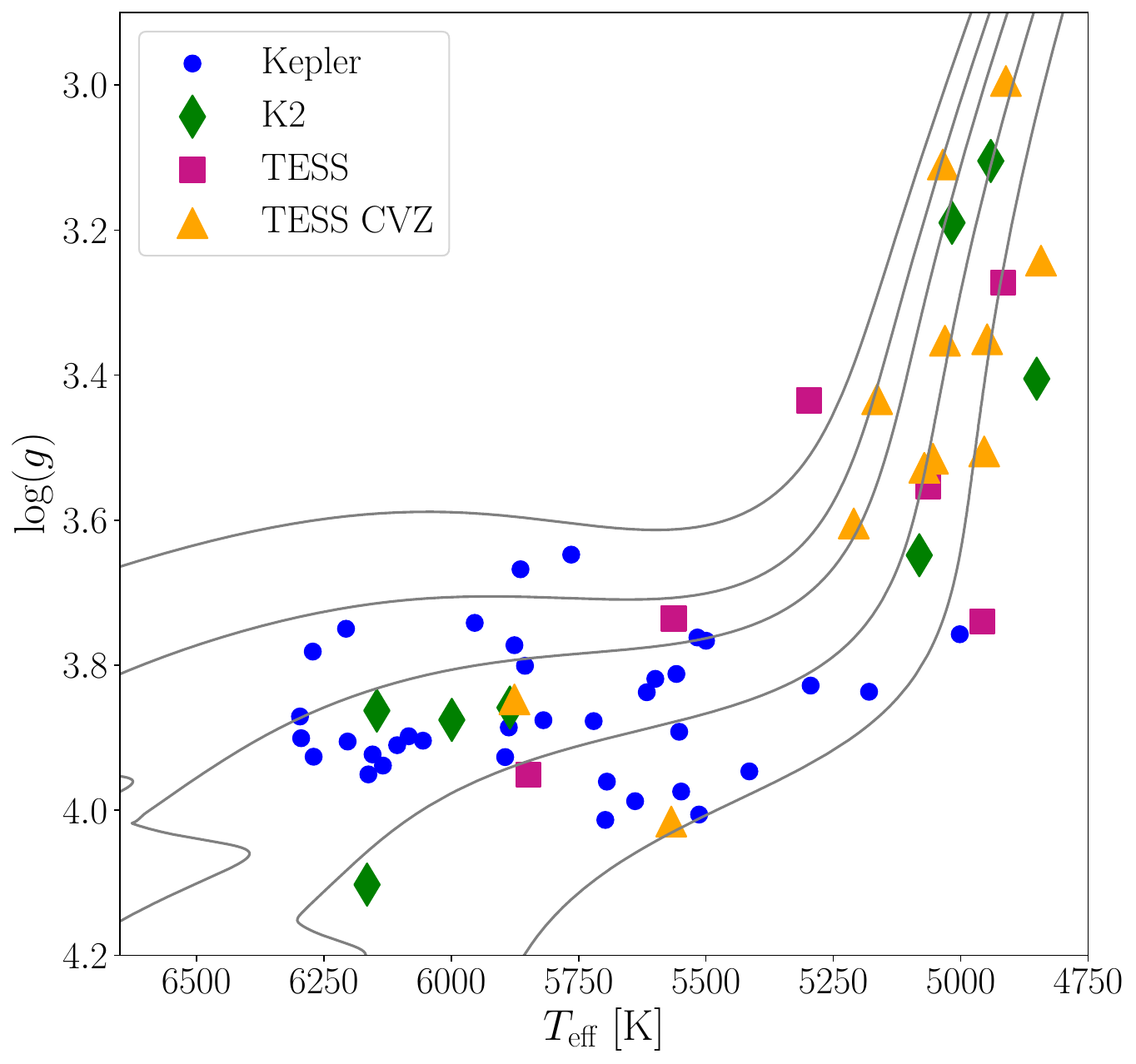}
    \caption{\edit{1}{Our sample of subgiants and early first-ascent red giant branch} stars \edit{1}{studied} in this work \edit{1}{shown} on a Kiel ($\log({g})$ vs. $T_{\rm eff}$) diagram. The different symbols refer to which mission observed the \edit{1}{star}. The background gray curves show solar-abundance  evolutionary tracks with masses between 1.0 and 1.8 $M_{\odot}$ in steps of $0.2 M_{\odot}$. The tracks are for modes without core overshoot.}
    \label{fig:sample}
\end{figure}

We study a total of 62 stars in this work, 36 observed during the \textit{Kepler} mission, 8 observed during the K2 mission, and 18 observed with TESS, of which 12 were observed in the TESS Southern Continuous Viewing Zone (CVZ) during its first year of operations. \edit{1}{We selected our sample of stars to include subgiants and early first-ascent red giant branch stars whose frequency échelle diagrams display avoided crossings, indicating the presence of mixed-character modes whose frequencies sample the core/envelope boundary.} \autoref{fig:sample} shows a Kiel diagram ($\log({g})$ vs. Effective Temperature) of our sample. The different missions observed the stars for varying lengths of time, consequently, the quality of the power spectra used to determine the oscillation frequencies varies tremendously.

\subsection{Kepler Stars}
The nominal \textit{Kepler} mission ran for just over 4 years, giving the best possible scenario for determining the asteroseismic mode frequencies we require for our fitting procedure. For this work, like in \citet{Ong2021c}, we study a sample of stars observed with short cadence which was already examined using a grid-based modeling approach \citep{LiT2020, LiY2020}. The mode frequencies used in our analysis were measured in \citet{LiY2020}. The global asteroseismic parameters, $\Delta \nu$ and $\nu_{\text{max}}$, for these targets were derived in \citet{Serenelli2017}, while the spectroscopic observables $T_{\text{eff}}$ and [$\mathrm{Fe/H}$] were taken from Table 1 of \citet{LiT2020}. When available, we also used stellar luminosity ($L$) measurements derived from the Gaia mission \citet{GaiaDR2}.

\subsection{K2 Stars}
The time series photometry available from the K2 mission are only about 75 days long in each campaign \citep{K2mission}, with 60 second sampling at short cadence. This results in significantly degraded frequency resolution, and, since the photometric noise was also higher due to decreased pointing stability, only 8 K2 subgiants and early first-ascent red giant branch stars studied in \citet{Ong2021c} showed significant oscillation power excess. The oscillation frequencies and spectroscopic properties for these stars are the same as those used in \citet{Ong2021c} (see section 3.3), although the mode frequencies for some of these stars were reanalyzed in \citet{Gonzalez-Cuesta_2023}.

\subsection{TESS Stars}
During its nominal mission, stars observed by TESS can, in the worst case, be observed for only a single sector (27 days) which is not much longer than the average oscillation mode lifetime. Therefore, in the frequency domain, the limited spectral resolution of the power spectrum is comparable to the mode line widths. In addition, the TESS pixels are much larger than Kepler's meaning the oscillation mode frequencies derived from TESS photometry are more sensitive to noise and suffer from more contamination when compared with \textit{Kepler}-derived mode frequencies. Since the detection and extraction of oscillation mode frequencies is not the focus of this work, our TESS sample of stars is relatively small and made from studies already communicated through the asteroseismology community. The spectroscopic and asteroseismic observables of our sample of 6 stars observed by the nominal TESS mission ($\beta$ Hyi, $\delta$ Eri, $\eta$ Cep, $\nu$ Ind, TOI 197, and HD 38529) is further described in section 3.2 of \citet{Ong2021c}.

In addition to these 6 stars, our sample includes 12 \edit{1}{subgiants and early first-ascent red giant branch stars} observed by TESS in its Southern Continuous Viewing Zone. The oscillation mode frequencies for these stars were fitted against the power spectra using 7 different peakbagging pipelines (J. M. J. Ong, in preparation). The spectroscopic properties ($T_{\text{eff}}$ and [$\mathrm{Fe/H}$]) were taken from the 17th data release of the Sloan Digital Sky Survey \citep{SloanDR17}, while luminosity measurements were found with SED measurements in conjunction with GAIA parallaxes. 


\section{Our Grid of Models}
\label{sec:modelgrid}
There are two main classes of techniques to model the interiors of \edit{1}{subgiants and early first-ascent red giant branch stars} using asteroseismic and spectroscopic data. One that uses large scale grid searches to study many targets \citep[as in ][]{McKeever2019, Jorgensen2020, LiT2020, Nsamba2021, Ong2021a, Ong2021c}, and another, where the parameter space of the grid is adapted to each target individually \citep[as in ][]{Ball2018, Ball2020, Huber2019, Chaplin2020, Noll2021}. These two approaches are often combined, with the results of coarser grid searches used to restrict the parameter space for further individual study. While boutique modelling involving optimization based parameter searches or individual dense grids created separately for each target is a good method for determining the properties of individual stars, it can be very slow and computationally expensive. Instead, we used a large scale grid search method for this study. 

\begin{figure}
    \centering
    \includegraphics[width=.45\textwidth]{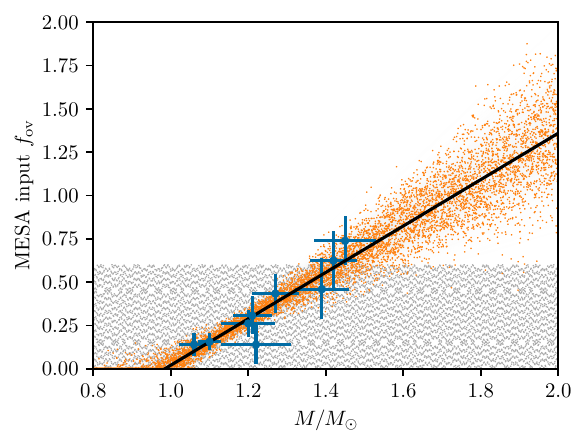}
    \caption{Comparison of sampling of the mass-overshoot plane in the model grid used in this work (gray points --- Sobol sampling in a finite range) against that of the grid used in \cite{Ong2021c} (orange points --- random sampling over the mass-overshoot relation of \citealt{Viani2020}, shown with the blue points).}
    \label{fig:grid-mass-overshoot}
\end{figure}

The grid we use in our fitting procedure is largely based on that used in \cite{Ong2021c}. It consists of \edit{1}{stellar} models created with MESA version r12778 \citep{Paxton2011, Paxton2013, Paxton2015, Paxton2018, Paxton2019}. These models use the solar chemical abundance mixture detailed in \citet{GS98}, an Eddington gray atmospheric boundary condition, and the mixing-length prescription of \citet{CoxGiuli1968}. Element diffusion was handled using the formulation of \citet{Thoul1994}, and we included a mass-dependent scaling prefactor \citep[see][]{Viani_2018}. The parameters of the models in the grid, Sobol-sampled in almost an identical fashion to that used in \cite{Ong2021c}, were the initial stellar mass ($M\in[0.8 M_{\odot}, 2.0 M_{\odot}$), the mixing length parameter ($\alpha_{\text{mlt}} \in [1.3, 2.2]$), initial metallicity ($[\mathrm{Fe/H}]_0 \in [-1, 0.5])$, and initial helium abundance ($Y_0 \in [0.176, 0.32]$). We construct models with sub-primordial ($Y_{\text{primordial}} \sim 0.248$) to avoid edge-effects from our grid's construction affecting the fitting results, but we down-weight the likelihoods of stellar models with, $Y_0<Y_{\text{primordial}}$, as we explain in \autoref{sec:methods}.

Convective overshoot from the stellar core was incorporated according to MESA’s implementation of overmixing with a step profile \citep[cf. §2 of][]{Lindsay2022}. Unlike the grid used in \cite{Ong2021c}, we Sobol-sample the step overmixing parameter, $\alpha_{\text{ov}}$, between $0$ and $0.6$, as shown in \autoref{fig:grid-mass-overshoot}. The result of MESA's implementation of step overmixing is that the well-mixed stellar core extends beyond the Schwarzschild boundary by a distance $r_{\text{ov}}$, given by

\begin{equation}
r_{\text{ov}} = \left\{
        \begin{array}{ll}
            \alpha_{\text{ov}} H_p & \quad \text{if } H_p \leq r_{\text{cz}} \\
            \alpha_{\text{ov}} r_{\text{cz}} &  \quad \text{if } H_p > r_{\text{cz}}
        \end{array}
    \right. ,
\end{equation}
where $H_p$ is the pressure scale height at the convective boundary and $r_{\text{cz}}$ is the radius of the core convection zone. This ensures that  when the convective core is very small, the overshooting region does not become unphysically large, as would be the case for overshooting by  $\alpha_{\text{ov}}H_p$. Another feature of the MESA implementation of step overmixing is that the overshoot region does not start exactly at the Schwarzschild boundary (estimated by where $\nabla_{\text{radiative}} = \nabla_{\text{adiabatic}}$) but rather begins at a location $f_0  H_p$ into the convective core from the Schwarzschild boundary. This is because the mixing coefficient approaches 0 at the Schwarzschild boundary. In our grid, $f_0$ is set to 0.005. In order to avoid confusion when referencing the step overshooting parameter, and to make it easier to compare our results with studies that used other stellar evolutionary codes, we save the \edit{1}{two} different overshoot parameters in our grid.

\begin{enumerate}
    \item Input Overshoot, $\alpha_{\text{ov}}$: This is the value entered into the MESA controls inlist when creating the stellar model tracks. We varied this parameter in our grid from 0 to 0.6.  
    \item Effective Overshoot, $\alpha_{\text{ov, eff}}$: The effective overshoot parameter is calculated directly from the MESA-generated profile file for the time step at which the MESA-defined convective core mass was at its maximum. We define the effective overshoot parameter as $\alpha_{\text{ov, eff}} = (r_{\text{well-mixed}} - r_{\text{cz}})/H_p$ where $r_{\text{well-mixed}}$ is the well-mixed core boundary, defined by looking for the central-most grid point where the gradient of the mean molecular weight gradient, $\frac{\partial \nabla_\mu}{\partial r}$, spikes to at least $|\frac{\partial \nabla_\mu}{\partial r}| = 0.001 \text{cm}^{-1}$. 
\end{enumerate}

For the subsequent analysis of the effective overshoot parameter, $\alpha_{\text{ov, eff}}$ is set to 0 in our grid if the corresponding model tracks maintain a convective core for less than \edit{1}{30\% of their main-sequence lifetime.} This cutoff ensures that the overshoot parameters for models which do not maintain a convective core \edit{1}{for a significant amount of time} is 0. \edit{1}{The 30\% limit was chosen to ensure that models with varying initial compositions, but with masses $\lesssim 1.0 M_{\odot}$, all maintain $\alpha_{\text{ov, eff}} = 0$ (see \autoref{fig:Lindsay_mass_overshoot}).} 

\edit{1}{The models we use in this work use MESA's `basic.net' nuclear network, which considers lithium and beryllium to be at equilibrium and therefore is known to overestimate the modelled size of the convective during the main sequence, when compared with a stellar model calculated using a full nuclear network \citep[see][]{NollDeheuvels2023}. We tested how using a full nuclear network (MESA's `h\_burn.net') would affect our calculation of the effective overshoot parameter ($\alpha_{\text{ov, eff}}$) for a given model track and found that since $\alpha_{\text{ov, eff}}$ is calculated using the difference between the well-mixed core boundary and the convection zone boundary, and both $r_{\text{well-mixed}}$ and $r_{\text{cz}}$ are \edit{2}{decreased} by the usage of a full nuclear network, our determination of $\alpha_{\text{ov, eff}}$ for a given model track is only modestly altered by the choice of nuclear networks.} \edit{3}{The choice of nuclear network changes the overall size of the convective core, since with `basic.net' the core abundance of lithium and beryllium is assumed to be at equilibrium, which in turn means the overall core energy production from the proton-proton chain could be incorrect. Thus, the choice of network will change the oscillation frequencies of the stellar models, and may therefore alter the inferred stellar parameters obtained through asteroseismic modelling.}

The grid input parameters which we vary, $M$, $\alpha_{\text{mlt}}$, $[\mathrm{Fe/H}]_0$, $Y_0$, and $\alpha_{\text{ov}}$, are distributed uniformly using joint Sobol sequences of length 16382 over the ranges described previously. \edit{1}{Since each of the 16382 tracks we calculate has a slightly different combination of input parameter values, we are able to obtain more precise estimates of our output parameters, when compared to if we used an equi-sampled grid.} Each of these tracks were evolved using MESA from the pre-main sequence until the point where $\Delta \nu = 9 \mu\text{Hz}$ \citep[following][]{Ong2021c}. For each model in our grid, the oscillation mode frequencies are calculated using GYRE version 6.0 \citep{Townsend2013}. The radial and quadrupole ($\ell = 0$ and $\ell = 2$) p-mode frequencies are calculated within $\pm 6 \Delta \nu$ of $\nu_{\text{max}}$. For the dipole ($\ell = 1$ modes, we calculated both $\pi$-mode and $\gamma$-mode frequencies and mixed-mode coupling matrices according to the mode isolation construction of \citet{OngBasu2020}. $\pi$-modes refer to the pure p-modes whose frequencies only depend on the p-mode cavity of the stellar model, while $\gamma$-modes refer to the pure g-modes whose frequencies only depend on the g-mode cavity; mixed modes are those linear combinations of $\pi$ and $\gamma$ modes that are also eigenfunctions of the wave operator. Following \citet{Ong2021c}, we compute $\gamma$-mode frequencies and matrix elements for $\gamma$ modes from a lower-bound frequency of $\nu_{\text{max}} - 7 \Delta \nu$ up to the $n_g$ = 1 $\gamma$-mode. The $\ell = 0 $ and $\ell = 2$ modes, as well as the $\ell = 1$ $\pi-$ modes were corrected for inadequate modelling of the near surface layers (surface effects or surface term), before comparison with the observed mode frequencies, as discussed further in \autoref{sec:methods}.

\section{Modelling Procedure}
\label{sec:methods}

Many grid-based pipelines have been used to analyze asteroseismic data from spaced based missions, including those described in \citet{Basu2010, Gai2011, Campante2019, Stello2022}, among others. \citet{Cunha2021} compared different asteroseismic pipelines using artificial asteroseismic data. 

For this work, we start with a given set of observables and corresponding uncertainties for a target star. In this work, these include the spectroscopic parameters effective temperature ($T_{\text{eff}}$), metallicity ([$\mathrm{Fe/H}$]), and when available, Luminosity ($L$). These variables are used to define the spectroscopic likelihood. The global asteroseismic observables, $\Delta \nu$ and $\nu_{\text{max}}$, are also used just in the down-selection of the large grid. 

\subsection{Down-selecting the Complete Grid}
The first step in our modelling procedure is to search for all models in the complete grid with properties close to the target star's observed values of $T_{\text{eff}}$, $[\mathrm{Fe/H}]$, $\Delta \nu$, and $\nu_{\text{max}}$. We calculate a penalty function based on individual $\chi^2$ values for each observed parameter, $P$, given by,

\begin{equation}
    \label{eq:chi2}
    \chi_{P}^2 = \frac{(P_{\text{obs}} - P_{\text{model}})^2}{\sigma_{P_{\text{obs}}}^2},
\end{equation}

and the total $\chi^2$ is given by 
\begin{equation}
    \chi_{\text{downselect}}^2 = \chi_{T_{\text{eff}}}^2 + \chi_{[\mathrm{Fe/H}]}^2 + \chi_{\Delta \nu}^2 + \chi_{\nu_{\text{max}}}^2.
\end{equation}
If luminosity is available for the target star, we include $\chi_{L}^2$ in the calculation of $\chi_{\text{downselect}}$. We down-select the complete grid to models satisfying $\chi_{\text{downselect}}^2 < 10^2$. This greatly reduces computational cost by restricting the number of models requiring an expensive seismic likelihood evaluation from more than 14 million (the whole grid) to around ten thousand models per target star. 

\subsection{Spectroscopic Likelihoods}
To quantify how well the models in the down-selected grid matches the spectroscopic observables of the target star, we calculate the following spectroscopic cost function, 
\begin{equation}
    \label{eq:spec_chi2}
    \chi^2_{\text{spec}} = \frac{1}{3} \bigg(\chi_{T_{\text{eff}}}^2 +  \chi_{[\mathrm{Fe/H}]}^2 + \chi_{L}^2 \bigg)
\end{equation}

following \cref{eq:chi2}. For the 6 target stars (KIC6442183, KIC11137075, KIC11414712, $\delta$ Eri, EPIC212478598, and EPIC246305274) for which we could not find reliable luminosity measurements, we omitted $\chi_{L}^2$ from the calculation of $\chi^2_{spec}$, obtaining
\begin{equation}
    \label{eq:spec_chi2_no_L}
    \chi^2_{\text{spec, no $L$}} = \frac{1}{2} \bigg(\chi_{T_{\text{eff}}}^2 +  \chi_{[\mathrm{Fe/H}]}^2 \bigg)
\end{equation}

\edit{1}{We divide $\chi^2_{\text{spec}}$ by the number of spectroscopic parameters we use in order to ensure that the combination of spectroscopic constraints are weighted the same as the asteroseismic constraints, described in the next subsection.}

The spectroscopic likelihood for every model in the down-selected grid is finally calculated as 
\begin{equation}
    \label{eq:spec_like}
    \mathcal{L}_{\text{spec}} = \exp{\left(-\frac{\chi^2_{\text{spec}}}{2}\right)}.
\end{equation}


\subsection{Seismic Likelihoods}
In order to calculate a seismic cost function, we must quantify how well a stellar model's oscillation mode frequencies match the observed frequencies of a given target. To do this, the model's set of oscillation modes must be compared to the set of observed modes, which entails matching each of the observed oscillation mode frequencies to a corresponding model mode frequency. The model modes also must be corrected for the surface effect, a frequency-dependent error in stellar model oscillation mode frequencies caused by our inability to model the near-surface layers of a star in one dimension. 

We match the model oscillation mode frequencies to the observed frequencies in two different ways, depending on the angular degree ($\ell$) of the mode. The radial and quadrupole modes ($\ell = 0$ and 2) are matched based on their inferred values of radial order $n_p$. We infer the $n_p$ values of the observed modes based on the observed mode frequency, the observed $\Delta \nu$ as $n_{\text{p, obs}} = (\nu_{\text{obs}}/\Delta_\nu) - (\ell /2) $. The $n_p$ values for a models' oscillation frequencies are returned from GYRE \citep{Townsend2013} along with the mode frequencies. The matched $\ell = 0$ and 2 modes are then used to determine the coefficients of the two-term surface term from \citet{BallGizon2014} by minimizing the quantity,
\begin{equation}
    \label{eq:minimize_this}
    \sum_{\ell\in \{0,2\}}\sum_{n_p=0}^N \frac{\nu_{\text{obs, n}_p\ell} - (\nu_{\text{model, n}_p\ell} + \delta \nu_{\text{surf, n}_p\ell})}{\sigma_{\nu, \text{obs, n}_p\ell}}
\end{equation}
where $n$ denotes the radial order of the mode, $\ell$ denotes the angular degree of the mode, $N$ is the total number of modes, $\nu_{\text{obs, n}\ell}$ is the observed mode frequency,  $\sigma_{\nu, \text{obs, n}\ell}$ is that mode's associated frequency error, $\nu_{\text{model, n}\ell}$ is the uncorrected model mode frequency, while $\delta \nu_{\text{surf, n}\ell}$ is the two term parametric correction of \citet{BallGizon2014}. This correction is given by, 
\begin{equation}
    \label{eq:surface_term}
    \delta \nu_{\text{surf, n}_p\ell} = \frac{1}{I_{\text{n}_p\ell}} \left(a_{-1} \bigg[\frac{\nu_{\text{n}_p \ell}}{\nu_0}\bigg]^{-1} + a_{3} \bigg(\frac{\nu_{\text{n}_p \ell}}{\nu_0}\bigg)^{3} \right), 
\end{equation}
where $\nu_{\text{n} \ell}$ is the model mode frequency, $I_{\text{n}\ell}$ is the model mode inertia, $\nu_0$ is set to $\nu_{\text{max}}$, and the coefficients ($a_{-1}$ and $a_{3}$) are chosen to minimize the quantity in \cref{eq:minimize_this}. 

The above procedure gives us surface corrected model mode frequencies for  $\ell = 0$ and $\ell = 2$ modes, and we can now define $\nu_{\text{model, corr}} = \nu_{\text{model, uncorr}} + \delta \nu_{\text{surf}}$. For the $\ell = 1$ modes though, following \citet{Ong2021b}, we apply the surface correction to the dipole $\pi$ modes, and recover surface-corrected model mixed modes by coupling these surface corrected $\pi$ modes to the $\gamma$ modes, which remain unaffected by surface effects \citep{OngBasu2020}. Surface-corrected dipole ($\ell = 1$) mode frequencies are matched to the observed mode frequencies using an iterative nearest-neighbor search.


\begin{figure}
    \centering
    \includegraphics[width=0.45\textwidth]{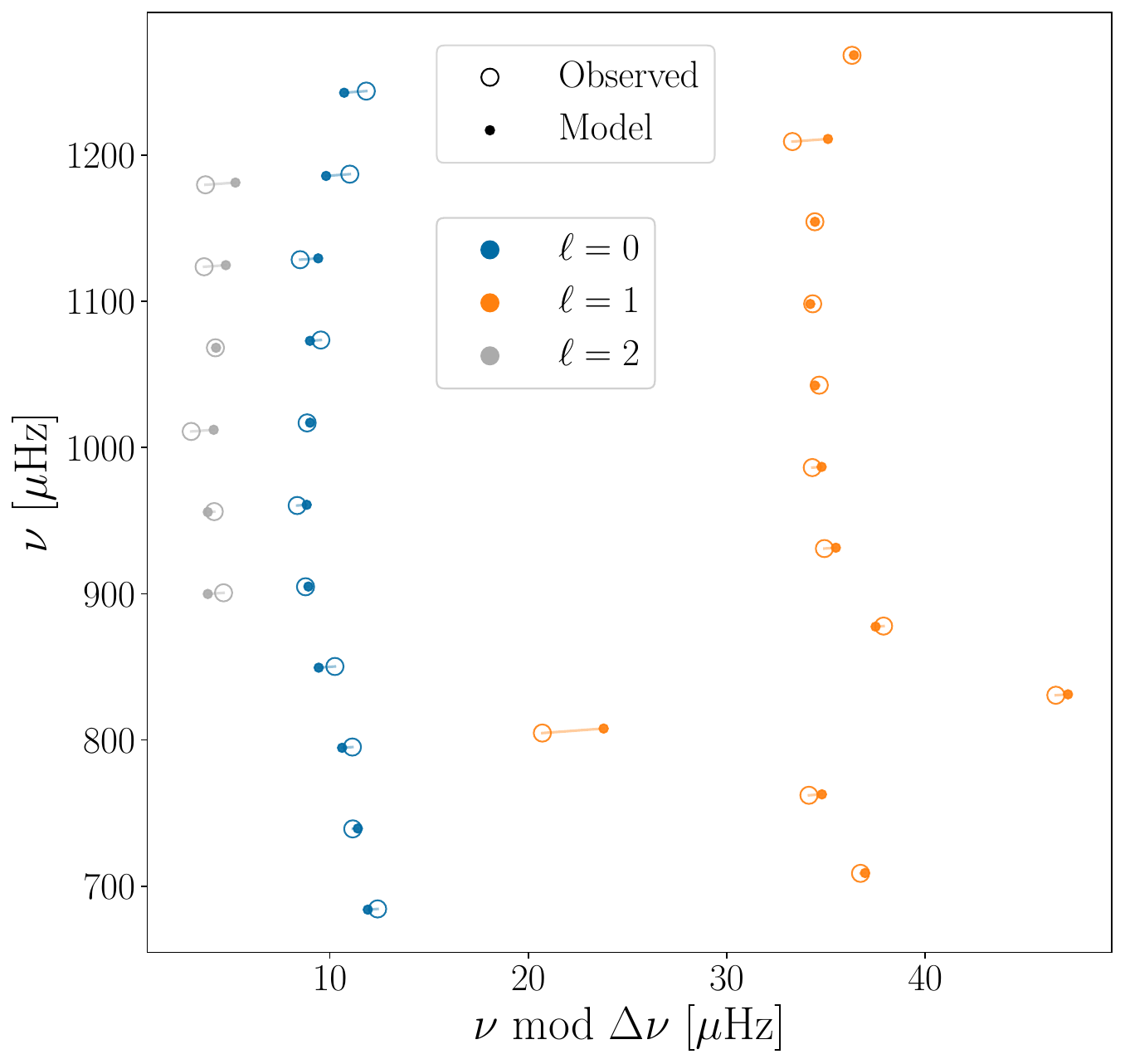}
    \caption{Échelle diagram showing the best mode match for the subgiant target KIC4346201 \edit{1}{used to determine $\sigma_{\nu\text{, eff}}$. The best seismic model is determined} by minimizing \cref{eq:seismic_cost} for all models in the down-selected grid (with $\sigma_{\nu\text{, eff}}$ set to 0). The observed modes are shown with open circles, and the surface-corrected modes of the model are shown with dots. Each observed mode is connected to its corresponding model mode with a line. The colors of the points and lines indicate the angular degree of the modes ($\ell = 0$ in blue, $\ell = 1$ in orange, and $\ell = 2$ in gray.)}
    \label{fig:modematch}
\end{figure}

\begin{figure*}[ht!]
    \centering
    \includegraphics[width=0.8\textwidth]{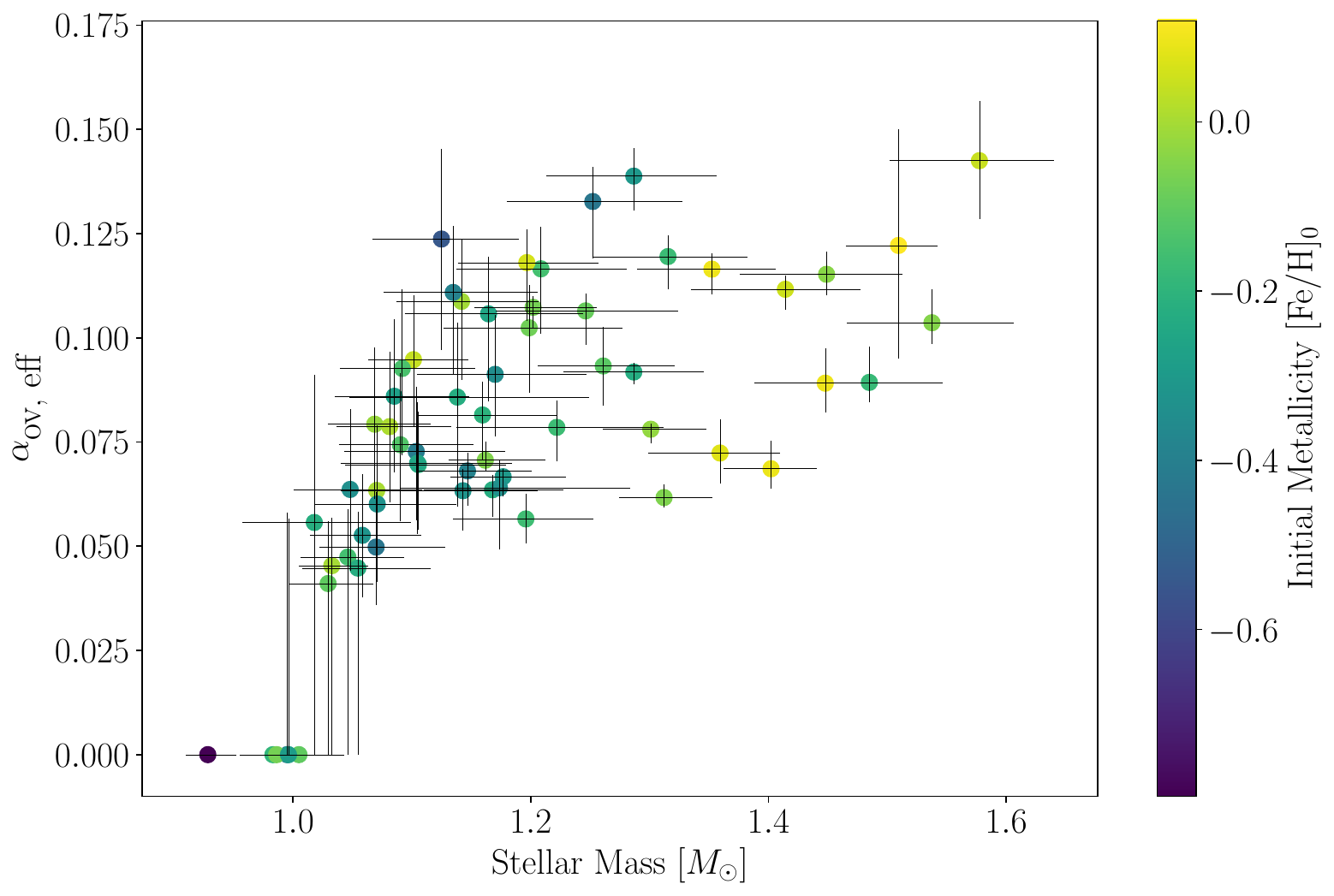}
    \caption{Effective overshoot parameter $\alpha_{\text{ov, eff}}$ versus stellar mass for our sample of 62 \edit{1}{subgiants and early first-ascent red giant branch stars}. The points are colored by the modelling results for initial metallicity. Each point's position and color represents the 50th percentiles of the given parameter for each target. The error bars are given by the 16th and 84th percentiles of each parameter's posterior distribution. }
    \label{fig:Lindsay_mass_overshoot}
\end{figure*}

Next, for each combination of model in the down-selected grid and target, we calculate a seismic cost function with the form,
\begin{equation}
    \label{eq:seismic_cost}
    \chi^2_{\text{seis}} = \mathcal{W}_{\nu} \frac{1}{N_\nu}\sum^{N_{\nu}}_{n} \left(\frac{\nu_{\text{obs}, n} - \nu_{\text{model}, n}}{\sqrt{\sigma^2_{\nu_{\text{obs}, n}} + \sigma^2_{\nu, \text{eff}}}}\right)^2,
\end{equation}
\edit{1}{Following appendix B1 of \citet{Cunha2021} and \citet{Ong2021c}, $\chi^2_{\text{seis}}$ is weighted by $1/N_{\nu}$, the total number of matched oscillation modes in the set. We also weight $\chi^2_{\text{seis}}$ by $\mathcal{W}_{\nu} $, a surface-term-weight, due to limitations involved with correcting stellar model frequencies for near-surface effects using only a power-law based frequency shift \citep{BallGizon2014}.} The surface term makes the model mode frequencies larger than the observed frequencies, and this difference increases with frequency. The model of the surface term shown in \cref{eq:surface_term} does not take this into account, and it is possible that after the surface term correction, a model that has frequencies lower than the observed frequencies would end up with a smaller $\chi^2_{\text{seis}}$ than a model with the expected behavior of the surface term. To account for this, we add the \edit{1}{surface-term-weight,} $\mathcal{W}_{\nu}$, in \cref{eq:seismic_cost} to give a larger weight to models whose uncorrected frequencies were larger than the observed frequencies. To do this, we assign $\mathcal{W}_{\nu}$ a value of 0.5 if all model mode frequencies are higher than their corresponding observed frequencies. If this condition is not satisfied, we set $\mathcal{W}_{\nu}$ to 1.

$\sigma_{\nu\text{, eff}}$ \edit{1}{in \cref{eq:seismic_cost}} accounts for the systematic error in the modeling due to grid undersampling \citep[following][]{LiT2020, Ong2021c}. We first calculate the seismic cost function (\cref{eq:seismic_cost}) with $\sigma_{\nu\text{, eff}}$ set to 0 to identify the best seismic fit model \edit{1}{for each target}. \edit{2}{Using this best seismic fit model, we set $\sigma_{\nu\text{, eff}}$ in \cref{eq:seismic_cost} to the root-mean-squared difference between the observed mode frequencies and the best fitting model's surface corrected mode frequencies. The values of $\sigma_{\nu\text{, eff}}$ we find for each target in this work are reported in \autoref{table:sigma_nu_eff_kepler} and \autoref{table:sigma_nu_eff_tess}. This gives an indication about the asteroseismic goodness-of-fit that is reached for each star.} 

\edit{1}{As an illustrative example for how we calculate $\sigma_{\nu\text{, eff}}$, the best mode match for the target KIC4346201 is shown in \autoref{fig:modematch}. Note that the parameters of this best seismic fit model are not the best-fit parameters we report in our results, but are rather incorporated (along with all the parameters of all the other models in the grid) into the results through taking the likelihood weighted means of our parameters of interest, as described in the next subsection. The fact that the best seismic fit model's frequencies represent the best seismic match to the observed mode frequencies simply means that the seismic likelihood of this model is higher than the others. Additionally, we can see in \autoref{fig:modematch} that the most g-mode-dominated dipole frequency (left-most $\ell = 1$ mode) does not agree with its corresponding model mode frequency, an issue which could be eliminated by using an optimization approach to fitting the modes \citep{Noll2021, NollDeheuvels2023}, rather than the grid-based approach we take in this work. }

For models with interiors that match observed stars, it is a known property of the surface term that the frequency differences between the observed and model p-modes should be largest at high frequencies, and lowest at low frequencies. \edit{1}{We follow \citet{BasuKinnane2018}, \citet{Ong2021a}, and the appendix B1 and B2 procedures of \citet{Cunha2021} in accounting for this by adding another penalty function to $\chi^2_{\text{seis}}$ with the form,}
\begin{equation}
    \label{eq:low_n_cost}
    \chi^2_{\text{low }n} = \frac{1}{10}  \sum_{\ell\in \{0,2\}} \frac{1}{4} \sum_{n = 0}^3 \left(\frac{\nu_{\text{obs}, n} - \nu_{\text{model}, n}}{\sqrt{\sigma^2_{\nu_{\text{obs}, n}} + \sigma^2_{\nu, \text{eff}}}}\right)^2.
\end{equation}
\edit{1}{This term is calculated using the 4 lowest frequency radial ($\ell = 0$) modes and the 4 lowest frequency quadrupole ($\ell = 2$) modes. }The factor $1/4$ comes from the number of modes that we are summing over, and the factor $1/10$ gives a lower weight to $\chi^2_{\text{low }n}$ compared to $\chi^2_{\text{seis}}$, calculated in \cref{eq:seismic_cost} (following the procedure of \citet{Ong2021c}). \edit{1}{We do this because we do not want to double count the 4 modes with the lowest frequencies, $\chi^2_{\text{low }n}$ is incorporated for the purpose of regularization, and is not meant to influence the overall posterior distribution. }

For each of the 62 targets in our sample, we calculate the two terms ($\chi^2_{\text{seis}}$ and $\chi^2_{\text{low }n}$) of the seismic cost function. We then calculate a seismic likelihood ($\mathcal{L}_{\text{seis}}$) for each model in the target star's associated down-selected grid as

\begin{equation}
    \label{eq:seismic_like}
    \mathcal{L}_{\text{seis}} = \exp{\left[-\frac{\left(\chi^2_{\text{seis}} + \chi^2_{\text{low }n}\right)}{2}\right]}.
\end{equation}


\subsection{Estimating Stellar Parameters}

For each of the stars in our sample, we determine total likelihoods, $\mathcal{L}_{\text{tot}}$, for each model in the down-selected grid by combining the spectroscopic and seismic normalized likelihoods as
\begin{equation}
    \label{eq:total_like}
    \mathcal{L}_{\text{tot}} = t_{\text{model}} \times \mathcal{W}_{\text{He}} \times \mathcal{L}_{\text{seis}} \times \mathcal{L}_{\text{spec}}
\end{equation}
where $t_{\text{model}}$ is a time interval defined below, and $\mathcal{W}_{\text{He}}$ is a helium weight. $t_{\text{model}}$ is the length of time (in seconds) each model spends during that model's specific MESA time step, \edit{1}{and we take it into account here because, unlike in \citet{Cunha2021}, not every model in our grid has the same time-step length.} The inclusion of $t_{\text{model}}$ down-weights the likelihoods of models in portions of the model grid which are more densely sampled in time. The Helium weight, $\mathcal{W}_{\text{He}}$ is given by
\begin{equation}
\mathcal{W}_{\text{He}} = \left\{
        \begin{array}{ll}
            1 & \quad \text{if } Y_0 > Y_{\text{p}} \\
            \exp\left[ -\left(\frac{(Y_0-Y_{\text{p}})}{0.016}\right)^2\right] &  \quad \text{if } Y_0 \leq Y_{\text{p}}
        \end{array}
    \right. ,
\end{equation}
with $Y_{\text{p}} = 0.248$ \citep{Steigman2010}. The total likelihood of a given model incorporates this weight term in order to penalize models with initial helium abundance lower than the primordial value of helium, \edit{1}{see \citet{SilvaAguirre2017}}. The total likelihoods are normalized \edit{1}{by dividing each likelihood by the sum of all the models' total likelihoods}, leaving us with an associated likelihood value, $\mathcal{L}_{\text{tot}}$ for every model in the target star's down-selected grid. These likelihoods are used to infer our results, stellar parameters, especially the amount of convective core overshooting, for the 62 subgiant and early first-ascent red giant branch stars in our sample. 



\section{Results and Discussion}
\label{sec:results}

\subsection{Best Fit Parameters for our Targets}
To obtain the best fit parameters and parameter errors for each of the \edit{1}{targets} in our sample (\autoref{sec:sample}), we calculate the likelihood weighted means of each parameter saved in the grid discussed in \autoref{sec:modelgrid} 10,000 times for different realizations of the spectroscopic parameters. The likelihood weighted mean for a given parameter, $\bar{P}$, is calculated as,

\begin{equation}
    \label{eq:like_weighted_mean}
    \bar{P} = \frac{\sum_{i=0}^{N} (\mathcal{L}_{\text{tot, norm}_i} P_{\text{model}_i} ) }{\sum_{i=0}^{N}(\mathcal{L}_{\text{tot, norm}_i} )},
\end{equation}
where $N$ is the number of models in a target's downselected grid and $P_{\text{model}}$ is the model parameter. 

We draw 10,000 sample values of each \edit{1}{spectroscopic} parameter from a normal distribution with the mean equal to the literature-reported value of the parameter and standard deviation equal to the reported error, as was done in \citet{Ong2021c}. Each sampling of the spectroscopic parameters results in a different likelihood weighted mean for our model output parameters, and hence,  performing a Monte--Carlo over the spectroscopic parameters allows us to determine posterior distributions of our output parameters. The 16th, 50th, and 84th percentiles of the posterior distributions for stellar mass ($M$), radius ($R$), luminosity ($L$), temperature ($T_{\text{eff}}$), age, initial metallicity ([FeH]$_0$), initial helium abundance ($Y_0$), mixing length ($\alpha_{\text{mlt}}$), and effective overshoot parameter ($\alpha_{\text{ov, eff}}$) are listed in tables  \autoref{table:Kepler} and \autoref{table:TESS} for the stars observed by \textit{Kepler} and TESS respectively. 

\edit{1}{Previous modelling studies of main sequence stars and subgiants \citep{Lebreton2014, LiT2020, Noll2021} have shown that a high level of degeneracy exists between the initial helium abundnace and mass, due to an apparent anti-correlation between mass and $Y_0$. Analysis of the posterior distributions for our inferred stellar parameters, however, does not show this strong degeneracy between mass and $Y_0$. This is likely due to our inclusion of luminosity in our likelihood function, which is not included in \citet{Lebreton2014} and \citet{LiT2020}. }



\subsection{The relationship between $\alpha_{\text{ov, eff}}$ and Mass}

\autoref{fig:Lindsay_mass_overshoot} shows the results for the effective overshoot parameter ($\alpha_{\text{ov, eff}}$) as a function of the stellar mass. The errors bars for both parameters show the 16th and 84th percentiles of the output parameter posterior distributions. We can see that for sufficiently low mass stars ($M\leq1.0 M_{\odot}$) the modelled $\alpha_{\text{ov, eff}}$ is 0. This is because the stellar models at this mass do not maintain a convective core for a sufficiently long amount of time \edit{1}{($>30\%$ of their main sequence lifetimes.)} At slightly higher masses, from  $M \approx 1.0 M_{\odot}$ to $M \approx 1.15 M_{\odot}$, the value of the effective overshoot parameter  increases with stellar mass from $	\alpha_{\text{ov, eff}}\lesssim 0.05$ to $\alpha_{\text{ov, eff}} \approx 0.1$ -- $0.15$. At masses higher than $1.2 M_{\odot}$, our results for $\alpha_{\text{ov, eff}}$ are seemingly independent of stellar mass. The points in \autoref{fig:Lindsay_mass_overshoot} are colored by our initial metallicity ([$\mathrm{Fe/H}]_0$) results. We do not see a strong relationship between our results for $\alpha_{\text{ov, eff}}$ and [$\mathrm{Fe/H}]_0$.

\begin{figure}
    \centering
    \includegraphics[width=0.45\textwidth]{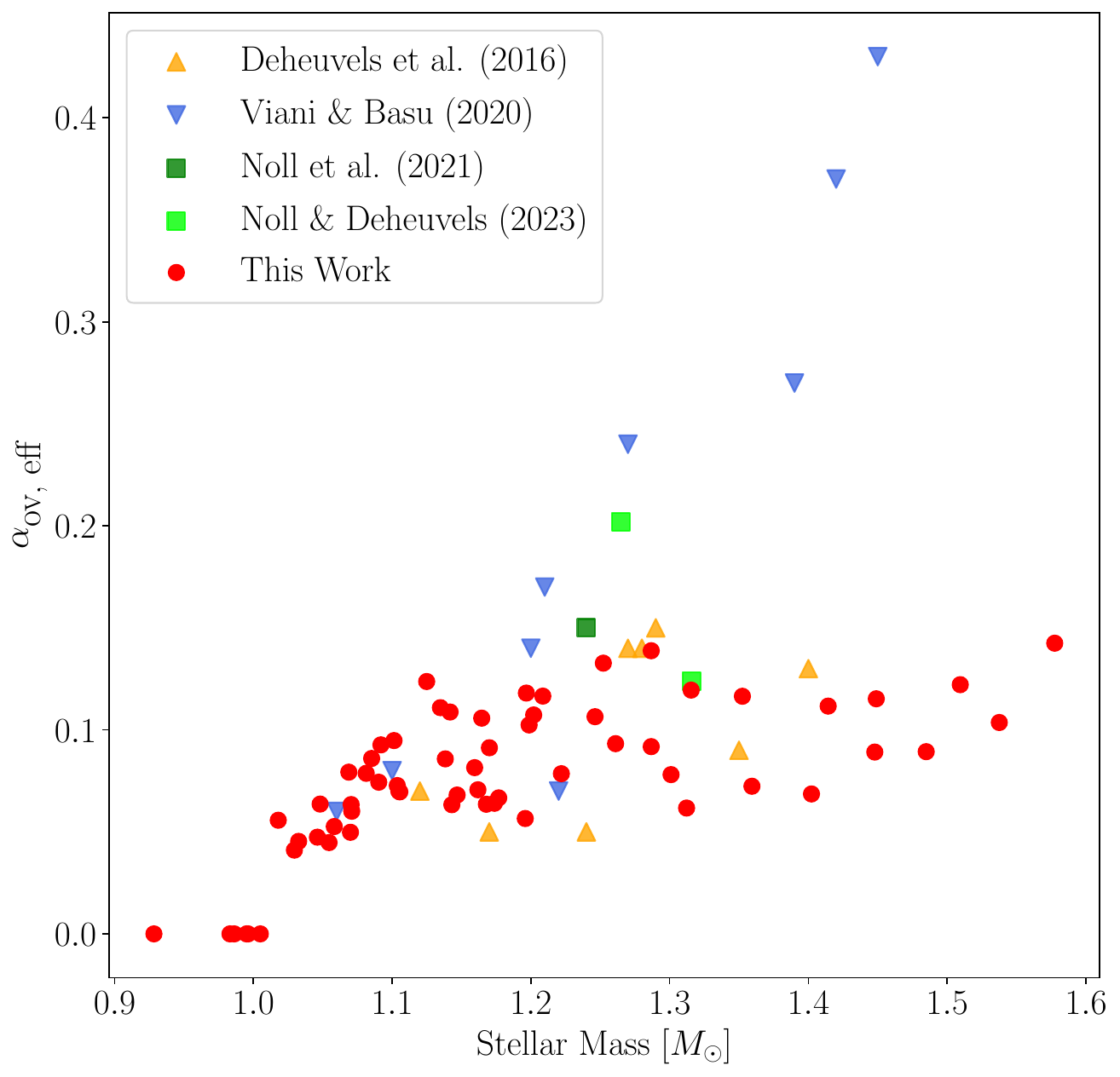}
    \caption{$\alpha_{\text{ov, eff}}$ versus stellar mass results (red points) in comparison with other asteroseismic studies of convective core overshoot.   }
    \label{fig:Asteroseismic_mass_overshoot}
\end{figure}

\begin{figure}
    \centering
    \includegraphics[width=0.45\textwidth]{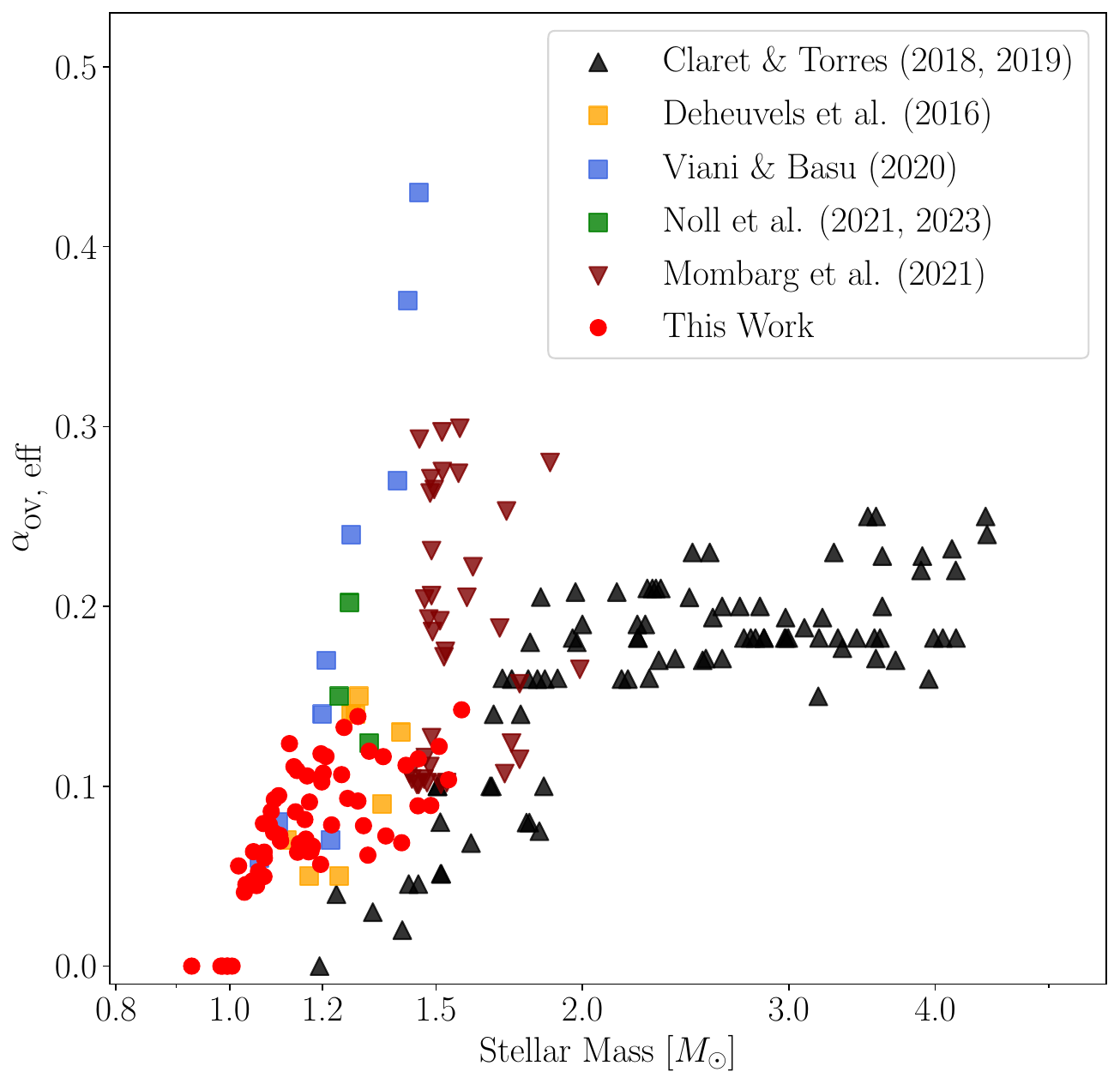}
    \caption{Our $\alpha_{\text{ov, eff}}$ versus stellar mass results (red points) in comparison with other studies of convective core overshoot, including those shown in \autoref{fig:Asteroseismic_mass_overshoot} as well as the eclipsing binary work of \citet{ClaretandTorres2018} and \citet{ClaretTorres2019}, shown in black triangles. The machine learning results for $\gamma$ Doradus stars from \citet{Mombarg2021} are shown in maroon triangles. }
    \label{fig:Complete_mass_overshoot}
\end{figure}

To place our results shown in \autoref{fig:Lindsay_mass_overshoot} into context, we compare them with other studies of convective core overshooting \edit{1}{in \autoref{fig:Asteroseismic_mass_overshoot} and \autoref{fig:Complete_mass_overshoot}.} \edit{1}{We note that in \autoref{fig:Asteroseismic_mass_overshoot} and \autoref{fig:Complete_mass_overshoot}, we plot our values of $\alpha_{\text{ov, eff}}$ along side the input overshoot values the modelling work in \citet{Deheuvels2016}, \citet{ClaretandTorres2018}, \citet{ClaretTorres2019}, \citet{Viani2020}, \citet{Noll2021}, \citet{Mombarg2021} and \citet{NollDeheuvels2023}. Our construction of $\alpha_{\text{ov, eff}}$ (see \autoref{sec:modelgrid}) is meant to correspond to the input overshooting values in the other stellar evolution codes. \edit{2}{We have defined $\alpha_{\text{ov, eff}}$ (which in general is time-dependent) such that the thickness of the overshoot region is equal to $\alpha_{\text{ov, eff}}$ times $H_p$ no matter what $\alpha_{\text{ov}}$ is. We have however restricted our attention to its value at when the convective core is at its maximum size. Since the overshoot region thickness is the physical quantity ultimately being modified in the stellar model by overshooting, it is $\alpha_{\text{ov, eff}}$ rather than input $\alpha_{\text{ov}}$ which is constrained by the mode frequencies.} }

\edit{1}{We employ this construction of $\alpha_{\text{ov, eff}}$ since some other \edit{2}{studies \citep{ClaretandTorres2018, ClaretTorres2019}} calculate the size of the overshoot region as $\alpha_{\text{ov}}$ times the pressure scale height at the convective boundary ($H_{p}$), instead of calculating the overshoot region size as $\alpha_{\text{ov}}$ multiplied by the minimum of $H_{\text{P}}$ and the radius of the convective core (as is done in MESA, see \autoref{sec:modelgrid} and \edit{2}{\citet{Deheuvels2016}}). Additionally MESA's overshooting regions begin slightly interior to the convective boundary, unlike in the other stellar modelling codes used in the other works, such as YREC \citep{Demarque2008}.}

Studies that have used asteroseismic data to estimate overshooting in main-sequence convective cores have been carried out by \citet{Deheuvels2016}, \citet{Viani2020}, and \citet{NollDeheuvels2023}, directly measuring the extent of the well-mixed convective core in a total of 19 main sequence stars (9 by \citet{Viani2020}, 8 by \citet{Deheuvels2016}, and 2 by \citet{NollDeheuvels2023}). In addition, \citet{Noll2021} modelled one subgiant star and produced an estimate for that star's convective core overshooting parameter. \autoref{fig:Asteroseismic_mass_overshoot} shows our $\alpha_{\text{ov, eff}}$ vs stellar mass results alongside the results of \citet{Deheuvels2016}, \citet{Viani2020}, \citet{Noll2021}, and \citet{NollDeheuvels2023}.

Overall, the results from this work agrees with the previous overshooting versus mass results from \citet{Deheuvels2016}, \citet{Noll2021}, and \citet{NollDeheuvels2023}, as well as with the results at the low mass end of the sample from \citet{Viani2020}. The spread in our reported values of $\alpha_{\text{ov, eff}}$ for stars in the mass range of $1.1 M_{\odot} \leq M \leq 1.4$ is also similar to the spread in the overshoot amplitude results of \citet{Deheuvels2016}, for the same range of masses. 

On the higher mass end of their sample ($M \geq 1.3 M_{\odot}$), \citet{Viani2020} reported significantly larger values of effective overshoot parameter, when compared with our results for stars in our sample with comparable masses. We identify two possible reasons for this discrepancy. First, the models calculated in \citet{Viani2020} had set the temperature gradient, $\nabla$ to $\nabla_{\text{ad}}$ in the convective overshooting regions above the models' convective cores. This is analogous to the "Step Penetrative Overshoot" we describe in section 2 of \citet{Lindsay2022} or the "convective penetration" described in \citet{AndersPedersen2023}. As shown in figure 8 of \citet{Viani2020}, the effective overshoot parameter results are larger, for a given stellar mass, if the stellar models used in the analysis maintained an adiabatic temperature gradient in the overshooting region. This temperature gradient difference alone cannot fully explain the large discrepancy between our results and the results of \citet{Viani2020}. The second major difference between the two modelling methods is that our modelling procedure sampled $\alpha_{\text{ov, input}}$ far more densely (see \autoref{fig:grid-mass-overshoot}) when compared to the procedure of \citet{Viani2020}, which only allowed about 8 different $\alpha_{\text{ov, input}}$ values per target star. The denser sampling of both $\alpha_{\text{ov, input}}$ and initial mass in our work results in more precise results of stellar mass and overshoot parameter, whereas the sparse sampling of $\alpha_{\text{ov, input}}$ in \citet{Viani2020}. means that one model with a high overshooting parameter could dominate the likelihood function and significantly increase the output $\alpha_{\text{ov, eff}}$ result. It should also be noted that the \citet{Viani2020} models did not include the gravitational settling of heavy elements, and this too could increase their estimate of overshoot, however, that is unlikely to explain the large difference. 

\citet{Mombarg2019} and \citet{Mombarg2021} studied convective core overshoot properties in intermediate mass stars through asteroseismology of $\gamma$ Doradus stars. \citet{Mombarg2021} used machine learning techniques to study core overshooting in the same sample of stars from \citet{Mombarg2019}, and we show how our results compare to theirs in \autoref{fig:Complete_mass_overshoot}. We note that the masses of the stars in their sample are higher than most of the stars in our sample. The reported exponential overshoot parameters, $f_{\text{ov}}$ from \citet{Mombarg2021} are multiplied by 10 before plotting in \autoref{fig:Complete_mass_overshoot}, as \citet{Mombarg2021} uses the conversion factor of 10 in their analysis. We also include the stellar mass and overshooting amplitude results of \citet{ClaretandTorres2018} and \citet{ClaretTorres2019} in \autoref{fig:Complete_mass_overshoot}. The stars studied in these works were higher-mass main-sequence stars in double line eclipsing binary systems, with $M\geq 1.2 M_{\odot}$. \citet{ClaretandTorres2018} and \citet{ClaretTorres2019} modelled the stars' masses and core overshoot parameters by comparing the observed masses, radii, and effective temperatures of the stars to stellar models made with varying amounts of convective core overshoot. In order to convert the exponential overshoot parameters, $f_{\text{ov}}$, reported in \citet{ClaretTorres2019} to step overshooting parameters, $\alpha_{\text{ov}}$, we follow \citet{ClaretTorres2019} and multiply the reported values of $f_{\text{ov}}$ by 11.36 ($11.36 f_{\text{ov}} = \alpha_{\text{ov}}$) before plotting those points in \autoref{fig:Complete_mass_overshoot}. When a star is reported in both \citet{ClaretandTorres2018} and \citet{ClaretTorres2019}, we plot the value from \citet{ClaretTorres2019}. 

Comparing our results (red points in \autoref{fig:Complete_mass_overshoot}) to the results of \citet{ClaretandTorres2018}, \citet{ClaretTorres2019}, and \citet{Mombarg2021} shows significant differences in stellar mass/overshooting amplitude relationship. The combined results of \citet{ClaretandTorres2018} and \citet{ClaretTorres2019} show the overshoot amplitude increasing with stellar mass from $\alpha_{\text{ov, eff}} \approx 0$ at $M = 1.2 M_{\odot}$ to $\alpha_{\text{ov, eff}} \gtrsim 0.15$ at $M = 2.0 M_{\odot}$. On the other hand, for the same mass range, \citet{Mombarg2021} finds that the step overshooting parameters for their sample of $\gamma$ Doradus stars are larger in magnitude ($\alpha_{\text{ov, eff}} > 0.1$) and less dependent on stellar mass. 

Although our sample of stars do not go past masses of $M \approx 1.5 M_{\odot}$, we report higher values of $\alpha_{\text{ov, eff}}$ for stars in the mass range of 1.2 to 1.5 $M_{\odot}$ range when compared to the eclipsing binary work of \citet{ClaretandTorres2018}, \citet{ClaretTorres2019}. The majority of the $\gamma$ Doradus stars studied by \citet{Mombarg2021} appear to have higher values of $\alpha_{\text{ov, eff}}$ compared to the results from our sample at similar masses. This is not an apples to apples comparison though, since we are using detailed asteroseismic data of individual oscillation modes while \citet{ClaretandTorres2018} and \citet{ClaretTorres2019} produced their overshooting results based on the masses, radii, and effective temperatures of binary stars, without asteroseismology, while  \citet{Mombarg2021} trained a machine learning model to return stellar masses and overshoot parameters based on observed period spacings between adjacent modes. In addition, the grid of models used in this work sampled the input parameters more densely, and covered a wider range of parameters when compared to the grids of models used in the eclipsing binary and $\gamma$ Doradus work. Finally, the models calculated by \citet{ClaretandTorres2018}, \citet{ClaretTorres2019}, and \citet{Mombarg2021} implemented penetrative overshoot which, as discussed previously, would cause higher output values of $\alpha_{\text{ov, eff}}$, when compared with an analysis done with models with a radiative temperature gradient in the overshooting region. 

\edit{3}{We note that our reported relationship between overshooting amplitude and stellar mass also depends on stellar modelling choices beyond the input parameters described in \autoref{sec:modelgrid}, such as the choice of nuclear network. For example, \citet{NollDeheuvels2023} find that changing the choice of nuclear network could change the size of the convective core on the main sequence. Since, like overshooting amplitude, the convective core size is mass dependent, the choice of nuclear network used in the model grid could alter the resultant relationship determined by fitting asteroseismic observations to the model mode frequencies calculated from the models of that grid. Quantifying the magnitude of this dependence on nuclear network will be the aim of subsequent work.}


\section{Summary and Conclusion}
\label{sec:conclusion}
We have conducted a study analyzing 62 \edit{1}{subgiant and early first-ascent red giant branch stars} with high quality asteroseismic data from \textit{Kepler} and TESS. The goal of this work was to better understand convective overshooting above main sequence star convective cores. Previous studies have analyzed main sequence convective core properties using asteroseismology \edit{1}{\citep{Deheuvels2016, Noll2021, NollDeheuvels2023}}, but the total number of targets that are bright enough and in the mass range of interest is quite small. Additionally, \citet{Lindsay2023} showed that core regions of stars are poorly constrained by non-radial p-mode oscillations on the main sequence. Therefore, we analyze \edit{1}{subgiants and early first-ascent red giant branch stars}, which are much brighter, and whose dipolar mixed-mode oscillation frequencies sample the interior structures of subgiants. The main sequence structure of stars with convective cores alter the stellar structure, and these effects persist after the main sequence, when the star evolves into a subgiant (see \autoref{fig:propagation_evolution}). 

Our sample consisted of 44 stars observed during the \textit{Kepler} and K2 missions, as well as 18 stars observed with TESS. The spectroscopic parameters, as well the individual oscillation mode frequencies, were taken from various different literature sources (see \autoref{sec:sample}). To determine the amount of convective core overshooting, as well as the other properties of the stars in our sample, we implement a grid-based modelling technique, as opposed to performing a boutique modelling of every star in our sample. The grid we use densely \edit{1}{Sobol-}samples stellar mass, mixing length, initial helium abundance, metallicity, and the input overshoot parameter, $\alpha_{\text{ov, input}}$. We match the observed mode frequencies of each star in our sample to the surface corrected model mode frequencies \edit{1}{for the models in our grid}. Using the matched mode frequencies, we calculate a seismic likelihood (\cref{eq:seismic_like}) for each combination of star and model. These seismic likelihoods are combined with a spectroscopic likelihood (\cref{eq:spec_like}) to produce a combined 'total' likelihood (\cref{eq:total_like}), which we use to determine the likelihood weighted mean parameter values for all the different quantities in our model grid. The posterior distributions of our output quantities were determined by taking 10,000 samples of the spectroscopic observables and recalculating the likelihood weighted mean parameter values for each random sample. 

These modelling results of the stars observed by \textit{Kepler} are detailed in \autoref{table:Kepler} while \autoref{table:TESS} contains our results for the stars observed by TESS. Our results for stellar mass and effective overshoot parameter, $\alpha_{\text{ov, eff}}$, are visualized in \autoref{fig:Lindsay_mass_overshoot} which shows that $\alpha_{\text{ov, eff}}$ increases with stellar mass from $M = 1.0 M_{\odot}$ to $M = 1.2 M_{\odot}$. For stars with masses greater than 1.2 $M_{\odot}$ our results indicate a weak, or no correlation between stellar mass and overshooting amplitude. We compare these findings to previous work in \autoref{sec:results} while discussing how differences between our results and previous works can be explained by differences in modelling procedure. 

\section*{Acknowledgement}
CJL would like to thank Y. Asali, D. Bowman, M. Michielsen, J. Mombarg, J. Guerra, I. Pasha, M. G. Pedersen, J. van Saders, and R. Townsend for insightful and helpful discussions. We thank the Yale Center for Research Computing for guidance and use of the research computing infrastructure. SB and CJL acknowledges support from NSF grant AST-2205026. CJL also acknowledges support of a Gruber Science Fellowship. JMJO acknowledges support from NASA through the NASA Hubble Fellowship grants HST-HF2-51517.001-A, awarded by STScI, which is operated by the Association of Universities for Research in Astronomy, Incorporated, under NASA contract NAS5-26555.

\software{
MESA \citep{Paxton2011,Paxton2013,Paxton2015,Paxton2018,Paxton2019, Jermyn2023}, GYRE \citep{Townsend2013}},
SciPy \citep{scipy}, Pandas \citep{pandas}

\edit1{The MESA and GYRE inlists we used to generate our models and frequencies are archived on Zenodo and can be downloaded at \dataset[https://doi.org/10.5281/zenodo.10067668.]{https://doi.org/10.5281/zenodo.10067668}} The Python scripts used to evaluate the matrix elements (which are used to determine the dipolar mixed-mode frequencies) discussed in this work are available at https://gitlab.com/darthoctopus/mesatricks.

\bibliographystyle{aasjournal-compact}
\footnotesize
\bibliography{ref}

\begin{thebibliography}{}
\expandafter\ifx\csname natexlab\endcsname\relax\def\natexlab#1{#1}\fi
\providecommand{\url}[1]{\href{#1}{#1}}
\providecommand{\mhref}[2]{\href{#1}{\color{magenta}#2}}
\providecommand{\dodoi}[1]{doi:~\href{http://doi.org/#1}{\nolinkurl{#1}}}
\providecommand{\doeprint}[1]{\href{http://ascl.net/#1}{\nolinkurl{http://ascl.net/#1}}}
\providecommand{\doarXiv}[1]{\href{https://arxiv.org/abs/#1}{\nolinkurl{https://arxiv.org/abs/#1}}}

\bibitem[{{Abdurro'uf} {et~al.}(2022){Abdurro'uf}, {Accetta}, {Aerts}, {Silva
  Aguirre}, {Ahumada}, {Ajgaonkar}, {Filiz Ak}, {Alam}, {Allende Prieto},
  {Almeida}, {Anders}, {Anderson}, {Andrews}, {Anguiano}, {Aquino-Ort{\'\i}z},
  {Arag{\'o}n-Salamanca}, {Argudo-Fern{\'a}ndez}, {Ata}, {Aubert},
  {Avila-Reese}, {Badenes}, {Barb{\'a}}, {Barger}, {Barrera-Ballesteros},
  {Beaton}, {Beers}, {Belfiore}, {Bender}, {Bernardi}, {Bershady}, {Beutler},
  {Bidin}, {Bird}, {Bizyaev}, {Blanc}, {Blanton}, {Boardman}, {Bolton},
  {Boquien}, {Borissova}, {Bovy}, {Brandt}, {Brown}, {Brownstein}, {Brusa},
  {Buchner}, {Bundy}, {Burchett}, {Bureau}, {Burgasser}, {Cabang}, {Campbell},
  {Cappellari}, {Carlberg}, {Wanderley}, {Carrera}, {Cash}, {Chen}, {Chen},
  {Cherinka}, {Chiappini}, {Choi}, {Chojnowski}, {Chung}, {Clerc}, {Cohen},
  {Comerford}, {Comparat}, {da Costa}, {Covey}, {Crane}, {Cruz-Gonzalez},
  {Culhane}, {Cunha}, {Dai}, {Damke}, {Darling}, {Davidson}, {Davies},
  {Dawson}, {De Lee}, {Diamond-Stanic}, {Cano-D{\'\i}az}, {S{\'a}nchez},
  {Donor}, {Duckworth}, {Dwelly}, {Eisenstein}, {Elsworth}, {Emsellem},
  {Eracleous}, {Escoffier}, {Fan}, {Farr}, {Feng}, {Fern{\'a}ndez-Trincado},
  {Feuillet}, {Filipp}, {Fillingham}, {Frinchaboy}, {Fromenteau}, {Galbany},
  {Garc{\'\i}a}, {Garc{\'\i}a-Hern{\'a}ndez}, {Ge}, {Geisler}, {Gelfand},
  {G{\'e}ron}, {Gibson}, {Goddy}, {Godoy-Rivera}, {Grabowski}, {Green},
  {Greener}, {Grier}, {Griffith}, {Guo}, {Guy}, {Hadjara}, {Harding},
  {Hasselquist}, {Hayes}, {Hearty}, {Hern{\'a}ndez}, {Hill}, {Hogg},
  {Holtzman}, {Horta}, {Hsieh}, {Hsu}, {Hsu}, {Huber}, {Huertas-Company},
  {Hutchinson}, {Hwang}, {Ibarra-Medel}, {Chitham}, {Ilha}, {Imig}, {Jaekle},
  {Jayasinghe}, {Ji}, {Johnson}, {Jones}, {J{\"o}nsson}, {Katkov}, {Khalatyan},
  {Kinemuchi}, {Kisku}, {Knapen}, {Kneib}, {Kollmeier}, {Kong}, {Kounkel},
  {Kreckel}, {Krishnarao}, {Lacerna}, {Lane}, {Langgin}, {Lavender}, {Law},
  {Lazarz}, {Leung}, {Leung}, {Lewis}, {Li}, {Li}, {Lian}, {Liang}, {Lin},
  {Lin}, {Lin}, {Lintott}, {Long}, {Longa-Pe{\~n}a}, {L{\'o}pez-Cob{\'a}},
  {Lu}, {Lundgren}, {Luo}, {Mackereth}, {de la Macorra}, {Mahadevan},
  {Majewski}, {Manchado}, {Mandeville}, {Maraston}, {Margalef-Bentabol},
  {Masseron}, {Masters}, {Mathur}, {McDermid}, {Mckay}, {Merloni},
  {Merrifield}, {Meszaros}, {Miglio}, {Di Mille}, {Minniti}, {Minsley},
  {Monachesi}, {Moon}, {Mosser}, {Mulchaey}, {Muna}, {Mu{\~n}oz}, {Myers},
  {Myers}, {Nadathur}, {Nair}, {Nandra}, {Neumann}, {Newman}, {Nidever},
  {Nikakhtar}, {Nitschelm}, {O'Connell}, {Garma-Oehmichen}, {Luan Souza de
  Oliveira}, {Olney}, {Oravetz}, {Ortigoza-Urdaneta}, {Osorio}, {Otter},
  {Pace}, {Padilla}, {Pan}, {Pan}, {Parikh}, {Parker}, {Peirani}, {Pe{\~n}a
  Ram{\'\i}rez}, {Penny}, {Percival}, {Perez-Fournon}, {Pinsonneault},
  {Poidevin}, {Poovelil}, {Price-Whelan}, {B{\'a}rbara de Andrade Queiroz},
  {Raddick}, {Ray}, {Rembold}, {Riddle}, {Riffel}, {Riffel}, {Rix}, {Robin},
  {Rodr{\'\i}guez-Puebla}, {Roman-Lopes}, {Rom{\'a}n-Z{\'u}{\~n}iga}, {Rose},
  {Ross}, {Rossi}, {Rubin}, {Salvato}, {S{\'a}nchez}, {S{\'a}nchez-Gallego},
  {Sanderson}, {Santana Rojas}, {Sarceno}, {Sarmiento}, {Sayres}, {Sazonova},
  {Schaefer}, {Schiavon}, {Schlegel}, {Schneider}, {Schultheis}, {Schwope},
  {Serenelli}, {Serna}, {Shao}, {Shapiro}, {Sharma}, {Shen}, {Shetrone}, {Shu},
  {Simon}, {Skrutskie}, {Smethurst}, {Smith}, {Sobeck}, {Spoo}, {Sprague},
  {Stark}, {Stassun}, {Steinmetz}, {Stello}, {Stone-Martinez},
  {Storchi-Bergmann}, {Stringfellow}, {Stutz}, {Su}, {Taghizadeh-Popp},
  {Talbot}, {Tayar}, {Telles}, {Teske}, {Thakar}, {Theissen}, {Tkachenko},
  {Thomas}, {Tojeiro}, {Hernandez Toledo}, {Troup}, {Trump}, {Trussler},
  {Turner}, {Tuttle}, {Unda-Sanzana}, {V{\'a}zquez-Mata}, {Valentini},
  {Valenzuela}, {Vargas-Gonz{\'a}lez}, {Vargas-Maga{\~n}a}, {Alfaro},
  {Villanova}, {Vincenzo}, {Wake}, {Warfield}, {Washington}, {Weaver},
  {Weijmans}, {Weinberg}, {Weiss}, {Westfall}, {Wild}, {Wilde}, {Wilson},
  {Wilson}, {Wilson}, {Wolf}, {Wood-Vasey}, {Yan}, {Zamora}, {Zasowski},
  {Zhang}, {Zhao}, {Zheng}, {Zheng}, \& {Zhu}}]{SloanDR17}
{Abdurro'uf}, {Accetta}, K., {Aerts}, C., {et~al.} 2022,
  {\mhref{http://doi.org/10.3847/1538-4365/ac4414}{\apjs}},
  {\href{https://ui.adsabs.harvard.edu/abs/2022ApJS..259...35A}{259}}{\href{https://ui.adsabs.harvard.edu/abs/2022ApJS..259...35A}{,
  35}}

\bibitem[{{Aizenman} {et~al.}(1977){Aizenman}, {Smeyers}, \&
  {Weigert}}]{Aizenman1977}
{Aizenman}, M., {Smeyers}, P., \& {Weigert}, A. 1977, \aap,
  {\href{https://ui.adsabs.harvard.edu/abs/1977A&A....58...41A}{58}}{\href{https://ui.adsabs.harvard.edu/abs/1977A&A....58...41A}{,
  41}}

\bibitem[{{Anders} \& {Pedersen}(2023)}]{AndersPedersen2023}
{Anders}, E.~H., \& {Pedersen}, M.~G. 2023,
  {\mhref{http://doi.org/10.3390/galaxies11020056}{Galaxies}},
  {\href{https://ui.adsabs.harvard.edu/abs/2023Galax..11...56A}{11}}{\href{https://ui.adsabs.harvard.edu/abs/2023Galax..11...56A}{,
  56}}

\bibitem[{{Aparicio} {et~al.}(1990){Aparicio}, {Bertelli}, {Chiosi}, \&
  {Garcia-Pelayo}}]{Aparicio1990}
{Aparicio}, A., {Bertelli}, G., {Chiosi}, C., \& {Garcia-Pelayo}, J.~M. 1990,
  \aap,
  {\href{https://ui.adsabs.harvard.edu/abs/1990A&A...240..262A}{240}}{\href{https://ui.adsabs.harvard.edu/abs/1990A&A...240..262A}{,
  262}}

\bibitem[{{Baglin} {et~al.}(2006){Baglin}, {Auvergne}, {Barge}, {Deleuil},
  {Catala}, {Michel}, {Weiss}, \& {COROT Team}}]{Baglin2006}
{Baglin}, A., {Auvergne}, M., {Barge}, P., {et~al.} 2006, in ESA Special
  Publication, Vol. 1306, The CoRoT Mission Pre-Launch Status - Stellar
  Seismology and Planet Finding, ed. M.~{Fridlund}, A.~{Baglin}, J.~{Lochard},
  \& L.~{Conroy}, 33

\bibitem[{{Ball} \& {Gizon}(2014)}]{BallGizon2014}
{Ball}, W.~H., \& {Gizon}, L. 2014,
  {\mhref{http://doi.org/10.1051/0004-6361/201424325}{\aap}},
  {\href{https://ui.adsabs.harvard.edu/abs/2014A&A...568A.123B}{568}}{\href{https://ui.adsabs.harvard.edu/abs/2014A&A...568A.123B}{,
  A123}}

\bibitem[{{Ball} {et~al.}(2018){Ball}, {Theme{\ss}l}, \& {Hekker}}]{Ball2018}
{Ball}, W.~H., {Theme{\ss}l}, N., \& {Hekker}, S. 2018,
  {\mhref{http://doi.org/10.1093/mnras/sty1141}{\mnras}},
  {\href{https://ui.adsabs.harvard.edu/abs/2018MNRAS.478.4697B}{478}}{\href{https://ui.adsabs.harvard.edu/abs/2018MNRAS.478.4697B}{,
  4697}}

\bibitem[{{Ball} {et~al.}(2020){Ball}, {Chaplin}, {Nielsen},
  {Gonz{\'a}lez-Cuesta}, {Mathur}, {Santos}, {Garc{\'\i}a}, {Buzasi}, {Mosser},
  {Deal}, {Stokholm}, {Mosumgaard}, {Silva Aguirre}, {Nsamba}, {Campante},
  {Cunha}, {Ong}, {Basu}, {{\"O}rtel}, {{\c{C}}elik Orhan}, {Y{\i}ld{\i}z},
  {Stassun}, {Kane}, \& {Huber}}]{Ball2020}
{Ball}, W.~H., {Chaplin}, W.~J., {Nielsen}, M.~B., {et~al.} 2020,
  {\mhref{http://doi.org/10.1093/mnras/staa3190}{\mnras}},
  {\href{https://ui.adsabs.harvard.edu/abs/2020MNRAS.499.6084B}{499}}{\href{https://ui.adsabs.harvard.edu/abs/2020MNRAS.499.6084B}{,
  6084}}

\bibitem[{{Basu} {et~al.}(2010){Basu}, {Chaplin}, \& {Elsworth}}]{Basu2010}
{Basu}, S., {Chaplin}, W.~J., \& {Elsworth}, Y. 2010,
  {\mhref{http://doi.org/10.1088/0004-637X/710/2/1596}{\apj}},
  {\href{https://ui.adsabs.harvard.edu/abs/2010ApJ...710.1596B}{710}}{\href{https://ui.adsabs.harvard.edu/abs/2010ApJ...710.1596B}{,
  1596}}

\bibitem[{{Basu} \& {Kinnane}(2018)}]{BasuKinnane2018}
{Basu}, S., \& {Kinnane}, A. 2018,
  {\mhref{http://doi.org/10.3847/1538-4357/aae922}{\apj}},
  {\href{https://ui.adsabs.harvard.edu/abs/2018ApJ...869....8B}{869}}{\href{https://ui.adsabs.harvard.edu/abs/2018ApJ...869....8B}{,
  8}}

\bibitem[{{Bertelli} {et~al.}(1992){Bertelli}, {Mateo}, {Chiosi}, \&
  {Bressan}}]{Bertelli1992}
{Bertelli}, G., {Mateo}, M., {Chiosi}, C., \& {Bressan}, A. 1992,
  {\mhref{http://doi.org/10.1086/171163}{\apj}},
  {\href{https://ui.adsabs.harvard.edu/abs/1992ApJ...388..400B}{388}}{\href{https://ui.adsabs.harvard.edu/abs/1992ApJ...388..400B}{,
  400}}

\bibitem[{{Borucki} {et~al.}(2010){Borucki}, {Koch}, {Basri}, {Batalha},
  {Brown}, {Caldwell}, {Caldwell}, {Christensen-Dalsgaard}, {Cochran},
  {DeVore}, {Dunham}, {Dupree}, {Gautier}, {Geary}, {Gilliland}, {Gould},
  {Howell}, {Jenkins}, {Kondo}, {Latham}, {Marcy}, {Meibom}, {Kjeldsen},
  {Lissauer}, {Monet}, {Morrison}, {Sasselov}, {Tarter}, {Boss}, {Brownlee},
  {Owen}, {Buzasi}, {Charbonneau}, {Doyle}, {Fortney}, {Ford}, {Holman},
  {Seager}, {Steffen}, {Welsh}, {Rowe}, {Anderson}, {Buchhave}, {Ciardi},
  {Walkowicz}, {Sherry}, {Horch}, {Isaacson}, {Everett}, {Fischer}, {Torres},
  {Johnson}, {Endl}, {MacQueen}, {Bryson}, {Dotson}, {Haas}, {Kolodziejczak},
  {Van Cleve}, {Chandrasekaran}, {Twicken}, {Quintana}, {Clarke}, {Allen},
  {Li}, {Wu}, {Tenenbaum}, {Verner}, {Bruhweiler}, {Barnes}, \&
  {Prsa}}]{Kepler_inst}
{Borucki}, W.~J., {Koch}, D., {Basri}, G., {et~al.} 2010,
  {\mhref{http://doi.org/10.1126/science.1185402}{Science}},
  {\href{https://ui.adsabs.harvard.edu/abs/2010Sci...327..977B}{327}}{\href{https://ui.adsabs.harvard.edu/abs/2010Sci...327..977B}{,
  977}}

\bibitem[{{Bressan} {et~al.}(2012){Bressan}, {Marigo}, {Girardi}, {Salasnich},
  {Dal Cero}, {Rubele}, \& {Nanni}}]{Bressan2012}
{Bressan}, A., {Marigo}, P., {Girardi}, L., {et~al.} 2012,
  {\mhref{http://doi.org/10.1111/j.1365-2966.2012.21948.x}{\mnras}},
  {\href{https://ui.adsabs.harvard.edu/abs/2012MNRAS.427..127B}{427}}{\href{https://ui.adsabs.harvard.edu/abs/2012MNRAS.427..127B}{,
  127}}

\bibitem[{{Campante} {et~al.}(2019){Campante}, {Corsaro}, {Lund}, {Mosser},
  {Serenelli}, {Veras}, {Adibekyan}, {Antia}, {Ball}, {Basu}, {Bedding},
  {Bossini}, {Davies}, {Delgado Mena}, {Garc{\'\i}a}, {Handberg}, {Hon},
  {Kane}, {Kawaler}, {Kuszlewicz}, {Lucas}, {Mathur}, {Nardetto}, {Nielsen},
  {Pinsonneault}, {Reffert}, {Silva Aguirre}, {Stassun}, {Stello}, {Stock},
  {Vrard}, {Y{\i}ld{\i}z}, {Chaplin}, {Huber}, {Bean}, {{\c{C}}elik Orhan},
  {Cunha}, {Christensen-Dalsgaard}, {Kjeldsen}, {Metcalfe}, {Miglio},
  {Monteiro}, {Nsamba}, {{\"O}rtel}, {Pereira}, {Sousa}, {Tsantaki}, \&
  {Turnbull}}]{Campante2019}
{Campante}, T.~L., {Corsaro}, E., {Lund}, M.~N., {et~al.} 2019,
  {\mhref{http://doi.org/10.3847/1538-4357/ab44a8}{\apj}},
  {\href{https://ui.adsabs.harvard.edu/abs/2019ApJ...885...31C}{885}}{\href{https://ui.adsabs.harvard.edu/abs/2019ApJ...885...31C}{,
  31}}

\bibitem[{{Chaplin} {et~al.}(2020){Chaplin}, {Serenelli}, {Miglio}, {Morel},
  {Mackereth}, {Vincenzo}, {Kjeldsen}, {Basu}, {Ball}, {Stokholm}, {Verma},
  {Mosumgaard}, {Silva Aguirre}, {Mazumdar}, {Ranadive}, {Antia}, {Lebreton},
  {Ong}, {Appourchaux}, {Bedding}, {Christensen-Dalsgaard}, {Creevey},
  {Garc{\'\i}a}, {Handberg}, {Huber}, {Kawaler}, {Lund}, {Metcalfe}, {Stassun},
  {Bazot}, {Beck}, {Bell}, {Bergemann}, {Buzasi}, {Benomar}, {Bossini},
  {Bugnet}, {Campante}, {Orhan}, {Corsaro}, {Gonz{\'a}lez-Cuesta}, {Davies},
  {Di Mauro}, {Egeland}, {Elsworth}, {Gaulme}, {Ghasemi}, {Guo}, {Hall},
  {Hasanzadeh}, {Hekker}, {Howe}, {Jenkins}, {Jim{\'e}nez}, {Kiefer},
  {Kuszlewicz}, {Kallinger}, {Latham}, {Lundkvist}, {Mathur}, {Montalb{\'a}n},
  {Mosser}, {Bed{\'o}n}, {Nielsen}, {{\"O}rtel}, {Rendle}, {Ricker},
  {Rodrigues}, {Roxburgh}, {Safari}, {Schofield}, {Seager}, {Smalley},
  {Stello}, {Szab{\'o}}, {Tayar}, {Theme{\ss}l}, {Thomas}, {Vanderspek}, {van
  Rossem}, {Vrard}, {Weiss}, {White}, {Winn}, \& {Y{\i}ld{\i}z}}]{Chaplin2020}
{Chaplin}, W.~J., {Serenelli}, A.~M., {Miglio}, A., {et~al.} 2020,
  {\mhref{http://doi.org/10.1038/s41550-019-0975-9}{Nature Astronomy}},
  {\href{https://ui.adsabs.harvard.edu/abs/2020NatAs...4..382C}{4}}{\href{https://ui.adsabs.harvard.edu/abs/2020NatAs...4..382C}{,
  382}}

\bibitem[{{Claret}(2007)}]{Claret2007}
{Claret}, A. 2007, {\mhref{http://doi.org/10.1051/0004-6361:20078024}{\aap}},
  {\href{https://ui.adsabs.harvard.edu/abs/2007A&A...475.1019C}{475}}{\href{https://ui.adsabs.harvard.edu/abs/2007A&A...475.1019C}{,
  1019}}

\bibitem[{{Claret} {et~al.}(2021){Claret}, {Gim{\'e}nez}, {Baroch}, {Ribas},
  {Morales}, \& {Anglada-Escud{\'e}}}]{Claret2021}
{Claret}, A., {Gim{\'e}nez}, A., {Baroch}, D., {et~al.} 2021,
  {\mhref{http://doi.org/10.1051/0004-6361/202141484}{\aap}},
  {\href{https://ui.adsabs.harvard.edu/abs/2021A&A...654A..17C}{654}}{\href{https://ui.adsabs.harvard.edu/abs/2021A&A...654A..17C}{,
  A17}}

\bibitem[{{Claret} \& {Torres}(2016)}]{ClaretTorres2016}
{Claret}, A., \& {Torres}, G. 2016,
  {\mhref{http://doi.org/10.1051/0004-6361/201628779}{\aap}},
  {\href{https://ui.adsabs.harvard.edu/abs/2016A&A...592A..15C}{592}}{\href{https://ui.adsabs.harvard.edu/abs/2016A&A...592A..15C}{,
  A15}}

\bibitem[{{Claret} \& {Torres}(2018)}]{ClaretandTorres2018}
---. 2018, {\mhref{http://doi.org/10.3847/1538-4357/aabd35}{\apj}},
  {\href{https://ui.adsabs.harvard.edu/abs/2018ApJ...859..100C}{859}}{\href{https://ui.adsabs.harvard.edu/abs/2018ApJ...859..100C}{,
  100}}

\bibitem[{{Claret} \& {Torres}(2019)}]{ClaretTorres2019}
---. 2019, {\mhref{http://doi.org/10.3847/1538-4357/ab1589}{\apj}},
  {\href{https://ui.adsabs.harvard.edu/abs/2019ApJ...876..134C}{876}}{\href{https://ui.adsabs.harvard.edu/abs/2019ApJ...876..134C}{,
  134}}

\bibitem[{{Constantino} \& {Baraffe}(2018)}]{Constantino2018}
{Constantino}, T., \& {Baraffe}, I. 2018,
  {\mhref{http://doi.org/10.1051/0004-6361/201833568}{\aap}},
  {\href{https://ui.adsabs.harvard.edu/abs/2018A&A...618A.177C}{618}}{\href{https://ui.adsabs.harvard.edu/abs/2018A&A...618A.177C}{,
  A177}}

\bibitem[{{Costa} {et~al.}(2019){Costa}, {Girardi}, {Bressan}, {Marigo},
  {Rodrigues}, {Chen}, {Lanza}, \& {Goudfrooij}}]{Costa2019}
{Costa}, G., {Girardi}, L., {Bressan}, A., {et~al.} 2019,
  {\mhref{http://doi.org/10.1093/mnras/stz728}{\mnras}},
  {\href{https://ui.adsabs.harvard.edu/abs/2019MNRAS.485.4641C}{485}}{\href{https://ui.adsabs.harvard.edu/abs/2019MNRAS.485.4641C}{,
  4641}}

\bibitem[{{Cox} \& {Giuli}(1968)}]{CoxGiuli1968}
{Cox}, J.~P., \& {Giuli}, R.~T. 1968, {Principles of stellar structure}

\bibitem[{{Cunha} {et~al.}(2021){Cunha}, {Roxburgh}, {Aguirre B{\o}rsen-Koch},
  {Ball}, {Basu}, {Chaplin}, {Goupil}, {Nsamba}, {Ong}, {Reese}, {Verma},
  {Belkacem}, {Campante}, {Christensen-Dalsgaard}, {Clara}, {Deheuvels},
  {Monteiro}, {Noll}, {Ouazzani}, {R{\o}rsted}, {Stokholm}, \&
  {Winther}}]{Cunha2021}
{Cunha}, M.~S., {Roxburgh}, I.~W., {Aguirre B{\o}rsen-Koch}, V., {et~al.} 2021,
  {\mhref{http://doi.org/10.1093/mnras/stab2886}{\mnras}},
  {\href{https://ui.adsabs.harvard.edu/abs/2021MNRAS.508.5864C}{508}}{\href{https://ui.adsabs.harvard.edu/abs/2021MNRAS.508.5864C}{,
  5864}}

\bibitem[{{Deheuvels} {et~al.}(2016){Deheuvels}, {Brand{\~a}o}, {Silva
  Aguirre}, {Ballot}, {Michel}, {Cunha}, {Lebreton}, \&
  {Appourchaux}}]{Deheuvels2016}
{Deheuvels}, S., {Brand{\~a}o}, I., {Silva Aguirre}, V., {et~al.} 2016,
  {\mhref{http://doi.org/10.1051/0004-6361/201527967}{\aap}},
  {\href{https://ui.adsabs.harvard.edu/abs/2016A&A...589A..93D}{589}}{\href{https://ui.adsabs.harvard.edu/abs/2016A&A...589A..93D}{,
  A93}}

\bibitem[{{Deheuvels} \& {Michel}(2011)}]{Deheuvels2011}
{Deheuvels}, S., \& {Michel}, E. 2011,
  {\mhref{http://doi.org/10.1051/0004-6361/201117232}{\aap}},
  {\href{https://ui.adsabs.harvard.edu/abs/2011A&A...535A..91D}{535}}{\href{https://ui.adsabs.harvard.edu/abs/2011A&A...535A..91D}{,
  A91}}

\bibitem[{{Demarque} {et~al.}(2008){Demarque}, {Guenther}, {Li}, {Mazumdar}, \&
  {Straka}}]{Demarque2008}
{Demarque}, P., {Guenther}, D.~B., {Li}, L.~H., {Mazumdar}, A., \& {Straka},
  C.~W. 2008, {\mhref{http://doi.org/10.1007/s10509-007-9698-y}{\apss}},
  {\href{https://ui.adsabs.harvard.edu/abs/2008Ap&SS.316...31D}{316}}{\href{https://ui.adsabs.harvard.edu/abs/2008Ap&SS.316...31D}{,
  31}}

\bibitem[{{Demarque} {et~al.}(1994){Demarque}, {Sarajedini}, \&
  {Guo}}]{Demarque1994}
{Demarque}, P., {Sarajedini}, A., \& {Guo}, X.~J. 1994,
  {\mhref{http://doi.org/10.1086/174052}{\apj}},
  {\href{https://ui.adsabs.harvard.edu/abs/1994ApJ...426..165D}{426}}{\href{https://ui.adsabs.harvard.edu/abs/1994ApJ...426..165D}{,
  165}}

\bibitem[{{Demarque} {et~al.}(2004){Demarque}, {Woo}, {Kim}, \&
  {Yi}}]{Demarque2004}
{Demarque}, P., {Woo}, J.-H., {Kim}, Y.-C., \& {Yi}, S.~K. 2004,
  {\mhref{http://doi.org/10.1086/424966}{\apjs}},
  {\href{https://ui.adsabs.harvard.edu/abs/2004ApJS..155..667D}{155}}{\href{https://ui.adsabs.harvard.edu/abs/2004ApJS..155..667D}{,
  667}}

\bibitem[{{Gai} {et~al.}(2011){Gai}, {Basu}, {Chaplin}, \&
  {Elsworth}}]{Gai2011}
{Gai}, N., {Basu}, S., {Chaplin}, W.~J., \& {Elsworth}, Y. 2011,
  {\mhref{http://doi.org/10.1088/0004-637X/730/2/63}{\apj}},
  {\href{https://ui.adsabs.harvard.edu/abs/2011ApJ...730...63G}{730}}{\href{https://ui.adsabs.harvard.edu/abs/2011ApJ...730...63G}{,
  63}}

\bibitem[{{Gaia Collaboration} {et~al.}(2018){Gaia Collaboration}, {Brown},
  {Vallenari}, {Prusti}, {de Bruijne}, {Babusiaux}, {Bailer-Jones}, {Biermann},
  {Evans}, {Eyer}, {Jansen}, {Jordi}, {Klioner}, {Lammers}, {Lindegren},
  {Luri}, {Mignard}, {Panem}, {Pourbaix}, {Randich}, {Sartoretti}, {Siddiqui},
  {Soubiran}, {van Leeuwen}, {Walton}, {Arenou}, {Bastian}, {Cropper},
  {Drimmel}, {Katz}, {Lattanzi}, {Bakker}, {Cacciari}, {Casta{\~n}eda},
  {Chaoul}, {Cheek}, {De Angeli}, {Fabricius}, {Guerra}, {Holl}, {Masana},
  {Messineo}, {Mowlavi}, {Nienartowicz}, {Panuzzo}, {Portell}, {Riello},
  {Seabroke}, {Tanga}, {Th{\'e}venin}, {Gracia-Abril}, {Comoretto},
  {Garcia-Reinaldos}, {Teyssier}, {Altmann}, {Andrae}, {Audard},
  {Bellas-Velidis}, {Benson}, {Berthier}, {Blomme}, {Burgess}, {Busso},
  {Carry}, {Cellino}, {Clementini}, {Clotet}, {Creevey}, {Davidson}, {De
  Ridder}, {Delchambre}, {Dell'Oro}, {Ducourant},
  {Fern{\'a}ndez-Hern{\'a}ndez}, {Fouesneau}, {Fr{\'e}mat}, {Galluccio},
  {Garc{\'\i}a-Torres}, {Gonz{\'a}lez-N{\'u}{\~n}ez}, {Gonz{\'a}lez-Vidal},
  {Gosset}, {Guy}, {Halbwachs}, {Hambly}, {Harrison}, {Hern{\'a}ndez},
  {Hestroffer}, {Hodgkin}, {Hutton}, {Jasniewicz}, {Jean-Antoine-Piccolo},
  {Jordan}, {Korn}, {Krone-Martins}, {Lanzafame}, {Lebzelter}, {L{\"o}ffler},
  {Manteiga}, {Marrese}, {Mart{\'\i}n-Fleitas}, {Moitinho}, {Mora}, {Muinonen},
  {Osinde}, {Pancino}, {Pauwels}, {Petit}, {Recio-Blanco}, {Richards},
  {Rimoldini}, {Robin}, {Sarro}, {Siopis}, {Smith}, {Sozzetti}, {S{\"u}veges},
  {Torra}, {van Reeven}, {Abbas}, {Abreu Aramburu}, {Accart}, {Aerts},
  {Altavilla}, {{\'A}lvarez}, {Alvarez}, {Alves}, {Anderson}, {Andrei},
  {Anglada Varela}, {Antiche}, {Antoja}, {Arcay}, {Astraatmadja}, {Bach},
  {Baker}, {Balaguer-N{\'u}{\~n}ez}, {Balm}, {Barache}, {Barata}, {Barbato},
  {Barblan}, {Barklem}, {Barrado}, {Barros}, {Barstow}, {Bartholom{\'e}
  Mu{\~n}oz}, {Bassilana}, {Becciani}, {Bellazzini}, {Berihuete}, {Bertone},
  {Bianchi}, {Bienaym{\'e}}, {Blanco-Cuaresma}, {Boch}, {Boeche}, {Bombrun},
  {Borrachero}, {Bossini}, {Bouquillon}, {Bourda}, {Bragaglia}, {Bramante},
  {Breddels}, {Bressan}, {Brouillet}, {Br{\"u}semeister}, {Brugaletta},
  {Bucciarelli}, {Burlacu}, {Busonero}, {Butkevich}, {Buzzi}, {Caffau},
  {Cancelliere}, {Cannizzaro}, {Cantat-Gaudin}, {Carballo}, {Carlucci},
  {Carrasco}, {Casamiquela}, {Castellani}, {Castro-Ginard}, {Charlot},
  {Chemin}, {Chiavassa}, {Cocozza}, {Costigan}, {Cowell}, {Crifo}, {Crosta},
  {Crowley}, {Cuypers}, {Dafonte}, {Damerdji}, {Dapergolas}, {David}, {David},
  {de Laverny}, {De Luise}, {De March}, {de Martino}, {de Souza}, {de Torres},
  {Debosscher}, {del Pozo}, {Delbo}, {Delgado}, {Delgado}, {Di Matteo},
  {Diakite}, {Diener}, {Distefano}, {Dolding}, {Drazinos}, {Dur{\'a}n},
  {Edvardsson}, {Enke}, {Eriksson}, {Esquej}, {Eynard Bontemps}, {Fabre},
  {Fabrizio}, {Faigler}, {Falc{\~a}o}, {Farr{\`a}s Casas}, {Federici},
  {Fedorets}, {Fernique}, {Figueras}, {Filippi}, {Findeisen}, {Fonti},
  {Fraile}, {Fraser}, {Fr{\'e}zouls}, {Gai}, {Galleti}, {Garabato},
  {Garc{\'\i}a-Sedano}, {Garofalo}, {Garralda}, {Gavel}, {Gavras}, {Gerssen},
  {Geyer}, {Giacobbe}, {Gilmore}, {Girona}, {Giuffrida}, {Glass}, {Gomes},
  {Granvik}, {Gueguen}, {Guerrier}, {Guiraud}, {Guti{\'e}rrez-S{\'a}nchez},
  {Haigron}, {Hatzidimitriou}, {Hauser}, {Haywood}, {Heiter}, {Helmi}, {Heu},
  {Hilger}, {Hobbs}, {Hofmann}, {Holland}, {Huckle}, {Hypki}, {Icardi},
  {Jan{\ss}en}, {Jevardat de Fombelle}, {Jonker}, {Juh{\'a}sz}, {Julbe},
  {Karampelas}, {Kewley}, {Klar}, {Kochoska}, {Kohley}, {Kolenberg},
  {Kontizas}, {Kontizas}, {Koposov}, {Kordopatis}, {Kostrzewa-Rutkowska},
  {Koubsky}, {Lambert}, {Lanza}, {Lasne}, {Lavigne}, {Le Fustec}, {Le
  Poncin-Lafitte}, {Lebreton}, {Leccia}, {Leclerc}, {Lecoeur-Taibi},
  {Lenhardt}, {Leroux}, {Liao}, {Licata}, {Lindstr{\o}m}, {Lister}, {Livanou},
  {Lobel}, {L{\'o}pez}, {Managau}, {Mann}, {Mantelet}, {Marchal}, {Marchant},
  {Marconi}, {Marinoni}, {Marschalk{\'o}}, {Marshall}, {Martino}, {Marton},
  {Mary}, {Massari}, {Matijevi{\v{c}}}, {Mazeh}, {McMillan}, {Messina},
  {Michalik}, {Millar}, {Molina}, {Molinaro}, {Moln{\'a}r}, {Montegriffo},
  {Mor}, {Morbidelli}, {Morel}, {Morris}, {Mulone}, {Muraveva}, {Musella},
  {Nelemans}, {Nicastro}, {Noval}, {O'Mullane}, {Ord{\'e}novic},
  {Ord{\'o}{\~n}ez-Blanco}, {Osborne}, {Pagani}, {Pagano}, {Pailler},
  {Palacin}, {Palaversa}, {Panahi}, {Pawlak}, {Piersimoni}, {Pineau}, {Plachy},
  {Plum}, {Poggio}, {Poujoulet}, {Pr{\v{s}}a}, {Pulone}, {Racero}, {Ragaini},
  {Rambaux}, {Ramos-Lerate}, {Regibo}, {Reyl{\'e}}, {Riclet}, {Ripepi}, {Riva},
  {Rivard}, {Rixon}, {Roegiers}, {Roelens}, {Romero-G{\'o}mez}, {Rowell},
  {Royer}, {Ruiz-Dern}, {Sadowski}, {Sagrist{\`a} Sell{\'e}s}, {Sahlmann},
  {Salgado}, {Salguero}, {Sanna}, {Santana-Ros}, {Sarasso}, {Savietto},
  {Schultheis}, {Sciacca}, {Segol}, {Segovia}, {S{\'e}gransan}, {Shih},
  {Siltala}, {Silva}, {Smart}, {Smith}, {Solano}, {Solitro}, {Sordo}, {Soria
  Nieto}, {Souchay}, {Spagna}, {Spoto}, {Stampa}, {Steele},
  {Steidelm{\"u}ller}, {Stephenson}, {Stoev}, {Suess}, {Surdej}, {Szabados},
  {Szegedi-Elek}, {Tapiador}, {Taris}, {Tauran}, {Taylor}, {Teixeira},
  {Terrett}, {Teyssandier}, {Thuillot}, {Titarenko}, {Torra Clotet}, {Turon},
  {Ulla}, {Utrilla}, {Uzzi}, {Vaillant}, {Valentini}, {Valette}, {van Elteren},
  {Van Hemelryck}, {van Leeuwen}, {Vaschetto}, {Vecchiato}, {Veljanoski},
  {Viala}, {Vicente}, {Vogt}, {von Essen}, {Voss}, {Votruba}, {Voutsinas},
  {Walmsley}, {Weiler}, {Wertz}, {Wevers}, {Wyrzykowski}, {Yoldas},
  {{\v{Z}}erjal}, {Ziaeepour}, {Zorec}, {Zschocke}, {Zucker}, {Zurbach}, \&
  {Zwitter}}]{GaiaDR2}
{Gaia Collaboration}, {Brown}, A.~G.~A., {Vallenari}, A., {et~al.} 2018,
  {\mhref{http://doi.org/10.1051/0004-6361/201833051}{\aap}},
  {\href{https://ui.adsabs.harvard.edu/abs/2018A&A...616A...1G}{616}}{\href{https://ui.adsabs.harvard.edu/abs/2018A&A...616A...1G}{,
  A1}}

\bibitem[{{Gonz{\'a}lez-Cuesta} {et~al.}(2023){Gonz{\'a}lez-Cuesta}, {Mathur},
  {Garc{\'\i}a}, {P{\'e}rez Hern{\'a}ndez}, {Delsanti}, {Breton}, {Hedges},
  {Jim{\'e}nez}, {Della Gaspera}, {El-Issami}, {Fox}, {Godoy-Rivera}, {Pitot},
  \& {Proust}}]{Gonzalez-Cuesta_2023}
{Gonz{\'a}lez-Cuesta}, L., {Mathur}, S., {Garc{\'\i}a}, R.~A., {et~al.} 2023,
  {\mhref{http://doi.org/10.1051/0004-6361/202244577}{\aap}},
  {\href{https://ui.adsabs.harvard.edu/abs/2023A&A...674A.106G}{674}}{\href{https://ui.adsabs.harvard.edu/abs/2023A&A...674A.106G}{,
  A106}}

\bibitem[{{Grevesse} \& {Sauval}(1998)}]{GS98}
{Grevesse}, N., \& {Sauval}, A.~J. 1998,
  {\mhref{http://doi.org/10.1023/A:1005161325181}{\ssr}},
  {\href{https://ui.adsabs.harvard.edu/abs/1998SSRv...85..161G}{85}}{\href{https://ui.adsabs.harvard.edu/abs/1998SSRv...85..161G}{,
  161}}

\bibitem[{{Hekker} \& {Christensen-Dalsgaard}(2017)}]{2017A&ARv..25....1H}
{Hekker}, S., \& {Christensen-Dalsgaard}, J. 2017,
  {\mhref{http://doi.org/10.1007/s00159-017-0101-x}{\aapr}},
  {\href{https://ui.adsabs.harvard.edu/abs/2017A&ARv..25....1H}{25}}{\href{https://ui.adsabs.harvard.edu/abs/2017A&ARv..25....1H}{,
  1}}

\bibitem[{{Higl} {et~al.}(2021){Higl}, {M{\"u}ller}, \& {Weiss}}]{Higl2021}
{Higl}, J., {M{\"u}ller}, E., \& {Weiss}, A. 2021,
  {\mhref{http://doi.org/10.1051/0004-6361/202039532}{\aap}},
  {\href{https://ui.adsabs.harvard.edu/abs/2021A&A...646A.133H}{646}}{\href{https://ui.adsabs.harvard.edu/abs/2021A&A...646A.133H}{,
  A133}}

\bibitem[{{Howell} {et~al.}(2014){Howell}, {Sobeck}, {Haas}, {Still},
  {Barclay}, {Mullally}, {Troeltzsch}, {Aigrain}, {Bryson}, {Caldwell},
  {Chaplin}, {Cochran}, {Huber}, {Marcy}, {Miglio}, {Najita}, {Smith},
  {Twicken}, \& {Fortney}}]{K2mission}
{Howell}, S.~B., {Sobeck}, C., {Haas}, M., {et~al.} 2014,
  {\mhref{http://doi.org/10.1086/676406}{\pasp}},
  {\href{https://ui.adsabs.harvard.edu/abs/2014PASP..126..398H}{126}}{\href{https://ui.adsabs.harvard.edu/abs/2014PASP..126..398H}{,
  398}}

\bibitem[{{Huber} {et~al.}(2019){Huber}, {Chaplin}, {Chontos}, {Kjeldsen},
  {Christensen-Dalsgaard}, {Bedding}, {Ball}, {Brahm}, {Espinoza}, {Henning},
  {Jord{\'a}n}, {Sarkis}, {Knudstrup}, {Albrecht}, {Grundahl}, {Fredslund
  Andersen}, {Pall{\'e}}, {Crossfield}, {Fulton}, {Howard}, {Isaacson},
  {Weiss}, {Handberg}, {Lund}, {Serenelli}, {R{\o}rsted Mosumgaard},
  {Stokholm}, {Bieryla}, {Buchhave}, {Latham}, {Quinn}, {Gaidos}, {Hirano},
  {Ricker}, {Vanderspek}, {Seager}, {Jenkins}, {Winn}, {Antia}, {Appourchaux},
  {Basu}, {Bell}, {Benomar}, {Bonanno}, {Buzasi}, {Campante}, {{\c{C}}elik
  Orhan}, {Corsaro}, {Cunha}, {Davies}, {Deheuvels}, {Grunblatt}, {Hasanzadeh},
  {Di Mauro}, {Garc{\'\i}a}, {Gaulme}, {Girardi}, {Guzik}, {Hon}, {Jiang},
  {Kallinger}, {Kawaler}, {Kuszlewicz}, {Lebreton}, {Li}, {Lucas}, {Lundkvist},
  {Mann}, {Mathis}, {Mathur}, {Mazumdar}, {Metcalfe}, {Miglio}, {Monteiro},
  {Mosser}, {Noll}, {Nsamba}, {Ong}, {{\"O}rtel}, {Pereira}, {Ranadive},
  {R{\'e}gulo}, {Rodrigues}, {Roxburgh}, {Silva Aguirre}, {Smalley},
  {Schofield}, {Sousa}, {Stassun}, {Stello}, {Tayar}, {White}, {Verma},
  {Vrard}, {Y{\i}ld{\i}z}, {Baker}, {Bazot}, {Beichmann}, {Bergmann}, {Bugnet},
  {Cale}, {Carlino}, {Cartwright}, {Christiansen}, {Ciardi}, {Creevey},
  {Dittmann}, {Do Nascimento}, {Van Eylen}, {F{\"u}r{\'e}sz}, {Gagn{\'e}},
  {Gao}, {Gazeas}, {Giddens}, {Hall}, {Hekker}, {Ireland}, {Latouf}, {LeBrun},
  {Levine}, {Matzko}, {Natinsky}, {Page}, {Plavchan}, {Mansouri-Samani},
  {McCauliff}, {Mullally}, {Orenstein}, {Garcia Soto}, {Paegert}, {van Saders},
  {Schnaible}, {Soderblom}, {Szab{\'o}}, {Tanner}, {Tinney}, {Teske}, {Thomas},
  {Trampedach}, {Wright}, {Yuan}, \& {Zohrabi}}]{Huber2019}
{Huber}, D., {Chaplin}, W.~J., {Chontos}, A., {et~al.} 2019,
  {\mhref{http://doi.org/10.3847/1538-3881/ab1488}{\aj}},
  {\href{https://ui.adsabs.harvard.edu/abs/2019AJ....157..245H}{157}}{\href{https://ui.adsabs.harvard.edu/abs/2019AJ....157..245H}{,
  245}}

\bibitem[{{Jermyn} {et~al.}(2023){Jermyn}, {Bauer}, {Schwab}, {Farmer}, {Ball},
  {Bellinger}, {Dotter}, {Joyce}, {Marchant}, {Mombarg}, {Wolf}, {Sunny Wong},
  {Cinquegrana}, {Farrell}, {Smolec}, {Thoul}, {Cantiello}, {Herwig}, {Toloza},
  {Bildsten}, {Townsend}, \& {Timmes}}]{Jermyn2023}
{Jermyn}, A.~S., {Bauer}, E.~B., {Schwab}, J., {et~al.} 2023,
  {\mhref{http://doi.org/10.3847/1538-4365/acae8d}{\apjs}},
  {\href{https://ui.adsabs.harvard.edu/abs/2023ApJS..265...15J}{265}}{\href{https://ui.adsabs.harvard.edu/abs/2023ApJS..265...15J}{,
  15}}

\bibitem[{{J{\o}rgensen} {et~al.}(2020){J{\o}rgensen}, {Montalb{\'a}n},
  {Miglio}, {Rendle}, {Davies}, {Buldgen}, {Scuflaire}, {Noels}, {Gaulme}, \&
  {Garc{\'\i}a}}]{Jorgensen2020}
{J{\o}rgensen}, A. C.~S., {Montalb{\'a}n}, J., {Miglio}, A., {et~al.} 2020,
  {\mhref{http://doi.org/10.1093/mnras/staa1480}{\mnras}},
  {\href{https://ui.adsabs.harvard.edu/abs/2020MNRAS.495.4965J}{495}}{\href{https://ui.adsabs.harvard.edu/abs/2020MNRAS.495.4965J}{,
  4965}}

\bibitem[{{Lebreton} \& {Goupil}(2014)}]{Lebreton2014}
{Lebreton}, Y., \& {Goupil}, M.~J. 2014,
  {\mhref{http://doi.org/10.1051/0004-6361/201423797}{\aap}},
  {\href{https://ui.adsabs.harvard.edu/abs/2014A&A...569A..21L}{569}}{\href{https://ui.adsabs.harvard.edu/abs/2014A&A...569A..21L}{,
  A21}}

\bibitem[{{Li} {et~al.}(2020{\natexlab{a}}){Li}, {Bedding},
  {Christensen-Dalsgaard}, {Stello}, {Li}, \& {Keen}}]{LiT2020}
{Li}, T., {Bedding}, T.~R., {Christensen-Dalsgaard}, J., {et~al.}
  2020{\natexlab{a}}, {\mhref{http://doi.org/10.1093/mnras/staa1350}{\mnras}},
  {\href{https://ui.adsabs.harvard.edu/abs/2020MNRAS.495.3431L}{495}}{\href{https://ui.adsabs.harvard.edu/abs/2020MNRAS.495.3431L}{,
  3431}}

\bibitem[{{Li} {et~al.}(2020{\natexlab{b}}){Li}, {Bedding}, {Li}, {Bi},
  {Stello}, {Zhou}, \& {White}}]{LiY2020}
{Li}, Y., {Bedding}, T.~R., {Li}, T., {et~al.} 2020{\natexlab{b}},
  {\mhref{http://doi.org/10.1093/mnras/staa1335}{\mnras}},
  {\href{https://ui.adsabs.harvard.edu/abs/2020MNRAS.495.2363L}{495}}{\href{https://ui.adsabs.harvard.edu/abs/2020MNRAS.495.2363L}{,
  2363}}

\bibitem[{{Lindsay} {et~al.}(2022){Lindsay}, {Ong}, \& {Basu}}]{Lindsay2022}
{Lindsay}, C.~J., {Ong}, J.~M.~J., \& {Basu}, S. 2022,
  {\mhref{http://doi.org/10.3847/1538-4357/ac67ed}{\apj}},
  {\href{https://ui.adsabs.harvard.edu/abs/2022ApJ...931..116L}{931}}{\href{https://ui.adsabs.harvard.edu/abs/2022ApJ...931..116L}{,
  116}}

\bibitem[{{Lindsay} {et~al.}(2023){Lindsay}, {Ong}, \& {Basu}}]{Lindsay2023}
---. 2023, {\mhref{http://doi.org/10.3847/1538-4357/acccf5}{\apj}},
  {\href{https://ui.adsabs.harvard.edu/abs/2023ApJ...950...19L}{950}}{\href{https://ui.adsabs.harvard.edu/abs/2023ApJ...950...19L}{,
  19}}

\bibitem[{{Lund} {et~al.}(2017){Lund}, {Silva Aguirre}, {Davies}, {Chaplin},
  {Christensen-Dalsgaard}, {Houdek}, {White}, {Bedding}, {Ball}, {Huber},
  {Antia}, {Lebreton}, {Latham}, {Handberg}, {Verma}, {Basu}, {Casagrande},
  {Justesen}, {Kjeldsen}, \& {Mosumgaard}}]{Lund2017}
{Lund}, M.~N., {Silva Aguirre}, V., {Davies}, G.~R., {et~al.} 2017,
  {\mhref{http://doi.org/10.3847/1538-4357/835/2/172}{\apj}},
  {\href{https://ui.adsabs.harvard.edu/abs/2017ApJ...835..172L}{835}}{\href{https://ui.adsabs.harvard.edu/abs/2017ApJ...835..172L}{,
  172}}

\bibitem[{{Maeder} \& {Mermilliod}(1981)}]{Maeder1981}
{Maeder}, A., \& {Mermilliod}, J.~C. 1981, \aap,
  {\href{https://ui.adsabs.harvard.edu/abs/1981A&A....93..136M}{93}}{\href{https://ui.adsabs.harvard.edu/abs/1981A&A....93..136M}{,
  136}}

\bibitem[{{McKeever} {et~al.}(2019){McKeever}, {Basu}, \&
  {Corsaro}}]{McKeever2019}
{McKeever}, J.~M., {Basu}, S., \& {Corsaro}, E. 2019,
  {\mhref{http://doi.org/10.3847/1538-4357/ab0c04}{\apj}},
  {\href{https://ui.adsabs.harvard.edu/abs/2019ApJ...874..180M}{874}}{\href{https://ui.adsabs.harvard.edu/abs/2019ApJ...874..180M}{,
  180}}

\bibitem[{{Mombarg} {et~al.}(2021){Mombarg}, {Van Reeth}, \&
  {Aerts}}]{Mombarg2021}
{Mombarg}, J.~S.~G., {Van Reeth}, T., \& {Aerts}, C. 2021,
  {\mhref{http://doi.org/10.1051/0004-6361/202039543}{\aap}},
  {\href{https://ui.adsabs.harvard.edu/abs/2021A&A...650A..58M}{650}}{\href{https://ui.adsabs.harvard.edu/abs/2021A&A...650A..58M}{,
  A58}}

\bibitem[{{Mombarg} {et~al.}(2019){Mombarg}, {Van Reeth}, {Pedersen},
  {Molenberghs}, {Bowman}, {Johnston}, {Tkachenko}, \& {Aerts}}]{Mombarg2019}
{Mombarg}, J.~S.~G., {Van Reeth}, T., {Pedersen}, M.~G., {et~al.} 2019,
  {\mhref{http://doi.org/10.1093/mnras/stz501}{\mnras}},
  {\href{https://ui.adsabs.harvard.edu/abs/2019MNRAS.485.3248M}{485}}{\href{https://ui.adsabs.harvard.edu/abs/2019MNRAS.485.3248M}{,
  3248}}

\bibitem[{{Noll} \& {Deheuvels}(2023)}]{NollDeheuvels2023}
{Noll}, A., \& {Deheuvels}, S. 2023,
  {\mhref{http://doi.org/10.1051/0004-6361/202245710}{\aap}},
  {\href{https://ui.adsabs.harvard.edu/abs/2023A&A...676A..70N}{676}}{\href{https://ui.adsabs.harvard.edu/abs/2023A&A...676A..70N}{,
  A70}}

\bibitem[{{Noll} {et~al.}(2021){Noll}, {Deheuvels}, \& {Ballot}}]{Noll2021}
{Noll}, A., {Deheuvels}, S., \& {Ballot}, J. 2021,
  {\mhref{http://doi.org/10.1051/0004-6361/202040055}{\aap}},
  {\href{https://ui.adsabs.harvard.edu/abs/2021A&A...647A.187N}{647}}{\href{https://ui.adsabs.harvard.edu/abs/2021A&A...647A.187N}{,
  A187}}

\bibitem[{{Nsamba} {et~al.}(2021){Nsamba}, {Moedas}, {Campante}, {Cunha},
  {Hern{\'a}ndez}, {Su{\'a}rez}, {Monteiro}, {Fernandes}, {Jiang}, \&
  {Akinsanmi}}]{Nsamba2021}
{Nsamba}, B., {Moedas}, N., {Campante}, T.~L., {et~al.} 2021,
  {\mhref{http://doi.org/10.1093/mnras/staa3228}{\mnras}},
  {\href{https://ui.adsabs.harvard.edu/abs/2021MNRAS.500...54N}{500}}{\href{https://ui.adsabs.harvard.edu/abs/2021MNRAS.500...54N}{,
  54}}

\bibitem[{{Ong} \& {Basu}(2020)}]{OngBasu2020}
{Ong}, J.~M.~J., \& {Basu}, S. 2020,
  {\mhref{http://doi.org/10.3847/1538-4357/ab9ffb}{\apj}},
  {\href{https://ui.adsabs.harvard.edu/abs/2020ApJ...898..127O}{898}}{\href{https://ui.adsabs.harvard.edu/abs/2020ApJ...898..127O}{,
  127}}

\bibitem[{{Ong} {et~al.}(2021{\natexlab{a}}){Ong}, {Basu}, {Lund}, {Bieryla},
  {Viani}, \& {Latham}}]{Ong2021c}
{Ong}, J.~M.~J., {Basu}, S., {Lund}, M.~N., {et~al.} 2021{\natexlab{a}},
  {\mhref{http://doi.org/10.3847/1538-4357/ac1e8b}{\apj}},
  {\href{https://ui.adsabs.harvard.edu/abs/2021ApJ...922...18O}{922}}{\href{https://ui.adsabs.harvard.edu/abs/2021ApJ...922...18O}{,
  18}}

\bibitem[{{Ong} {et~al.}(2021{\natexlab{b}}){Ong}, {Basu}, \&
  {McKeever}}]{Ong2021a}
{Ong}, J.~M.~J., {Basu}, S., \& {McKeever}, J.~M. 2021{\natexlab{b}},
  {\mhref{http://doi.org/10.3847/1538-4357/abc7c1}{\apj}},
  {\href{https://ui.adsabs.harvard.edu/abs/2021ApJ...906...54O}{906}}{\href{https://ui.adsabs.harvard.edu/abs/2021ApJ...906...54O}{,
  54}}

\bibitem[{{Ong} {et~al.}(2021{\natexlab{c}}){Ong}, {Basu}, \&
  {Roxburgh}}]{Ong2021b}
{Ong}, J.~M.~J., {Basu}, S., \& {Roxburgh}, I.~W. 2021{\natexlab{c}},
  {\mhref{http://doi.org/10.3847/1538-4357/ac12ca}{\apj}},
  {\href{https://ui.adsabs.harvard.edu/abs/2021ApJ...920....8O}{920}}{\href{https://ui.adsabs.harvard.edu/abs/2021ApJ...920....8O}{,
  8}}

\bibitem[{{Paxton} {et~al.}(2011){Paxton}, {Bildsten}, {Dotter}, {Herwig},
  {Lesaffre}, \& {Timmes}}]{Paxton2011}
{Paxton}, B., {Bildsten}, L., {Dotter}, A., {et~al.} 2011,
  {\mhref{http://doi.org/10.1088/0067-0049/192/1/3}{\apjs}},
  {\href{https://ui.adsabs.harvard.edu/abs/2011ApJS..192....3P}{192}}{\href{https://ui.adsabs.harvard.edu/abs/2011ApJS..192....3P}{,
  3}}

\bibitem[{{Paxton} {et~al.}(2013){Paxton}, {Cantiello}, {Arras}, {Bildsten},
  {Brown}, {Dotter}, {Mankovich}, {Montgomery}, {Stello}, {Timmes}, \&
  {Townsend}}]{Paxton2013}
{Paxton}, B., {Cantiello}, M., {Arras}, P., {et~al.} 2013,
  {\mhref{http://doi.org/10.1088/0067-0049/208/1/4}{\apjs}},
  {\href{https://ui.adsabs.harvard.edu/abs/2013ApJS..208....4P}{208}}{\href{https://ui.adsabs.harvard.edu/abs/2013ApJS..208....4P}{,
  4}}

\bibitem[{{Paxton} {et~al.}(2015){Paxton}, {Marchant}, {Schwab}, {Bauer},
  {Bildsten}, {Cantiello}, {Dessart}, {Farmer}, {Hu}, {Langer}, {Townsend},
  {Townsley}, \& {Timmes}}]{Paxton2015}
{Paxton}, B., {Marchant}, P., {Schwab}, J., {et~al.} 2015,
  {\mhref{http://doi.org/10.1088/0067-0049/220/1/15}{\apjs}},
  {\href{https://ui.adsabs.harvard.edu/abs/2015ApJS..220...15P}{220}}{\href{https://ui.adsabs.harvard.edu/abs/2015ApJS..220...15P}{,
  15}}

\bibitem[{{Paxton} {et~al.}(2018){Paxton}, {Schwab}, {Bauer}, {Bildsten},
  {Blinnikov}, {Duffell}, {Farmer}, {Goldberg}, {Marchant}, {Sorokina},
  {Thoul}, {Townsend}, \& {Timmes}}]{Paxton2018}
{Paxton}, B., {Schwab}, J., {Bauer}, E.~B., {et~al.} 2018,
  {\mhref{http://doi.org/10.3847/1538-4365/aaa5a8}{\apjs}},
  {\href{https://ui.adsabs.harvard.edu/abs/2018ApJS..234...34P}{234}}{\href{https://ui.adsabs.harvard.edu/abs/2018ApJS..234...34P}{,
  34}}

\bibitem[{{Paxton} {et~al.}(2019){Paxton}, {Smolec}, {Schwab}, {Gautschy},
  {Bildsten}, {Cantiello}, {Dotter}, {Farmer}, {Goldberg}, {Jermyn}, {Kanbur},
  {Marchant}, {Thoul}, {Townsend}, {Wolf}, {Zhang}, \& {Timmes}}]{Paxton2019}
{Paxton}, B., {Smolec}, R., {Schwab}, J., {et~al.} 2019,
  {\mhref{http://doi.org/10.3847/1538-4365/ab2241}{\apjs}},
  {\href{https://ui.adsabs.harvard.edu/abs/2019ApJS..243...10P}{243}}{\href{https://ui.adsabs.harvard.edu/abs/2019ApJS..243...10P}{,
  10}}

\bibitem[{{Pietrinferni} {et~al.}(2004){Pietrinferni}, {Cassisi}, {Salaris}, \&
  {Castelli}}]{Pietrinferni2004}
{Pietrinferni}, A., {Cassisi}, S., {Salaris}, M., \& {Castelli}, F. 2004,
  {\mhref{http://doi.org/10.1086/422498}{\apj}},
  {\href{https://ui.adsabs.harvard.edu/abs/2004ApJ...612..168P}{612}}{\href{https://ui.adsabs.harvard.edu/abs/2004ApJ...612..168P}{,
  168}}

\bibitem[{{Pols} {et~al.}(1997){Pols}, {Tout}, {Schroder}, {Eggleton}, \&
  {Manners}}]{Pols1997}
{Pols}, O.~R., {Tout}, C.~A., {Schroder}, K.-P., {Eggleton}, P.~P., \&
  {Manners}, J. 1997, {\mhref{http://doi.org/10.1093/mnras/289.4.869}{\mnras}},
  {\href{https://ui.adsabs.harvard.edu/abs/1997MNRAS.289..869P}{289}}{\href{https://ui.adsabs.harvard.edu/abs/1997MNRAS.289..869P}{,
  869}}

\bibitem[{{Reback} {et~al.}(2021){Reback}, {Mendel}, {McKinney}, {Van Den
  Bossche}, {Augspurger}, {Cloud}, {Hawkins}, {Gfyoung}, {Sinhrks}, {Roeschke},
  {Klein}, {Petersen}, {Tratner}, {She}, {Ayd}, {Hoefler}, {Naveh}, {Garcia},
  {Schendel}, {Hayden}, {Saxton}, {Gorelli}, {Shadrach}, {Jancauskas},
  {McMaster}, {Li}, {Battiston}, {Seabold}, {Attack68}, \& {Dong}}]{pandas}
{Reback}, J., {Mendel}, J.~B., {McKinney}, W., {et~al.} 2021,
  {pandas-dev/pandas: Pandas 1.3.0}, v1.3.0,  Zenodo,
  \dodoi{10.5281/zenodo.3509134}

\bibitem[{{Ribas} {et~al.}(2000){Ribas}, {Jordi}, \& {Gim{\'e}nez}}]{Ribas2000}
{Ribas}, I., {Jordi}, C., \& {Gim{\'e}nez}, {\'A}. 2000,
  {\mhref{http://doi.org/10.1046/j.1365-8711.2000.04035.x}{\mnras}},
  {\href{https://ui.adsabs.harvard.edu/abs/2000MNRAS.318L..55R}{318}}{\href{https://ui.adsabs.harvard.edu/abs/2000MNRAS.318L..55R}{,
  L55}}

\bibitem[{{Ricker} {et~al.}(2015){Ricker}, {Winn}, {Vanderspek}, {Latham},
  {Bakos}, {Bean}, {Berta-Thompson}, {Brown}, {Buchhave}, {Butler}, {Butler},
  {Chaplin}, {Charbonneau}, {Christensen-Dalsgaard}, {Clampin}, {Deming},
  {Doty}, {De Lee}, {Dressing}, {Dunham}, {Endl}, {Fressin}, {Ge}, {Henning},
  {Holman}, {Howard}, {Ida}, {Jenkins}, {Jernigan}, {Johnson}, {Kaltenegger},
  {Kawai}, {Kjeldsen}, {Laughlin}, {Levine}, {Lin}, {Lissauer}, {MacQueen},
  {Marcy}, {McCullough}, {Morton}, {Narita}, {Paegert}, {Palle}, {Pepe},
  {Pepper}, {Quirrenbach}, {Rinehart}, {Sasselov}, {Sato}, {Seager},
  {Sozzetti}, {Stassun}, {Sullivan}, {Szentgyorgyi}, {Torres}, {Udry}, \&
  {Villasenor}}]{TESS_inst}
{Ricker}, G.~R., {Winn}, J.~N., {Vanderspek}, R., {et~al.} 2015,
  {\mhref{http://doi.org/10.1117/1.JATIS.1.1.014003}{Journal of Astronomical
  Telescopes, Instruments, and Systems}},
  {\href{https://ui.adsabs.harvard.edu/abs/2015JATIS...1a4003R}{1}}{\href{https://ui.adsabs.harvard.edu/abs/2015JATIS...1a4003R}{,
  014003}}

\bibitem[{{Rosenfield} {et~al.}(2017){Rosenfield}, {Girardi}, {Williams},
  {Johnson}, {Dolphin}, {Bressan}, {Weisz}, {Dalcanton}, {Fouesneau}, \&
  {Kalirai}}]{Rosenfield2017}
{Rosenfield}, P., {Girardi}, L., {Williams}, B.~F., {et~al.} 2017,
  {\mhref{http://doi.org/10.3847/1538-4357/aa70a2}{\apj}},
  {\href{https://ui.adsabs.harvard.edu/abs/2017ApJ...841...69R}{841}}{\href{https://ui.adsabs.harvard.edu/abs/2017ApJ...841...69R}{,
  69}}

\bibitem[{{Schroder} {et~al.}(1997){Schroder}, {Pols}, \&
  {Eggleton}}]{Schroder1997}
{Schroder}, K.-P., {Pols}, O.~R., \& {Eggleton}, P.~P. 1997,
  {\mhref{http://doi.org/10.1093/mnras/285.4.696}{\mnras}},
  {\href{https://ui.adsabs.harvard.edu/abs/1997MNRAS.285..696S}{285}}{\href{https://ui.adsabs.harvard.edu/abs/1997MNRAS.285..696S}{,
  696}}

\bibitem[{{Scuflaire}(1974)}]{Scuflaire1974}
{Scuflaire}, R. 1974, \aap,
  {\href{https://ui.adsabs.harvard.edu/abs/1974A&A....36..107S}{36}}{\href{https://ui.adsabs.harvard.edu/abs/1974A&A....36..107S}{,
  107}}

\bibitem[{{Serenelli} {et~al.}(2017){Serenelli}, {Johnson}, {Huber},
  {Pinsonneault}, {Ball}, {Tayar}, {Silva Aguirre}, {Basu}, {Troup}, {Hekker},
  {Kallinger}, {Stello}, {Davies}, {Lund}, {Mathur}, {Mosser}, {Stassun},
  {Chaplin}, {Elsworth}, {Garc{\'\i}a}, {Handberg}, {Holtzman}, {Hearty},
  {Garc{\'\i}a-Hern{\'a}ndez}, {Gaulme}, \& {Zamora}}]{Serenelli2017}
{Serenelli}, A., {Johnson}, J., {Huber}, D., {et~al.} 2017,
  {\mhref{http://doi.org/10.3847/1538-4365/aa97df}{\apjs}},
  {\href{https://ui.adsabs.harvard.edu/abs/2017ApJS..233...23S}{233}}{\href{https://ui.adsabs.harvard.edu/abs/2017ApJS..233...23S}{,
  23}}

\bibitem[{{Silva Aguirre} {et~al.}(2017){Silva Aguirre}, {Lund}, {Antia},
  {Ball}, {Basu}, {Christensen-Dalsgaard}, {Lebreton}, {Reese}, {Verma},
  {Casagrande}, {Justesen}, {Mosumgaard}, {Chaplin}, {Bedding}, {Davies},
  {Handberg}, {Houdek}, {Huber}, {Kjeldsen}, {Latham}, {White}, {Coelho},
  {Miglio}, \& {Rendle}}]{SilvaAguirre2017}
{Silva Aguirre}, V., {Lund}, M.~N., {Antia}, H.~M., {et~al.} 2017,
  {\mhref{http://doi.org/10.3847/1538-4357/835/2/173}{\apj}},
  {\href{https://ui.adsabs.harvard.edu/abs/2017ApJ...835..173S}{835}}{\href{https://ui.adsabs.harvard.edu/abs/2017ApJ...835..173S}{,
  173}}

\bibitem[{{Steigman}(2010)}]{Steigman2010}
{Steigman}, G. 2010,
  {\mhref{http://doi.org/10.1088/1475-7516/2010/04/029}{\jcap}},
  {\href{https://ui.adsabs.harvard.edu/abs/2010JCAP...04..029S}{2010}}{\href{https://ui.adsabs.harvard.edu/abs/2010JCAP...04..029S}{,
  029}}

\bibitem[{{Stello} {et~al.}(2022){Stello}, {Saunders}, {Grunblatt}, {Hon},
  {Reyes}, {Huber}, {Bedding}, {Elsworth}, {Garc{\'\i}a}, {Hekker},
  {Kallinger}, {Mathur}, {Mosser}, \& {Pinsonneault}}]{Stello2022}
{Stello}, D., {Saunders}, N., {Grunblatt}, S., {et~al.} 2022,
  {\mhref{http://doi.org/10.1093/mnras/stac414}{\mnras}},
  {\href{https://ui.adsabs.harvard.edu/abs/2022MNRAS.512.1677S}{512}}{\href{https://ui.adsabs.harvard.edu/abs/2022MNRAS.512.1677S}{,
  1677}}

\bibitem[{{Thoul} {et~al.}(1994){Thoul}, {Bahcall}, \& {Loeb}}]{Thoul1994}
{Thoul}, A.~A., {Bahcall}, J.~N., \& {Loeb}, A. 1994,
  {\mhref{http://doi.org/10.1086/173695}{\apj}},
  {\href{https://ui.adsabs.harvard.edu/abs/1994ApJ...421..828T}{421}}{\href{https://ui.adsabs.harvard.edu/abs/1994ApJ...421..828T}{,
  828}}

\bibitem[{{Townsend} \& {Teitler}(2013)}]{Townsend2013}
{Townsend}, R.~H.~D., \& {Teitler}, S.~A. 2013,
  {\mhref{http://doi.org/10.1093/mnras/stt1533}{\mnras}},
  {\href{https://ui.adsabs.harvard.edu/abs/2013MNRAS.435.3406T}{435}}{\href{https://ui.adsabs.harvard.edu/abs/2013MNRAS.435.3406T}{,
  3406}}

\bibitem[{{VandenBerg} {et~al.}(2006){VandenBerg}, {Bergbusch}, \&
  {Dowler}}]{VandenBerg2006}
{VandenBerg}, D.~A., {Bergbusch}, P.~A., \& {Dowler}, P.~D. 2006,
  {\mhref{http://doi.org/10.1086/498451}{\apjs}},
  {\href{https://ui.adsabs.harvard.edu/abs/2006ApJS..162..375V}{162}}{\href{https://ui.adsabs.harvard.edu/abs/2006ApJS..162..375V}{,
  375}}

\bibitem[{{Viani} \& {Basu}(2020)}]{Viani2020}
{Viani}, L.~S., \& {Basu}, S. 2020,
  {\mhref{http://doi.org/10.3847/1538-4357/abba17}{\apj}},
  {\href{https://ui.adsabs.harvard.edu/abs/2020ApJ...904...22V}{904}}{\href{https://ui.adsabs.harvard.edu/abs/2020ApJ...904...22V}{,
  22}}

\bibitem[{{Viani} {et~al.}(2018){Viani}, {Basu}, {Ong J.}, {Bonaca}, \&
  {Chaplin}}]{Viani_2018}
{Viani}, L.~S., {Basu}, S., {Ong J.}, M.~J., {Bonaca}, A., \& {Chaplin}, W.~J.
  2018, {\mhref{http://doi.org/10.3847/1538-4357/aab7eb}{\apj}},
  {\href{https://ui.adsabs.harvard.edu/abs/2018ApJ...858...28V}{858}}{\href{https://ui.adsabs.harvard.edu/abs/2018ApJ...858...28V}{,
  28}}

\bibitem[{{Virtanen} {et~al.}(2020){Virtanen}, {Gommers}, {Oliphant},
  {Haberland}, {Reddy}, {Cournapeau}, {Burovski}, {Peterson}, {Weckesser},
  {Bright}, {van der Walt}, {Brett}, {Wilson}, {Millman}, {Mayorov}, {Nelson},
  {Jones}, {Kern}, {Larson}, {Carey}, {Polat}, {Feng}, {Moore}, {VanderPlas},
  {Laxalde}, {Perktold}, {Cimrman}, {Henriksen}, {Quintero}, {Harris},
  {Archibald}, {Ribeiro}, {Pedregosa}, {van Mulbregt}, \& {SciPy 1. 0
  Contributors}}]{scipy}
{Virtanen}, P., {Gommers}, R., {Oliphant}, T.~E., {et~al.} 2020,
  {\mhref{http://doi.org/10.1038/s41592-019-0686-2}{Nature Methods}},
  {\href{https://ui.adsabs.harvard.edu/abs/2020NatMe..17..261V}{17}}{\href{https://ui.adsabs.harvard.edu/abs/2020NatMe..17..261V}{,
  261}}

\bibitem[{{Yu} {et~al.}(2018){Yu}, {Huber}, {Bedding}, {Stello}, {Hon},
  {Murphy}, \& {Khanna}}]{Yu2018_16000}
{Yu}, J., {Huber}, D., {Bedding}, T.~R., {et~al.} 2018,
  {\mhref{http://doi.org/10.3847/1538-4365/aaaf74}{\apjs}},
  {\href{https://ui.adsabs.harvard.edu/abs/2018ApJS..236...42Y}{236}}{\href{https://ui.adsabs.harvard.edu/abs/2018ApJS..236...42Y}{,
  42}}

\end{thebibliography}

\begin{table*}[ht!]
\caption{Modelling results for the \edit{1}{stars} in our sample observed by the \textit{Kepler} and K2 missions.}
\label{table:Kepler}
\centering
\begin{tabular}{lccccccccc}
\toprule
Target & M [$M_{\odot}$] & R [$R_{\odot}$] & L [$L_{\odot}$] & T$_{\text{eff}}$ [K] & Age [Gyr] &[FeH]$_0$    & Y$_0$ & $\alpha_{\text{mlt}}$    & $\alpha_{\text{ov, eff}}$ \\
\hline
KIC2991448    & $1.03^{+0.04}_{-0.03}$ & $1.72^{+0.02}_{-0.02}$ & $\phantom{1}$$2.69^{+0.17}_{-0.15}$ & $5639^{+82}_{-77}$ & $\phantom{1}$$8.53^{+0.62}_{-0.62}$ & $-0.11^{+0.09}_{-0.09}$              & $0.278^{+0.002}_{-0.002}$ & $1.752^{+0.034}_{-0.029}$ & $0.041^{+0.015}_{-0.041}$ \\
KIC3852594    & $1.18^{+0.05}_{-0.04}$ & $2.00^{+0.03}_{-0.03}$ & $\phantom{1}$$5.44^{+0.26}_{-0.25}$ & $6237^{+55}_{-62}$ & $\phantom{1}$$4.29^{+0.31}_{-0.32}$ & $-0.31^{+0.10}_{-0.09}$              & $0.276^{+0.002}_{-0.002}$ & $1.934^{+0.027}_{-0.034}$ & $0.067^{+0.002}_{-0.005}$ \\
KIC4346201    & $1.20^{+0.06}_{-0.06}$ & $1.92^{+0.03}_{-0.03}$ & $\phantom{1}$$4.60^{+0.30}_{-0.29}$ & $6101^{+76}_{-78}$ & $\phantom{1}$$4.72^{+0.54}_{-0.45}$ & $-0.17^{+0.12}_{-0.11}$              & $0.273^{+0.002}_{-0.002}$ & $1.817^{+0.043}_{-0.041}$ & $0.057^{+0.006}_{-0.006}$ \\
KIC5108214    & $1.45^{+0.05}_{-0.06}$ & $2.54^{+0.03}_{-0.04}$ & $\phantom{1}$$6.78^{+0.37}_{-0.36}$ & $5855^{+68}_{-69}$ & $\phantom{1}$$3.27^{+0.20}_{-0.17}$ & $\phantom{-}$$0.10^{+0.08}_{-0.10}$  & $0.274^{+0.002}_{-0.003}$ & $1.774^{+0.006}_{-0.004}$ & $0.089^{+0.008}_{-0.007}$ \\
KIC5607242    & $1.21^{+0.07}_{-0.07}$ & $2.40^{+0.05}_{-0.05}$ & $\phantom{1}$$4.66^{+0.36}_{-0.35}$ & $5483^{+81}_{-85}$ & $\phantom{1}$$5.14^{+0.65}_{-0.54}$ & $-0.15^{+0.12}_{-0.12}$              & $0.276^{+0.002}_{-0.002}$ & $1.792^{+0.006}_{-0.004}$ & $0.117^{+0.010}_{-0.016}$ \\
KIC5689820    & $1.10^{+0.05}_{-0.04}$ & $2.29^{+0.03}_{-0.03}$ & $\phantom{1}$$2.95^{+0.19}_{-0.17}$ & $5004^{+66}_{-59}$ & $\phantom{1}$$8.69^{+0.65}_{-0.75}$ & $\phantom{-}$$0.04^{+0.07}_{-0.08}$  & $0.278^{+0.002}_{-0.002}$ & $1.789^{+0.030}_{-0.040}$ & $0.095^{+0.015}_{-0.016}$ \\
KIC5955122    & $1.16^{+0.06}_{-0.05}$ & $2.08^{+0.04}_{-0.03}$ & $\phantom{1}$$4.58^{+0.30}_{-0.29}$ & $5863^{+75}_{-77}$ & $\phantom{1}$$5.28^{+0.51}_{-0.49}$ & $-0.21^{+0.12}_{-0.11}$              & $0.275^{+0.002}_{-0.002}$ & $1.786^{+0.020}_{-0.022}$ & $0.081^{+0.008}_{-0.011}$ \\
KIC6064910    & $1.29^{+0.06}_{-0.06}$ & $2.31^{+0.03}_{-0.04}$ & $\phantom{1}$$7.27^{+0.40}_{-0.41}$ & $6238^{+66}_{-66}$ & $\phantom{1}$$3.29^{+0.32}_{-0.26}$ & $-0.23^{+0.10}_{-0.10}$              & $0.283^{+0.002}_{-0.002}$ & $1.861^{+0.010}_{-0.013}$ & $0.092^{+0.002}_{-0.003}$ \\
KIC6370489    & $1.14^{+0.06}_{-0.06}$ & $2.00^{+0.04}_{-0.03}$ & $\phantom{1}$$5.06^{+0.32}_{-0.30}$ & $6123^{+66}_{-67}$ & $\phantom{1}$$4.77^{+0.51}_{-0.48}$ & $-0.32^{+0.11}_{-0.10}$              & $0.278^{+0.002}_{-0.002}$ & $1.858^{+0.025}_{-0.021}$ & $0.063^{+0.005}_{-0.010}$ \\
KIC6442183    & $0.98^{+0.03}_{-0.03}$ & $1.64^{+0.02}_{-0.02}$ & $\phantom{1}$$2.55^{+0.16}_{-0.15}$ & $5698^{+83}_{-86}$ & $\phantom{1}$$9.13^{+0.66}_{-0.67}$ & $-0.19^{+0.09}_{-0.10}$              & $0.278^{+0.003}_{-0.003}$ & $1.766^{+0.044}_{-0.044}$ & $0.000^{+0.000}_{-0.000}$ \\
KIC6693861    & $1.00^{+0.05}_{-0.04}$ & $2.03^{+0.03}_{-0.02}$ & $\phantom{1}$$3.63^{+0.24}_{-0.22}$ & $5602^{+77}_{-77}$ & $\phantom{1}$$7.61^{+0.67}_{-0.70}$ & $-0.42^{+0.11}_{-0.10}$              & $0.276^{+0.002}_{-0.002}$ & $1.765^{+0.013}_{-0.011}$ & $0.000^{+0.058}_{-0.000}$ \\
KIC6766513    & $1.31^{+0.04}_{-0.04}$ & $2.11^{+0.02}_{-0.02}$ & $\phantom{1}$$5.85^{+0.30}_{-0.29}$ & $6181^{+70}_{-70}$ & $\phantom{1}$$3.62^{+0.23}_{-0.21}$ & $-0.06^{+0.07}_{-0.07}$              & $0.275^{+0.002}_{-0.002}$ & $1.830^{+0.049}_{-0.045}$ & $0.062^{+0.003}_{-0.002}$ \\
KIC7174707    & $1.07^{+0.05}_{-0.04}$ & $2.07^{+0.03}_{-0.03}$ & $\phantom{1}$$2.77^{+0.21}_{-0.18}$ & $5176^{+79}_{-75}$ & $\phantom{1}$$8.79^{+0.72}_{-0.77}$ & $-0.02^{+0.08}_{-0.09}$              & $0.278^{+0.002}_{-0.002}$ & $1.776^{+0.021}_{-0.029}$ & $0.079^{+0.018}_{-0.018}$ \\
KIC7199397    & $1.32^{+0.07}_{-0.06}$ & $2.55^{+0.04}_{-0.04}$ & $\phantom{1}$$6.90^{+0.36}_{-0.36}$ & $5862^{+60}_{-63}$ & $\phantom{1}$$3.54^{+0.28}_{-0.25}$ & $-0.17^{+0.11}_{-0.11}$              & $0.277^{+0.003}_{-0.003}$ & $1.759^{+0.018}_{-0.012}$ & $0.119^{+0.005}_{-0.008}$ \\
KIC7668623    & $1.51^{+0.03}_{-0.04}$ & $2.37^{+0.02}_{-0.02}$ & $\phantom{1}$$7.82^{+0.32}_{-0.37}$ & $6279^{+59}_{-67}$ & $\phantom{1}$$2.77^{+0.13}_{-0.09}$ & $\phantom{-}$$0.12^{+0.06}_{-0.07}$  & $0.273^{+0.002}_{-0.002}$ & $1.890^{+0.023}_{-0.029}$ & $0.122^{+0.028}_{-0.027}$ \\
KIC7747078    & $1.06^{+0.05}_{-0.04}$ & $1.91^{+0.03}_{-0.03}$ & $\phantom{1}$$3.92^{+0.26}_{-0.24}$ & $5878^{+74}_{-75}$ & $\phantom{1}$$6.38^{+0.62}_{-0.61}$ & $-0.32^{+0.10}_{-0.10}$              & $0.278^{+0.002}_{-0.002}$ & $1.774^{+0.026}_{-0.019}$ & $0.053^{+0.015}_{-0.015}$ \\
KIC7976303    & $1.07^{+0.06}_{-0.05}$ & $1.97^{+0.04}_{-0.03}$ & $\phantom{1}$$4.70^{+0.29}_{-0.28}$ & $6053^{+67}_{-71}$ & $\phantom{1}$$5.51^{+0.57}_{-0.54}$ & $-0.44^{+0.11}_{-0.10}$              & $0.279^{+0.002}_{-0.002}$ & $1.850^{+0.031}_{-0.032}$ & $0.050^{+0.013}_{-0.014}$ \\
KIC8026226    & $1.48^{+0.06}_{-0.06}$ & $2.88^{+0.04}_{-0.04}$ & $10.92^{+0.41}_{-0.41}$             & $6187^{+56}_{-61}$ & $\phantom{1}$$2.33^{+0.13}_{-0.10}$ & $-0.18^{+0.11}_{-0.11}$              & $0.272^{+0.003}_{-0.002}$ & $1.802^{+0.008}_{-0.008}$ & $0.089^{+0.009}_{-0.005}$ \\
KIC8524425    & $1.07^{+0.04}_{-0.03}$ & $1.79^{+0.02}_{-0.02}$ & $\phantom{1}$$2.70^{+0.20}_{-0.18}$ & $5538^{+81}_{-79}$ & $\phantom{1}$$8.33^{+0.70}_{-0.67}$ & $\phantom{-}$$0.00^{+0.06}_{-0.07}$ & $0.279^{+0.002}_{-0.002}$ & $1.748^{+0.021}_{-0.015}$ & $0.063^{+0.012}_{-0.012}$ \\
KIC8702606    & $1.16^{+0.08}_{-0.07}$ & $2.41^{+0.06}_{-0.05}$ & $\phantom{1}$$4.75^{+0.35}_{-0.34}$ & $5497^{+75}_{-79}$ & $\phantom{1}$$5.29^{+0.68}_{-0.61}$ & $-0.26^{+0.14}_{-0.13}$              & $0.277^{+0.002}_{-0.002}$ & $1.777^{+0.010}_{-0.011}$ & $0.106^{+0.014}_{-0.021}$ \\
KIC8738809    & $1.36^{+0.05}_{-0.06}$ & $2.18^{+0.03}_{-0.03}$ & $\phantom{1}$$5.69^{+0.33}_{-0.35}$ & $6042^{+68}_{-76}$ & $\phantom{1}$$3.75^{+0.35}_{-0.28}$ & $\phantom{-}$$0.08^{+0.08}_{-0.10}$  & $0.277^{+0.002}_{-0.002}$ & $1.839^{+0.026}_{-0.023}$ & $0.072^{+0.008}_{-0.007}$ \\
KIC9512063    & $1.07^{+0.07}_{-0.05}$ & $2.01^{+0.04}_{-0.03}$ & $\phantom{1}$$4.14^{+0.28}_{-0.26}$ & $5805^{+77}_{-77}$ & $\phantom{1}$$6.23^{+0.64}_{-0.62}$ & $-0.33^{+0.13}_{-0.11}$              & $0.277^{+0.002}_{-0.002}$ & $1.782^{+0.032}_{-0.027}$ & $0.060^{+0.019}_{-0.019}$ \\
KIC10018963   & $1.17^{+0.06}_{-0.06}$ & $1.94^{+0.03}_{-0.03}$ & $\phantom{1}$$4.74^{+0.27}_{-0.27}$ & $6128^{+65}_{-68}$ & $\phantom{1}$$4.81^{+0.48}_{-0.45}$ & $-0.23^{+0.12}_{-0.11}$              & $0.276^{+0.002}_{-0.002}$ & $1.888^{+0.028}_{-0.030}$ & $0.064^{+0.004}_{-0.007}$ \\
KIC10147635   & $1.45^{+0.06}_{-0.07}$ & $2.70^{+0.04}_{-0.05}$ & $\phantom{1}$$8.17^{+0.41}_{-0.42}$ & $5947^{+63}_{-67}$ & $\phantom{1}$$2.93^{+0.20}_{-0.14}$ & $-0.04^{+0.11}_{-0.13}$              & $0.273^{+0.002}_{-0.002}$ & $1.791^{+0.021}_{-0.017}$ & $0.115^{+0.005}_{-0.005}$ \\
KIC10273246   & $1.40^{+0.04}_{-0.04}$ & $2.25^{+0.02}_{-0.02}$ & $\phantom{1}$$6.18^{+0.33}_{-0.32}$ & $6079^{+74}_{-73}$ & $\phantom{1}$$3.45^{+0.22}_{-0.21}$ & $\phantom{-}$$0.09^{+0.07}_{-0.07}$  & $0.275^{+0.002}_{-0.002}$ & $1.855^{+0.015}_{-0.018}$ & $0.069^{+0.007}_{-0.005}$ \\
KIC10593351   & $1.58^{+0.06}_{-0.08}$ & $3.12^{+0.04}_{-0.05}$ & $\phantom{1}$$9.71^{+0.41}_{-0.40}$ & $5775^{+59}_{-60}$ & $\phantom{1}$$2.52^{+0.11}_{-0.11}$ & $\phantom{-}$$0.04^{+0.11}_{-0.14}$  & $0.274^{+0.002}_{-0.002}$ & $1.763^{+0.010}_{-0.009}$ & $0.142^{+0.014}_{-0.014}$ \\
KIC10873176   & $1.15^{+0.05}_{-0.04}$ & $2.07^{+0.03}_{-0.03}$ & $\phantom{1}$$5.87^{+0.30}_{-0.25}$ & $6242^{+57}_{-55}$ & $\phantom{1}$$4.19^{+0.31}_{-0.35}$ & $-0.39^{+0.10}_{-0.08}$              & $0.286^{+0.003}_{-0.003}$ & $1.929^{+0.019}_{-0.023}$ & $0.068^{+0.004}_{-0.008}$ \\
KIC10920273   & $1.01^{+0.03}_{-0.02}$ & $1.79^{+0.02}_{-0.01}$ & $\phantom{1}$$2.48^{+0.17}_{-0.15}$ & $5419^{+79}_{-75}$ & $\phantom{1}$$9.68^{+0.55}_{-0.64}$ & $-0.10^{+0.06}_{-0.06}$              & $0.275^{+0.001}_{-0.001}$ & $1.700^{+0.047}_{-0.041}$ & $0.000^{+0.000}_{-0.000}$ \\
KIC10972873   & $1.05^{+0.05}_{-0.04}$ & $1.80^{+0.03}_{-0.02}$ & $\phantom{1}$$3.02^{+0.20}_{-0.19}$ & $5681^{+81}_{-84}$ & $\phantom{1}$$7.85^{+0.65}_{-0.64}$ & $-0.15^{+0.10}_{-0.10}$              & $0.276^{+0.002}_{-0.002}$ & $1.775^{+0.024}_{-0.022}$ & $0.047^{+0.011}_{-0.047}$ \\
KIC11026764   & $1.09^{+0.06}_{-0.05}$ & $2.01^{+0.04}_{-0.03}$ & $\phantom{1}$$3.42^{+0.25}_{-0.23}$ & $5538^{+76}_{-76}$ & $\phantom{1}$$7.05^{+0.70}_{-0.67}$ & $-0.13^{+0.11}_{-0.11}$              & $0.279^{+0.002}_{-0.002}$ & $1.771^{+0.009}_{-0.012}$ & $0.074^{+0.017}_{-0.018}$ \\
KIC11137075   & $0.99^{+0.03}_{-0.03}$ & $1.64^{+0.02}_{-0.02}$ & $\phantom{1}$$2.23^{+0.15}_{-0.14}$ & $5517^{+86}_{-84}$ & $10.22^{+0.68}_{-0.74}$             & $-0.07^{+0.08}_{-0.10}$              & $0.279^{+0.002}_{-0.003}$ & $1.703^{+0.042}_{-0.042}$ & $0.000^{+0.000}_{-0.000}$ \\
KIC11193681   & $1.35^{+0.05}_{-0.06}$ & $2.41^{+0.03}_{-0.04}$ & $\phantom{1}$$4.93^{+0.33}_{-0.32}$ & $5541^{+80}_{-73}$ & $\phantom{1}$$4.37^{+0.35}_{-0.28}$ & $\phantom{-}$$0.10^{+0.09}_{-0.11}$  & $0.273^{+0.002}_{-0.002}$ & $1.781^{+0.015}_{-0.016}$ & $0.116^{+0.004}_{-0.006}$ \\
KIC11395018   & $1.20^{+0.05}_{-0.05}$ & $2.15^{+0.03}_{-0.03}$ & $\phantom{1}$$4.32^{+0.32}_{-0.30}$ & $5686^{+83}_{-86}$ & $\phantom{1}$$5.37^{+0.50}_{-0.45}$ & $-0.07^{+0.10}_{-0.09}$              & $0.276^{+0.002}_{-0.003}$ & $1.816^{+0.026}_{-0.020}$ & $0.107^{+0.003}_{-0.005}$ \\
KIC11414712   & $1.14^{+0.11}_{-0.09}$ & $2.24^{+0.07}_{-0.06}$ & $\phantom{1}$$4.37^{+0.46}_{-0.40}$ & $5586^{+96}_{-92}$ & $\phantom{1}$$5.79^{+0.98}_{-0.88}$ & $-0.23^{+0.17}_{-0.16}$              & $0.277^{+0.001}_{-0.001}$ & $1.771^{+0.008}_{-0.005}$ & $0.086^{+0.018}_{-0.026}$ \\
KIC11771760   & $1.54^{+0.07}_{-0.07}$ & $3.03^{+0.05}_{-0.05}$ & $\phantom{1}$$9.80^{+0.41}_{-0.43}$ & $5871^{+60}_{-64}$ & $\phantom{1}$$2.46^{+0.12}_{-0.09}$ & $-0.05^{+0.12}_{-0.12}$              & $0.270^{+0.002}_{-0.002}$ & $1.765^{+0.012}_{-0.012}$ & $0.104^{+0.008}_{-0.005}$ \\
KIC12508433   & $1.20^{+0.06}_{-0.06}$ & $2.23^{+0.04}_{-0.04}$ & $\phantom{1}$$3.47^{+0.29}_{-0.26}$ & $5280^{+82}_{-77}$ & $\phantom{1}$$6.57^{+0.72}_{-0.66}$ & $\phantom{-}$$0.06^{+0.09}_{-0.10}$  & $0.276^{+0.002}_{-0.002}$ & $1.774^{+0.010}_{-0.010}$ & $0.118^{+0.008}_{-0.014}$ \\
EPIC212478598 & $1.00^{+0.03}_{-0.03}$ & $2.45^{+0.03}_{-0.03}$ & $\phantom{1}$$3.61^{+0.22}_{-0.21}$ & $5090^{+79}_{-82}$ & $\phantom{1}$$9.05^{+0.44}_{-0.53}$ & $-0.29^{+0.11}_{-0.11}$              & $0.275^{+0.003}_{-0.003}$ & $1.775^{+0.070}_{-0.076}$ & $0.000^{+0.057}_{-0.000}$ \\
EPIC212516207 & $1.16^{+0.05}_{-0.03}$ & $1.68^{+0.02}_{-0.02}$ & $\phantom{1}$$3.56^{+0.14}_{-0.14}$ & $6114^{+46}_{-55}$ & $\phantom{1}$$5.34^{+0.35}_{-0.38}$ & $-0.05^{+0.08}_{-0.07}$              & $0.276^{+0.002}_{-0.002}$ & $1.984^{+0.011}_{-0.019}$ & $0.071^{+0.005}_{-0.003}$ \\
EPIC212586030 & $1.14^{+0.06}_{-0.05}$ & $3.52^{+0.07}_{-0.06}$ & $\phantom{1}$$6.12^{+0.32}_{-0.31}$ & $4844^{+54}_{-52}$ & $\phantom{1}$$7.91^{+0.72}_{-0.83}$ & $-0.01^{+0.08}_{-0.10}$              & $0.278^{+0.002}_{-0.002}$ & $1.783^{+0.059}_{-0.066}$ & $0.109^{+0.015}_{-0.019}$ \\
EPIC212683142 & $1.26^{+0.06}_{-0.06}$ & $2.25^{+0.04}_{-0.03}$ & $\phantom{1}$$5.40^{+0.32}_{-0.33}$ & $5868^{+78}_{-77}$ & $\phantom{1}$$4.34^{+0.35}_{-0.31}$ & $-0.11^{+0.11}_{-0.10}$              & $0.273^{+0.002}_{-0.002}$ & $1.859^{+0.041}_{-0.036}$ & $0.093^{+0.009}_{-0.010}$ \\
EPIC246154489 & $1.13^{+0.07}_{-0.06}$ & $4.67^{+0.10}_{-0.09}$ & $12.22^{+0.45}_{-0.44}$             & $5000^{+44}_{-50}$ & $\phantom{1}$$5.89^{+0.72}_{-0.73}$ & $-0.39^{+0.10}_{-0.08}$              & $0.275^{+0.001}_{-0.001}$ & $1.802^{+0.052}_{-0.058}$ & $0.111^{+0.016}_{-0.020}$ \\
EPIC246184564 & $1.29^{+0.07}_{-0.07}$ & $5.55^{+0.11}_{-0.12}$ & $16.76^{+0.47}_{-0.46}$             & $4963^{+49}_{-44}$ & $\phantom{1}$$4.18^{+0.57}_{-0.49}$ & $-0.31^{+0.10}_{-0.11}$              & $0.274^{+0.002}_{-0.002}$ & $1.816^{+0.053}_{-0.058}$ & $0.139^{+0.007}_{-0.008}$ \\
EPIC246305274 & $1.17^{+0.11}_{-0.08}$ & $2.15^{+0.07}_{-0.05}$ & $\phantom{1}$$5.82^{+0.64}_{-0.51}$ & $6122^{+96}_{-91}$ & $\phantom{1}$$4.32^{+0.68}_{-0.68}$ & $-0.34^{+0.17}_{-0.14}$              & $0.281^{+0.001}_{-0.003}$ & $1.830^{+0.022}_{-0.028}$ & $0.064^{+0.007}_{-0.015}$ \\
EPIC246305350 & $1.30^{+0.05}_{-0.04}$ & $2.20^{+0.03}_{-0.02}$ & $\phantom{1}$$5.53^{+0.31}_{-0.32}$ & $5976^{+82}_{-89}$ & $\phantom{1}$$3.97^{+0.23}_{-0.22}$ & $-0.03^{+0.09}_{-0.07}$              & $0.275^{+0.004}_{-0.003}$ & $1.721^{+0.065}_{-0.062}$ & $0.078^{+0.001}_{-0.003}$ \\
\hline 
\end{tabular}
\end{table*}

\clearpage

\begin{table*}[ht!]
\caption{Modelling results for the \edit{1}{stars} in our sample observed by the TESS mission. }
\label{table:TESS}
\centering
\begin{tabular}{lccccccccc}
\toprule
Target & M [$M_{\odot}$] & R [$R_{\odot}$] & L [$L_{\odot}$] & T$_{\text{eff}}$ [K] & Age [Gyr] &[FeH]$_0$    & Y$_0$ & $\alpha_{\text{mlt}}$    & $\alpha_{\text{ov, eff}}$\\
\hline 
HD 38529      & $1.41^{+0.06}_{-0.08}$ & $2.68^{+0.07}_{-0.06}$ & $\phantom{1}$$6.07^{+0.39}_{-0.41}$ & $5535^{+80}_{-82}$ & $\phantom{1}$$3.51^{+0.36}_{-0.25}$ & $\phantom{-}$$0.05^{+0.11}_{-0.13}$  & $0.279^{+0.002}_{-0.002}$ & $1.795^{+0.012}_{-0.009}$ & $0.112^{+0.003}_{-0.005}$ \\
$\nu$ Ind         & $0.93^{+0.02}_{-0.02}$ & $2.97^{+0.03}_{-0.02}$ & $\phantom{1}$$6.16^{+0.28}_{-0.25}$ & $5277^{+48}_{-50}$ & $\phantom{1}$$8.05^{+0.50}_{-0.59}$ & $-0.80^{+0.05}_{-0.04}$              & $0.274^{+0.002}_{-0.002}$ & $1.771^{+0.055}_{-0.059}$ & $0.000^{+0.000}_{-0.000}$ \\
$\delta$ Eri         & $1.08^{+0.05}_{-0.04}$ & $2.30^{+0.04}_{-0.03}$ & $\phantom{1}$$2.87^{+0.21}_{-0.17}$ & $4966^{+75}_{-69}$ & $\phantom{1}$$9.02^{+0.61}_{-0.78}$ & $\phantom{-}$$0.01^{+0.09}_{-0.11}$  & $0.278^{+0.002}_{-0.003}$ & $1.722^{+0.054}_{-0.060}$ & $0.079^{+0.018}_{-0.018}$ \\
$\beta$ Hyi         & $1.05^{+0.06}_{-0.05}$ & $1.80^{+0.03}_{-0.03}$ & $\phantom{1}$$3.40^{+0.21}_{-0.20}$ & $5837^{+71}_{-74}$ & $\phantom{1}$$6.93^{+0.63}_{-0.63}$ & $-0.23^{+0.12}_{-0.11}$              & $0.279^{+0.002}_{-0.002}$ & $1.768^{+0.037}_{-0.032}$ & $0.045^{+0.013}_{-0.045}$ \\
$\eta$ Cep         & $1.02^{+0.08}_{-0.06}$ & $3.94^{+0.11}_{-0.09}$ & $\phantom{1}$$8.21^{+0.40}_{-0.38}$ & $4917^{+71}_{-65}$ & $\phantom{1}$$8.97^{+0.89}_{-1.18}$ & $-0.22^{+0.16}_{-0.14}$              & $0.279^{+0.004}_{-0.004}$ & $1.843^{+0.100}_{-0.128}$ & $0.056^{+0.035}_{-0.056}$ \\
TOI 197       & $1.11^{+0.07}_{-0.06}$ & $2.90^{+0.06}_{-0.06}$ & $\phantom{1}$$4.91^{+0.30}_{-0.27}$ & $5050^{+68}_{-67}$ & $\phantom{1}$$7.15^{+0.77}_{-0.89}$ & $-0.22^{+0.12}_{-0.12}$              & $0.276^{+0.002}_{-0.002}$ & $1.823^{+0.048}_{-0.058}$ & $0.070^{+0.013}_{-0.016}$ \\
TIC300088321  & $1.03^{+0.03}_{-0.03}$ & $1.66^{+0.02}_{-0.01}$ & $\phantom{1}$$2.39^{+0.16}_{-0.14}$ & $5571^{+81}_{-75}$ & $\phantom{1}$$9.11^{+0.66}_{-0.66}$ & $\phantom{-}$$0.00^{+0.06}_{-0.08}$ & $0.281^{+0.002}_{-0.002}$ & $1.730^{+0.036}_{-0.030}$ & $0.045^{+0.012}_{-0.045}$ \\
TIC29987134   & $1.22^{+0.09}_{-0.08}$ & $2.23^{+0.06}_{-0.06}$ & $\phantom{1}$$5.24^{+0.47}_{-0.44}$ & $5849^{+88}_{-90}$ & $\phantom{1}$$4.67^{+0.69}_{-0.57}$ & $-0.17^{+0.15}_{-0.15}$              & $0.277^{+0.001}_{-0.001}$ & $1.759^{+0.015}_{-0.011}$ & $0.079^{+0.006}_{-0.008}$ \\
TIC374858999  & $1.10^{+0.08}_{-0.06}$ & $2.88^{+0.06}_{-0.05}$ & $\phantom{1}$$5.41^{+0.35}_{-0.32}$ & $5192^{+74}_{-79}$ & $\phantom{1}$$6.30^{+0.79}_{-0.83}$ & $-0.41^{+0.13}_{-0.12}$              & $0.274^{+0.002}_{-0.002}$ & $1.806^{+0.030}_{-0.037}$ & $0.073^{+0.015}_{-0.016}$ \\
TIC349059821  & $1.05^{+0.06}_{-0.05}$ & $2.91^{+0.06}_{-0.04}$ & $\phantom{1}$$5.02^{+0.29}_{-0.27}$ & $5064^{+67}_{-67}$ & $\phantom{1}$$7.64^{+0.66}_{-0.79}$ & $-0.34^{+0.12}_{-0.12}$              & $0.275^{+0.002}_{-0.002}$ & $1.767^{+0.053}_{-0.061}$ & $0.064^{+0.019}_{-0.019}$ \\
TIC55270123   & $1.10^{+0.08}_{-0.06}$ & $3.05^{+0.07}_{-0.06}$ & $\phantom{1}$$5.44^{+0.34}_{-0.31}$ & $5048^{+72}_{-72}$ & $\phantom{1}$$7.09^{+0.80}_{-0.92}$ & $-0.25^{+0.13}_{-0.13}$              & $0.274^{+0.002}_{-0.002}$ & $1.825^{+0.047}_{-0.058}$ & $0.070^{+0.015}_{-0.017}$ \\
TIC167548586  & $1.20^{+0.08}_{-0.07}$ & $3.18^{+0.07}_{-0.07}$ & $\phantom{1}$$5.49^{+0.35}_{-0.32}$ & $4958^{+66}_{-62}$ & $\phantom{1}$$6.41^{+0.84}_{-0.89}$ & $-0.08^{+0.11}_{-0.12}$              & $0.275^{+0.002}_{-0.002}$ & $1.799^{+0.049}_{-0.057}$ & $0.102^{+0.010}_{-0.016}$ \\
TIC299899690  & $1.17^{+0.08}_{-0.07}$ & $3.42^{+0.08}_{-0.07}$ & $\phantom{1}$$7.30^{+0.57}_{-0.50}$ & $5139^{+73}_{-77}$ & $\phantom{1}$$5.74^{+0.74}_{-0.79}$ & $-0.35^{+0.12}_{-0.12}$              & $0.276^{+0.002}_{-0.002}$ & $1.843^{+0.041}_{-0.052}$ & $0.091^{+0.014}_{-0.015}$ \\
TIC350343922  & $1.09^{+0.06}_{-0.05}$ & $3.73^{+0.07}_{-0.06}$ & $\phantom{1}$$7.76^{+0.44}_{-0.40}$ & $4994^{+55}_{-58}$ & $\phantom{1}$$7.36^{+0.72}_{-0.84}$ & $-0.28^{+0.10}_{-0.08}$              & $0.278^{+0.002}_{-0.002}$ & $1.820^{+0.055}_{-0.064}$ & $0.086^{+0.019}_{-0.018}$ \\
TIC150442152  & $1.25^{+0.08}_{-0.08}$ & $4.10^{+0.09}_{-0.09}$ & $\phantom{1}$$9.00^{+0.43}_{-0.41}$ & $4944^{+54}_{-53}$ & $\phantom{1}$$5.59^{+0.83}_{-0.76}$ & $-0.09^{+0.11}_{-0.11}$              & $0.275^{+0.002}_{-0.002}$ & $1.867^{+0.045}_{-0.052}$ & $0.106^{+0.004}_{-0.008}$ \\
TIC150166759  & $1.09^{+0.06}_{-0.05}$ & $4.25^{+0.09}_{-0.07}$ & $\phantom{1}$$8.83^{+0.44}_{-0.42}$ & $4827^{+57}_{-54}$ & $\phantom{1}$$8.12^{+0.63}_{-0.75}$ & $-0.14^{+0.10}_{-0.11}$              & $0.278^{+0.002}_{-0.002}$ & $1.746^{+0.064}_{-0.069}$ & $0.093^{+0.019}_{-0.021}$ \\
TIC141757732  & $1.25^{+0.08}_{-0.07}$ & $5.36^{+0.11}_{-0.11}$ & $16.67^{+0.58}_{-0.56}$             & $5042^{+47}_{-49}$ & $\phantom{1}$$4.22^{+0.62}_{-0.54}$ & $-0.43^{+0.11}_{-0.11}$              & $0.275^{+0.001}_{-0.001}$ & $1.824^{+0.048}_{-0.052}$ & $0.133^{+0.008}_{-0.014}$ \\
TIC350335258  & $1.12^{+0.07}_{-0.06}$ & $5.77^{+0.13}_{-0.12}$ & $17.48^{+0.73}_{-0.73}$             & $4915^{+53}_{-50}$ & $\phantom{1}$$5.48^{+0.70}_{-0.67}$ & $-0.54^{+0.10}_{-0.10}$              & $0.274^{+0.001}_{-0.001}$ & $1.653^{+0.062}_{-0.059}$ & $0.124^{+0.022}_{-0.027}$
\\
\hline 
\end{tabular}
\end{table*}

\clearpage

\begin{table}[]
\caption{\edit{2}{$\sigma_{\nu\text{, eff}}$ values for all of the Kepler targets in our sample. A higher value of $\sigma_{\nu\text{, eff}}$ indicates that our model grid has a higher degree of effective undersampling for that target. } }
\label{table:sigma_nu_eff_kepler}
\centering
\begin{tabular}{lc}
\toprule
Target        & $\sigma_{\nu\text{, eff}}$ \\
\hline
KIC2991448    & 0.83                      \\
KIC3852594    & 1.15                      \\
KIC4346201    & 0.93                      \\
KIC5108214    & 1.52                      \\
KIC5607242    & 1.09                      \\
KIC5689820    & 1.01                      \\
KIC5955122    & 1.86                      \\
KIC6064910    & 1.67                      \\
KIC6370489    & 1.23                      \\
KIC6442183    & 1.78                      \\
KIC6693861    & 1.34                      \\
KIC6766513    & 1.55                      \\
KIC7174707    & 1.41                      \\
KIC7199397    & 1.89                      \\
KIC7668623    & 1.33                      \\
KIC7747078    & 2.17                      \\
KIC7976303    & 1.98                      \\
KIC8026226    & 2.54                      \\
KIC8524425    & 1.13                      \\
KIC8702606    & 6.08                      \\
KIC8738809    & 0.98                      \\
KIC9512063    & 1.17                      \\
KIC10018963   & 1.82                      \\
KIC10147635   & 1.66                      \\
KIC10273246   & 1.93                      \\
KIC10593351   & 1.13                      \\
KIC10873176   & 2.22                      \\
KIC10920273   & 0.39                      \\
KIC10972873   & 1.25                      \\
KIC11026764   & 2.29                      \\
KIC11137075   & 14.08                     \\
KIC11193681   & 1.89                      \\
KIC11395018   & 1.41                      \\
KIC11414712   & 2.26                      \\
KIC11771760   & 0.76                      \\
KIC12508433   & 1.74                      \\
EPIC212478598 & 0.70                       \\
EPIC212516207 & 1.74                      \\
EPIC212586030 & 1.23                      \\
EPIC212683142 & 0.88                      \\
EPIC246154489 & 0.34                      \\
EPIC246184564 & 0.24                      \\
EPIC246305274 & 1.43                      \\
EPIC246305350 & 0.36                      \\     
\hline 
\end{tabular}
\end{table}

\begin{table}[]
\caption{\edit{2}{$\sigma_{\nu\text{, eff}}$ values for all of the TESS targets in our sample. A higher value of $\sigma_{\nu\text{, eff}}$ indicates that our model grid has a higher degree of effective undersampling for that target.  } }
\label{table:sigma_nu_eff_tess}
\centering
\begin{tabular}{lc}
\toprule
Target        & $\sigma_{\nu\text{, eff}}$ \\
\hline
HD 38529      & 0.48                     \\
$\nu$ Ind       & 0.43                      \\
$\delta$ Eri         & 1.38                  \\
$\beta$ Hyi         & 1.82                     \\
$\eta$ Cep         & 0.12                      \\
TOI 197       & 1.06                      \\
TIC300088321  & 0.87                      \\
TIC29987134   & 3.89                      \\
TIC374858999  & 0.86                      \\
TIC349059821  & 1.53                      \\
TIC55270123   & 0.88                      \\
TIC167548586  & 0.95                      \\
TIC299899690  & 0.81                      \\
TIC350343922  & 1.77                      \\
TIC150442152  & 0.15                      \\
TIC150166759  & 0.37                      \\
TIC141757732  & 0.30                       \\
TIC350335258  & 0.26                     \\     
\hline 
\end{tabular}
\end{table}

\end{document}